\documentclass[11pt,a4paper,english,twosite]{report}

\usepackage{babel}

\usepackage[latin1]{inputenc}
\usepackage{amsmath,amssymb,amsfonts,textcomp}

\usepackage[top=3cm, bottom=2cm, left=2.5cm, right=2cm]{geometry}

\usepackage{graphicx}

\begin{document}
\selectlanguage{english}
\title{Magnetism from intermetallics and perovskite oxides}
\author{Richard Javier Caraballo Vivas}
\maketitle
\pagenumbering{roman}

\newpage

\vspace*{0.4\textheight}

\begin{flushleft}
 This is a version of the Thesis presented by RJCV at\\ 
 {\it Intituto de F\'isica} at {\it Universidade Federal Fluminense}, \\
 Niter\'oi-RJ Brazil at march 2017,\\
 on supervision of Dr. Mario de Souza Reis Jr. ({\it marior@if.uff.br}).\\
 For more information, you can contact the author by:
  \begin{verbatim}
  caraballorichard@gmail.com
  caraballorichard@cbpf.br
  caraballo@if.uff.br
  \end{verbatim}
\end{flushleft}

\newpage

\vspace*{0.75\textheight}
\begin{flushright}
  \emph{to Vanessa}
\end{flushright}

\newpage
\noindent{\LARGE\textbf{Acknowledgment}}

\begin{itemize}

\item I thank the Brazilian institutions CAPES, CNPQ and FAPERJ for the financial assistance in the development of this work.

\item I thank to Programa de P\'os-gradua\c c\~ao do Instituto de F\'isica da Universidade Federal Fluminense.

\item I thank for the access to laboratory:

	\begin{itemize}
	\item Laborat\'orio de Raios X da Universidade Federal Fluminense.
	\item Laborat\'orio Nacional de Luz Synchrotron.
	\item Laborat\'orio de espetroscopia M\"ossbauer da Universidade Federal Fluminense.
	\item Laborat\'orio de Materiais e Baixas Temperaturas da Universidade Estadual de Campinas.
	\end{itemize}	 
	
\item I thank to my supervisor Mario de Souza Reis Jr.

\item I thank the several collaborators that contributed to the elaboration of this work. 

\item I thank the components to experimental magnetism group of Intituto de Física da Universidade Federal Fluminense:
	\begin{itemize}
	\item Prof. Dr. Mario de Souza Reis Jr.
	\item Prof. Dr. Daniel Leandro Rocco.
	\item Prof. Dr. Dalber Ruben Sanchez Candela.
	\item Prof. Dr. St\'ephane Serge Yves Jér\^ome Soriano.
	\item Prof. Dr. Yutao Xing.
	\item Prof. Dr. Wallace de Castro Nunes.
	\end{itemize} 
	
\item I thank to all people for the help. Family, friends and others.

\end{itemize}
\newpage

\begin{center}
  \emph{\begin{large}Abstract\end{large}}\label{abstract}
\vspace{2pt}
\end{center}

\selectlanguage{english}
\noindent

The purpose of this thesis is to elaborate intermetallic alloys and perovskites oxides, in order to understand the magnetic properties of these samples. Specifically, the intermetallic alloys with boron YNi$_{4-x}$Co$_x$B samples and Co$_2$FeSi and Fe$_2$MnSi$_{1-x}$Ga$_x$ Heusler alloys were fabricated by arc melting furnace. The sol-gel method was implemented for the synthesis of perovskite oxides of cobalt Nd$_{0.5}$Sr$_{0.5}$CoO$_3$ and manganese La$_{0.6}$Sr$_{0.4}$MnO$_3$.

YNi$_{4-x}$Co$_x$B was studied in order to explore the magnetic anisotropy originated in $3d$ sublattice of these samples.  Here, we associate the anisotropy with the Co occupation, due to the non-magnetic nature of yttrium and the non-contribution to the anisotropy of Ni ions. From the occupation models for Co ions and magnetic measurements, we explained the magnetic behavior of these samples.

We develop investigations on structural, magnetic and half-metallic properties of Co$_2$FeSi and Fe$_2$MnSi$_{1-x}$Ga$_x$ Heusler alloys. The Co$_2$FeSi is a promising material for spintronic devices, due to its Curie temperature above 1000 K and magnetic moment close to 6$\mu_B$. However, these properties can be influenced by the atomic disorder. Through anomalous X-ray diffraction and M\"osbauer spectroscopy, we obtained the atomic disorder of the Co$_2$FeSi, and the density functional theory calculations provided information about the influence of the atomic disorder in their half-metallic properties. On the order hand,  we also explored the behavior of the magnetic and magnetocaloric properties with the increased of the valence electron number in the Fe$_2$MnSi$_{1-x}$Ga$_x$ Heusler alloys.

In addition, we explored the structural and magnetic properties of perovskite oxides of the $ABO_3$ type, mainly the  Nd$_{0.5}$Sr$_{0.5}$CoO$_3$ cobaltite and La$_{0.6}$Sr$_{0.4}$MnO$_3$ manganite. For Nd$_{0.5}$Sr$_{0.5}$CoO$_3$, we focus in establishing its spin state configuration, for the Co$^{3+}$ and Co$^{4+}$, finding for both intermediate states. Also, we clarified their intrigue magnetic order, obtaining a ferrimagnetic material. The La$_{0.6}$Sr$_{0.4}$MnO$_3$ nanoparticles were synthesized in order to explore the effect of size on their properties. We found that the reduction of the nanoparticles size tends to broaden the paramagnetic to ferromagnetic transition, as well as promoting magnetic hysteresis and causing a remarkable change to the magnetic saturation.

\par
\vspace{1em}

\newpage
\listoffigures

\newpage

\tableofcontents

\newpage

\thispagestyle{empty}
\pagenumbering{arabic}

	\chapter{Introduction}\label{intro}

Magnetism has been known for thousands of years. The manifestations in which it was formerly known are those corresponding to natural magnets or magnetic stones, such as magnetite (iron oxide). The ancient Greeks and Chinese were the first to have been known to use this mineral, which is the ability to attract other pieces of the same material and iron.

\begin{figure}
\centering
\includegraphics[width=15cm]{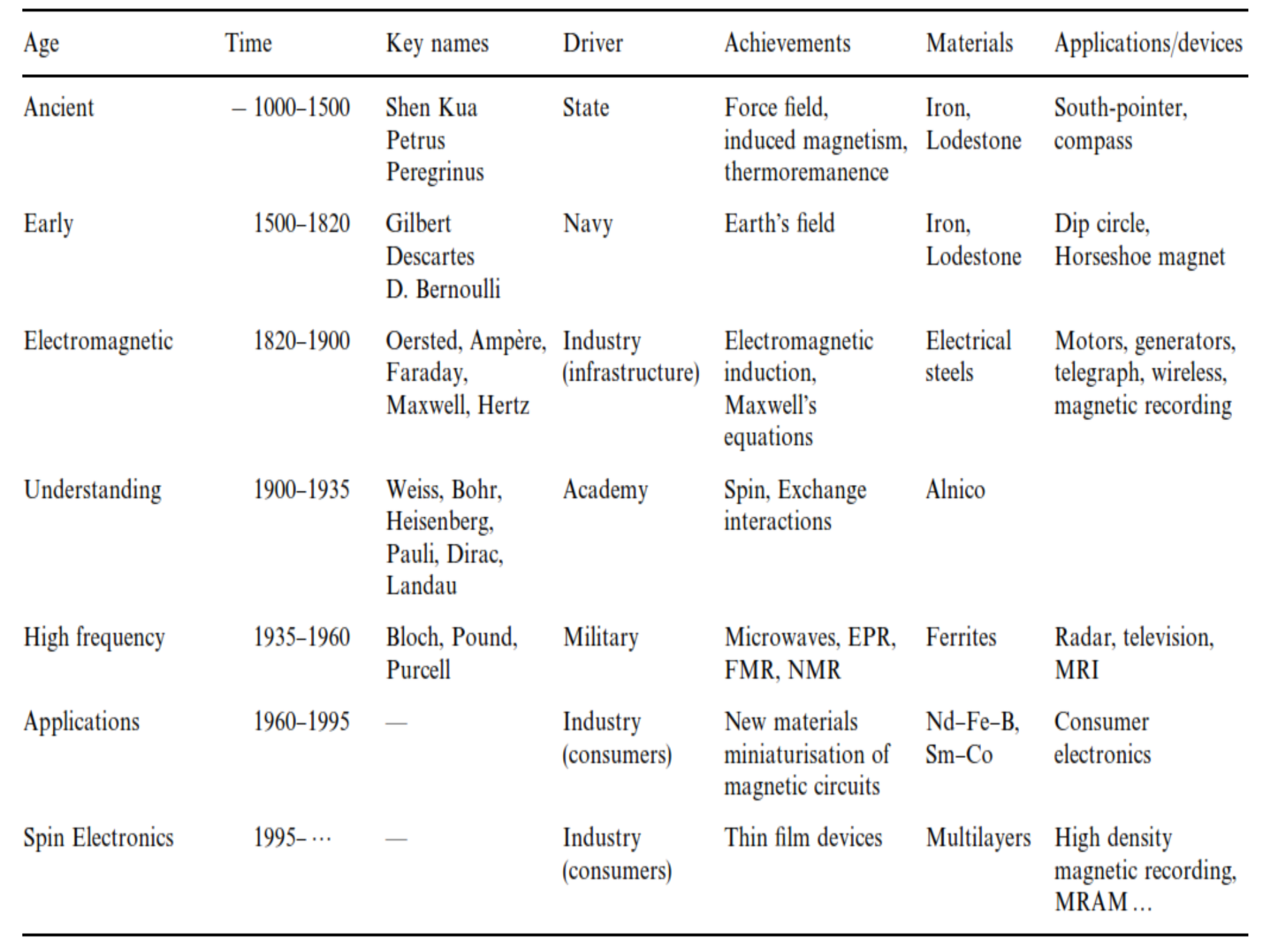}
\caption{Table for the seven ages of magnetism according to J. M. D. Coey \cite{coey2001magnetism}.}
\label{seven}
\end{figure}

Chronologically, in the 17th century, William Gilbert was the first to systematically investigate the phenomenon of magnetism using scientific methods. The first theoretical investigations were attributed to Carl Friedrich Gauss, while the first quantitative studies of magnetic phenomena were initiated in the 18th century by the French scientist, Charles Coulomb. The Danish physicist, Hans Christian Oersted, first suggested a link between electricity and magnetism, while the French scientist, Andre Marie Amp\`ere, and the Englishman, Michael Faraday, performed experiments involving the interactions of magnetic and electric fields with each other. In the 19th century, James Clerk Maxwell provided the theoretical foundation of the physics of electromagnetism. The modern understanding of magnetic phenomena in condensed matter originates from the work of Pierre and Marie Curie, who examined the effect of temperature on magnetic materials and observed that magnetism suddenly disappeared above a certain critical temperature in materials such as iron.  Pierre Weiss proposed a theory of magnetism based on an internal molecular field, proportional to the average magnetization that spontaneously aligns electronic micromagnets in magnetic matter. This is just to name a few historical examples. J. M. D. Coey \cite{coey2001magnetism} summarizes the history of magnetism in seven ages described in the table in Fig. \ref{seven}. In each of these ages it can be seen how the understanding and functionality of magnetism has evolved, mainly owing to the use and development of new technologies based on phenomena and materials that are, in turn, based on magnetism.

The greatest advance in technological development based on magnetism was in the 20th century with the manipulation of magnetic coercivity\cite{coey1999whither}, resulting in the combined control of magnetocrystalline anisotropy and microstructures. While, in the last years the  magnetic technological development is oriented mainly at  spintronics, which  is the study of the roles played by electron spin and the possible uses of their properties in order to develop devices in which it is not the electron charge but the electron spin that carries information. These devices, combining standard microelectronics with spin-dependent effects, arise from the interaction between the spin of the carrier and the magnetic properties of the material, as in the half-metallic ferromagnetic compounds \cite{hirohata2015roadmap,wolf}. The so-called half-metallic magnets have been proposed as good candidates for spintronic applications owing to their characteristic of exhibiting a hundred percent spin polarization at the Fermi level.

Then, the development of materials is important for the magnetic phenomena and knowledge of their magnetic mechanisms. Thus, we synthetized several intermetallic alloys and perovskites oxides, in order to explore their structural and magnetic properties, in addition were made studies magnetic phenomena  which provide information of these compounds for their use in technological devices. In Part \ref{parte1} of this thesis, we focus on basic magnetism , mainly describing several concepts to understand magnetic phenomena. The types of magnetic arrangement and their differences are explained in Chapter \ref{funda}. Furthermore, we approach the magnetocaloric effect and main phenomenology in Chapter \ref{fundaMCE}. Chapter \ref{capitulo4} is focused on sample preparation. 

Part \ref{parte2} of this Thesis, is focuses in the intermetallic compounds and are based our works in intermetallics with B (Chapter \ref{boratos} based on Ref. \cite{vivas2014competing}), and Heusler alloys in  Chapters \ref{capitulo6},\ref{cap7CFS} and \ref{capFMS} based on our investigation of Co$_2$FeSi and Fe$_2$MnSi$_{1-x}$Ga$_x$ \cite{pedro2015effects,vivas2016experimental}.

Among the most studied intermetallic compounds that present magnetic hardness are the Nd-Fe-B system \cite{kirchmayr1996permanent} and SmCo$_5$ \cite{ido1994new}, the latter being in the family of intermetallic compounds with B ($R_{n+1}M_{5+3n}$B$_{2n}$, with $R$ = rare earth, $M$ = $3d$ transition metal and $n=0, 1, 2, 3...\infty $). Magnetic anisotropy from the $3d$ sub-lattice is important for the magnetic hardness of these materials, since this strongly affects the shape of hysteresis loops which affect the design of most magnetic materials of industrial importance. Thus, we considered YNi$_{4-x}$Co$_x$B ($n=1$) alloys of the family of intermetallics with B (for $n=1$), and a non-magnetic rare-earth (yttrium) in order to be sure that the magnetic contributions are only due to the $3d$ sub-lattice, additionally Ni does not contribute to anisotropy. Thus, to explore these features, we developed a statistical and preferential model of Co occupation among the Wyckoff sites, in order to explore the $3d$ competition in the samples. We found that the preferential model agreed with the experimental measurements of the magnetization nature of $3d$ magnetic anisotropy. Thus, this result provides further knowledge in the area of hard magnets \cite{vivas2014competing}. It is explored in details in Chapter \ref{boratos}.
 
Chapter \ref{capitulo6} explores one of the most promising families of compounds with interesting materials for technological applications, the Heusler alloys. They have half-metallic properties, and such materials follow the Slater-Pauling rule that relates the magnetic moment to the valence electrons in the system. Generally, they are widely studied owing to their interesting structural and magnetic properties, such as magnetic shape memory ability, coupled magneto-structural phase transitions and half-metallicity. However, the atomic disorder produced by the interchange of X/Y atoms in Heusler alloys (X$_2$YZ) suppresses the half-metallic behavior of these compounds. For this,  in the Chapter \ref{cap7CFS}, we considered the Co$_2$-based Heusler alloys, mainly Co$_2$FeSi, which has the highest Curie temperature of these alloys at approximately 1100 K and a magnetic moment of approximately 6 $\mu_B$ \cite{inomata2006structural}.

The half-metallicity of the Heusler alloys has a close relationship with the valence electron number ($N_v$), which obeys the Slater-Pauling rule. Therefore, our approach seeks to understand the effect of $N_v$ on the magnetic properties of the Si-rich side of the half-metallic series, Fe$_2$MnSi$_{1-x}$Ga$_x$ Heusler alloys, such as the structural, magnetic, and finally magnetocaloric potentials \cite{pedro2015effects,vivas2016experimental}. These ideas are developments in the Chapter \ref{capFMS}.

In Part \ref{parte3}, we explored the peroviskite oxides. As with intermetallic alloys, perovskite ceramics ($AB$O$_3$) are compounds that have contributed much to the development of magnetism. Among the advantages they possess is the fact that they are oxidized and, therefore, can be prepared in the air without the undesirable influences of oxygen, in addition to being more economical compared to the intermetallic compounds. Finally, the perovskite oxides have interesting properties for technological developments based on magnetism, such as: spintronics \cite{ali2013robust,duan2015effects}, magnetic hyperthermia \cite{bubnovskaya2012nanohyperthermia}, and magnetic refrigeration \cite{wei2013review,phan2007review}, among others.

Cobaltite ($A$CoO$_3$) compounds show interesting magnetic and transport properties, owing to the strong relationship between the crystal structure and magnetism. However, the spin state of Co ions has an additional degree of freedom due to competition between Hund couple and crystal field splitting. Nd$_{0.5}$Sr$_{0.5}$CoO$_3$ has no well-established spin configuration, thus, we developed magnetization measurements in order to understand this aspect in detail. Consequently , we found that Co$^{3+}$ and Co$^{4+}$ are in an intermediate spin state and the Co and Nd magnetic sub-lattices couple antiferromagnetically below the Curie temperature $T_c$=215 K, down to very low temperatures \cite{reis2017spin}. In the Chapter \ref{cap11Coba} details of this study are given.

On the other hand, the structural and magnetic properties of $R_{1-x}T_{x}$MnO$_3$ manganites  are determined by lattice distortion on the unit cell, since the Mn-O-Mn bond is very sensitive to structural changes. Meanwhile, when the particle size is reduced to a few nanometers, the broadening of the paramagnetic to ferromagnetic transition, which decreases the saturation magnetization value, increases the magnetic hysteresis and appearance of superparamagnetic (SPM) behavior at very low particle sizes. Thus, we synthesized La$_{0.6}$Sr$_{0.4}$MnO$_3$ nanoparticles to explore the effect of size on the structural and magnetic properties of these samples. We found that the reduction of the nanoparticles size tends to broaden the paramagnetic to ferromagnetic transition, as well as promoting magnetic hysteresis and a remarkable change on the magnetic saturation \cite{andrade2014magnetic}. The Chapter \ref{manga10} is development these ideas. 

\vspace{2cm}

Finally, the main purpouse of this PhD thesis, is  understand the magnetic mechanism of the sample explained above. For this, we implemented several experimental techniques for the study of structural and magnetic properties. The possibles technological applications such as magnetic refrigeration and spintronics, also are explored. General conclusions of all this thesis are in the Chapter \ref{final}


\part{Background}\label{parte1}

	\chapter{Concepts of magnetism}\label{funda}

This chapter discusses the basic concepts of magnetism, which will be addressed during the rest of the text. Here, the main objective is to introduce the fundamental ideas about magnetism to the reader to provide clear understanding of this work. 

\section{Fundamental terms}

\subsection{Magnetic moment and magnetic dipole}

\begin{center}
\begin{figure}[h!]
\centering
\includegraphics[width=8cm]{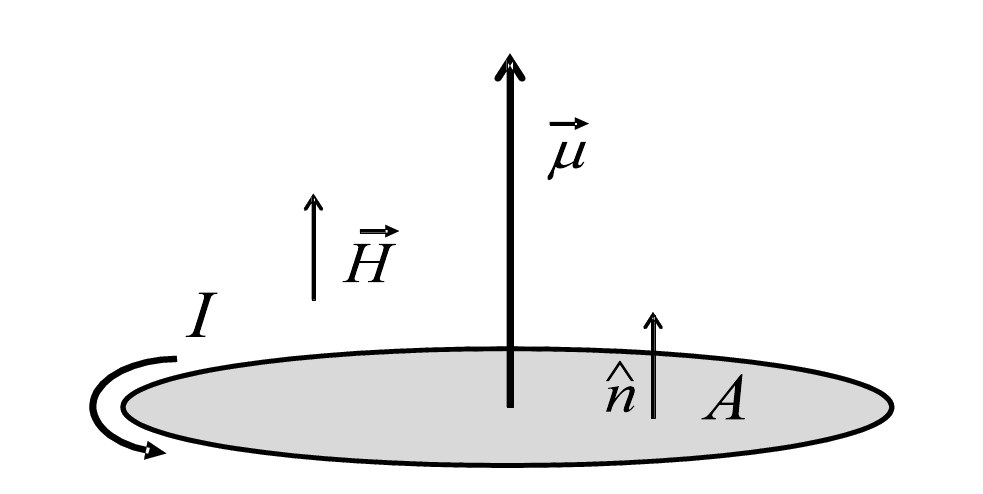}
\caption{Magnetic moment $\vec{\mu}$ due to the circulating current $I$ around an ring of area $A$. }
\label{mmoent}
\end{figure}
\end{center}

In classical electromagnetism the magnetic moment ($\vec{\mu}$) can be explained using the  Fig.\ref{mmoent}, where we assuming a   current around  an infinitely small loop with an area of $dA$ square meters (International System of Unit - SI). The corresponding magnetic moment $d\vec{\mu}$, is equal to \cite{getzlaff},
\begin{equation}
d\vec{\mu}= I dA \hat{n},
\end{equation}

\noindent where $I$ is the circulating current in amperes and $\hat{n}$ is an unitary vector normal to the ring area, which comes of the relation between $\vec{\mu}$ and $I$ by the right-hand corkscrew rule \cite{coeybook}. The magnetic moments of this small loop allows us to calculate the total $\vec{\mu}$, for a loop of a finite size:
\begin{equation}
\vec{\mu} = I \hat{n}\int dA = IA\hat{n}.
\end{equation}

The magnetic moment is measured in [Am$^2$] (SI), or [erg/G] (cgs). It is important to define that [erg/G] = [emu] ({\it electromagnetic unit}), because this quite common in the literature, mainly to express experimental results \cite{marioIntro}.

The magnetic dipole is equivalent to a magnetic moment of a current loop in the limit of a small area, but with a finite moment. The energy ($E$) of a magnetic moment is given by \cite{getzlaff}:
\begin{equation}
E = - \mu_0 \vec{\mu} \cdot \vec{H}= -\mu_0 \mu H \cos \theta,
\end{equation}

\noindent where $E$ is given by [J] ([erg]) for SI (cgs), $\theta$ being the angle between $\vec{\mu}$ and an external magnetic field $\vec{H}$ and $\mu_0 = 4 \pi \times 10^{-7}$ N $\cdot$ A$^{-2}$ (SI) is the magnetic permeability of free space \cite{getzlaff}.

\subsection{Magnetization}

The magnitude of the magnetization ($M$) can be defined as the total magnetic moment per unit volume \cite{marioIntro}:
\begin{equation}
M = \frac{\sum_i^N\mu_i}{V},
\end{equation}

\noindent where $M$ is given by [A/m] for SI or [emu/cm$^3$] for cgs. However, in the practice is more convenient to define the magnetization as the magnetic moment per mass.  Since, we do not need to know the sample volume, only its mass, which can be easily obtained \cite{buschow2003physics}. In this case the magnetization is given by  [A m$^2$ kg$^{-1}$] for SI or [emu/g] for cgs \cite{buschow2003physics}.

\subsection{Magnetic induction}

The magnetic response of a material when is applied an external magnetic field $\vec{H}$ is called the magnetic induction or magnetic flux density $\vec{B}$. The relationship between $\vec{H}$ and $\vec{B}$ is a characteristic property of the material itself. In the vacuum, we have a linear correlation between $\vec{B}$ and $\vec{H}$:

\begin{equation}
\vec{B} = \mu_0 \vec{H},
\end{equation}

\noindent considering SI. However, inside a magnetic material $\vec{B}$ and $\vec{H}$ may differ in magnitude and direction, due to of the magnetization $\vec{M}$ \cite{marioIntro}. Considering SI,

\begin{equation}
\vec{B}= \mu_0 (\vec{H}+\vec{M}).
\end{equation}

In the following sections, we refer to both as the magnetic field due to the common usage in the literature. In every situation, it can be understood, which term is meant.

\subsection{Magnetic susceptibility}

If we consider that the magnetization $\vec{M}$ is parallel to an external magnetic field $\vec{H}$:

\begin{equation}
\chi = \lim_{H \rightarrow 0} \frac{\partial M}{\partial H},
\end{equation} 

\noindent with $\chi$ being the magnetic susceptibility, for this case the material is considered a linear material. In this situation, a linear relationship between $B$ and $H$ remains:

\begin{equation}
B = \mu_0 (1 + \chi)H = \mu_m H,
\end{equation}

\noindent where $\mu_m$ represent the magnetic permeability of the material.

\section{Types of magnetic arrangement}

\subsection{Diamagnetism}

Diamagnetism is intrinsic to all materials, it is manifest when the electrons are under  an applied external magnetic field, then the precession around the nucleus changes the frequency to promote an extra magnetic field and shield the external one. Therefore, the diamagnetic susceptibility is negative \cite{marioIntro}

\begin{equation}
\chi^{\mbox{dia}}= \mbox{const.}<0
\end{equation}

A few examples of diamagnetic materials are:

\begin{itemize}
\item[$\circ$] Nearly all organic substances,
\item[$\circ$] Metals like Hg or noble metals like Cu,
\item[$\circ$] Superconductors below the critical temperature. These materials are ideal diamagnets.
\end{itemize}

\subsection{Paramagnetism}

Paramagnetism is a type of magnetism where there are no interactions between the magnetic moments. Thus, for creating an order of, is necessary the application of a magnetic field. Let us consider an ensemble of magnetic moments at a certain temperature, where they are directed randomly  and the magnetization is zero. The application of a magnetic field promotes a relative orientation of magnetic moments, increasing the value of magnetization. At high values of magnetization fields, all the magnetic moments are parallel to each other, and the magnetization reaches its maximum value (saturation $M_S$). For a certain value of the applied magnetic field, after decreasing temperature, this relative orientation of the magnetization and, at 0 K, the magnetization reaches its maximum value.  The magnetic susceptibility, obtained for low values of the applied magnetic field, is given by Curie law:

\begin{equation}
\chi^{\mbox{para}} = \frac{C}{T},
\end{equation}

\noindent where $C$ is the Curie constant, given by \cite{kittel},

\begin{equation}
C = \frac{N \mu_0 \mu_B^2 g^2 j(j+1)}{3k_B}
\end{equation}

\noindent where $N$ is the number of magnetic atoms per unit volume, $g$ is the Land\'e $g$-factor, $\mu _{B}$ is the Bohr magneton, $j$ is the total angular momentum and $k_{B}$ is Boltzmann constant. Then, the inverse magnetic susceptibility is a straight line, which pass through zero at 0 K.

\subsection*{Collective magnetism}

The other types of magnetic arrangements that we will discuss are the consequence of the interactions between the magnetic moments to obtain spontaneous magnetization, which results in collective magnetism (or cooperative systems). The  susceptibility of the collective magnetism  exhibits a functionality significantly more complicated compared to dia- and paramagnetism and consist in an interaction between permanent magnetic dipoles of the material \cite{getzlaff}.

For materials that present collective magnetism are characterized by a spontaneous magnetization bellow of the critical temperature $T_c$.  Collective magnetism is divided into several subclasses, such as ferromagnetism, antiferromagnetism, ferrimagnetism, among others.

\subsection{Ferromagnetism}

This fact allows that the system reaches the magnetization saturation value for magnetic field relatively small values  (those possible to be reached in a laboratory) depending on their magnetic anisotropy. Here, one magnetic moment depends on the neighbors to then creates the magnetic ordering; it is a long range interaction \cite{marioIntro}. 

Two parameters characterize the ferromagnetic ordering: ($i$) the critical temperature $T_C$ (Curie temperature), above which the system behaves like a paramagnetic system. Here, for a zero applied magnetic field, zero magnetization. Below $T_C$, the system has spontaneous magnetization, i.e., finite magnetization even without applied magnetic field. In the first approximation, $T_C$ is the measure of how strong is the interaction between magnetic moments. ($ii$) The saturation value of the magnetization $M_S$ (saturation magnetization) is analogously to the paramagnetic case \cite{marioIntro}. 

Then, we expect the magnetization as a function of temperature curve to be a finite value of magnetization that decreases by increasing the temperature, to a critical value $T_C$, above which there is no longer spontaneous magnetization. By concerning the magnetization as a function of the magnetic field, there are two situations: the first one is for temperatures above $T_C$. For this case, as mentioned, the system behaves like a paramagnetic specie, and then, we expect a curve similar to the paramagnetic case. For temperatures below $T_C$, it has a spontaneous magnetization (see figure \ref{magarr}).

The magnetic susceptibility is given by the Curie-Weiss law:

\begin{equation}
\chi = \frac{C}{T- \theta_p}
\end{equation}

\noindent where $\theta_p$ is the paramagnetic Curie temperature (for the mean field model, $\theta_p = T_C$). The inverse magnetic susceptibility is a straight line, and zero $1/\chi$ meets $T=\theta_p>0$ \cite{marioIntro}.

\subsection{Antiferromagnetism}

Antiferromagnetism is a cooperative ordering that can be understood by considering two magnetic sub-lattices: $M_A$ and $M_B$, of the same magnitude. Each one is ferromagnetic and behaves (approximately), according to the description mentioned above. The difference is that both sub-lattices are oriented in opposition, i.e., the total magnetization vanishes \cite{marioIntro}

\begin{equation}
M = M_A + M_B = 0.
\end{equation}

The paramagnetism like behavior appears above the critical temperature $T_N$ (Neel temperature), below which these sub-lattices are spontaneously ordered in opposition \cite{getzlaff}. This ordering is not a simple addition of two ferromagnetic sub-lattices, aligned in an antiparallel configuration, there is an interaction between these two sub-lattices, making this system a bit more complex that described here.

The magnetization as a function of the  magnetic field, at $T<<T_N$, one sub-lattice (say, $M_A$), is aligned with the external magnetic field and then does not change by increasing the field. The other sub-lattice ($M_B$), is opposite to the field and then will be flipped due to the increase of the magnetic field. For the case without external applied magnetic field, each sub-lattice has a ferromagnetic like dependence with temperature and the total magnetization is then zero. The magnetic susceptibility is also given by the Curie-Weiss law, however, $1/\chi$ is zero in the $T$ axis at $\theta_p < 0$ \cite{marioIntro}. This behavior is shown in Fig.\ref{magarr}.

\subsection{Ferrimagnetism}

Ferrimagnetism is quite similar to antiferromagnetic cooperative ordering, however, for the present case, the two sub-lattices have different values of the magnetization in opposition \cite{marioIntro},

\begin{equation}
|M_A|\neq|M_B|.
\end{equation}

The behavior of the magnetization as a function of the magnetic field (for low values of the temperature), is the same as before; however, the difference in the values of the magnetization of each sub-lattice, it is not zero at a zero magnetic field. Therefore, 

\begin{equation}
M = M_A + M_B \neq 0.
\end{equation}

Thus, these two sub-lattices are different, can cross themselves for a certain value of the temperature $T_{\mbox{comp}}$, and then promote a compensation, where the total magnetization is zero. The system loses the spontaneous ordering above $T_N$, analogously to the ferromagnetic case. Finally, the magnetic susceptibility only follows the Curie-Weiss law for a very high temperature. Close to $T_N$ the inverse magnetic susceptibility loses its linearity and assumes a hyperbolic-like behavior, with a downturn to zero $1/\chi$ \cite{marioIntro}.

\begin{center}
\begin{figure}[t!]
\centering
\includegraphics[width=16cm,height=15cm]{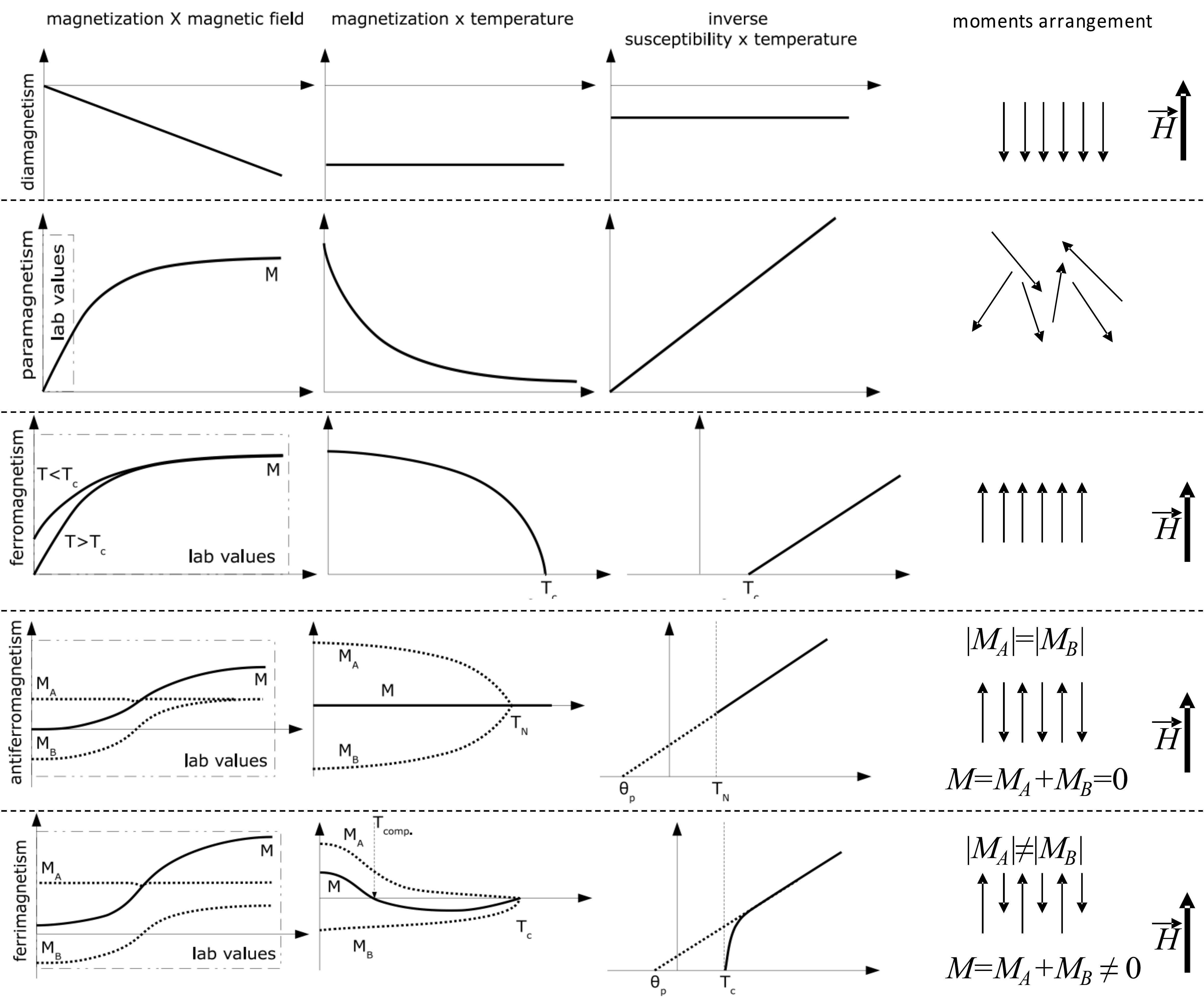}
\caption[Types of magnetic arrangement schemes]{Types of magnetic arrangement schemes. The graphs were extracted from Ref. \cite{marioIntro}.}
\label{magarr}
\end{figure}
\end{center}

Fig.\ref{magarr} shows the behaviors of all cooperative and non-cooperative systems mentioned above, from Ref. \cite{marioIntro}.

\section{Hysteresis cycles}

\begin{center}
\begin{figure}[h!]
\centering
\includegraphics[width=12cm]{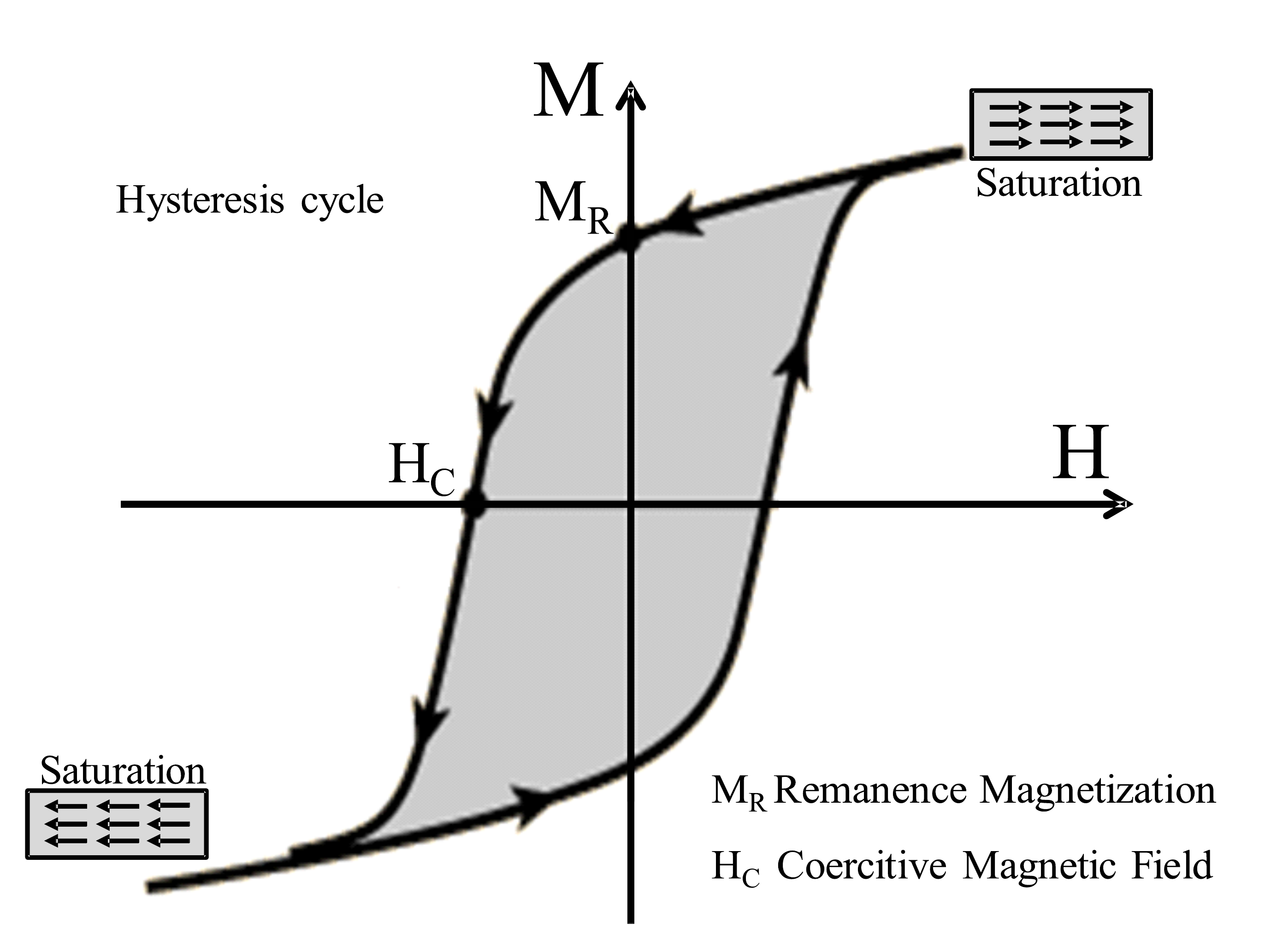}
\caption[Magnetic hysteresis in a ferromagnetic material]{Magnetic hysteresis in a ferromagnetic material. Intercepts points H$_C$ and M$_R$ are the coercivity and the remanence respectively.}
\label{hyster1}
\end{figure}
\end{center}

When a ferromagnetic material is magnetized in one direction until the saturation magnetization, it will not relax back to a zero magnetization when the imposed magnetizing field is removed. The amount of magnetization retained at a zero applied magnetic  field is called {\bf remanence}. Then, it must be driven back to zero by a field in the opposite direction; the amount of reverse applied magnetic  field required to driving to zero the magnetization  it is called {\bf coercivity}. The reverse magnetic field is applied until the saturation magnetization, then it is reversed again until saturation at first direction of applied magnetic field; thus its magnetization will trace out a loop called a hysteresis loop \cite{meyers1997introductory} (see Fig. \ref{hyster1}).

The absence of reversibility of the magnetization curve is the property called {\it hysteresis} and it can be related to several factors including the sample shape, surface roughness, microscopic defects and thermal history \cite{coeybook}.  This property of ferromagnetic materials is useful as a magnetic memory. Some compositions of ferromagnetic materials will retain an imposed magnetization indefinitely and are useful as permanent magnets \cite{meyers1997introductory}. The main cause of this phenomenon is magnetic anisotropy and will be explained in the next section.

\section{Magnetic anisotropy}

{\it The magnetic anisotropy is defined as the energy of the rotation of the magnetization direction from the easy into the hard direction} \cite{stohr2006magnetism}. 

Magnetic anisotropies may be generated by the electric field of a solid or crystal, by the shape of the magnetic body, or by mechanical strain or stress, all of which are characterized by polar vectors \cite{stohr2006magnetism}. Hence, they cannot define a unique direction of the magnetization, which is an axial vector. This is why no unique anisotropy direction can exist, but only a unique axis. Therefore, the energy density $E_{\mbox{ani}}$ connected with the magnetic anisotropy must be constant when the magnetization is inverted, which requires that it be an even function of the angle $\theta$ enclosed by $\vec{M}$ and the magnetic axes \cite{stohr2006magnetism},

\begin{equation}
E_{\mbox{ani}}= K_0 + K_1 \sin^2\theta + K_2 \sin^4\theta + K_3 \sin^6\theta + ...
\end{equation}

\noindent where $K_i$ ($i$ =1, 2, 3,...) are the anisotropy constants in the series expansion.  These are not usually defined in theoretical terms, but rather through measurements, depending on the magnetic material. The values of the constants are affected by the magnetic behavior of system and depended on the symmetry of the lattice \cite{getzlaff}. Table \ref{tablanisotroiasimetry} shows the different symmetry systems with their respective expressions of anisotropy energies ($E_{\mbox{ani}}$).

\begin{table}\begin{center}
\begin{tabular}{|c|c|}
\hline
&\\
Symmetry systems & $E_{\mbox{ani}}$\\
&\\
\hline
Uniaxial  & $K_1 \sin^2\theta$ \\
Hexagonal & $K_1 \sin^2\theta + K_2 \sin^4\theta + K_3 \sin^6\theta\cos 6\phi $\\
Tetragonal & $K_1 \sin^2\theta + K_2 \sin^4\theta + K_3 \sin^4\theta\sin 4\phi $\\
Rhombohedral & $K_1 \sin^2\theta + K_2 \sin^4\theta + K_3 \sin^3\theta\cos 3\phi $\\
Cubic &  $K_1 (\alpha^2 \beta^2 + \beta^2 \gamma^2 + \gamma^2 \alpha^2 )+ K_2 \alpha^2 \beta^2 \gamma^2$\\
\hline
\end{tabular}
\caption[Energy density expressions for each symmetry systems]{Energy density expressions for each symmetry systems from Refs. \cite{buschow2003physics,bertotti1998hysteresis}. For cubic system $\alpha= \cos\phi\sin\theta$, $\beta=\sin\phi\sin\theta$ and $\gamma = \cos\theta$, using spherical coordinates.}\label{tablanisotroiasimetry}
\end{center}

\end{table}

The cubic case is more complex because of its high symmetry. For uniaxial symmetry, when $K_1$ is positive, the easy direction is an axis ($z$ for instance), while, when has negative value, the perpendicular plane is the easy direction. As hexagonal, tetragonal and rhombohedral systems are considered, $K_1$ and $K_2$ constants play a relevant role in the anisotropy energy density. Here are distinguished cases for them \cite{buschow2003physics}:

\begin{itemize}

\item[($i$)] For $K_1=K_2=0$, the system is an isotropic ferromagnet.

\item[($ii$)] For $K_1>0$ and $K_2+K1>0$, we have an easy axis of magnetization for $\theta = 0$.

\item[($iii$)] For $K_1>0$ ($K_1< 0$) and $K_2+K_1<0$ ($2K_2+K_1>0$), the perpendicular plane to the $z$ axis is the easy magnetization plane.

\item[($iv$)] For $2K_2<K_1<0$, the easy axis will be reached for a $\theta$ value given by, $\sin^2 \theta = -K_1/2K_2$.

\end{itemize} 

The determination of $K_1$ and $K_2$ constants is possible by different methods, such as, measurement of the anisotropy magnetic field, area method, torque method (all explained in Ref. \cite{cullity2011introduction}), Sucksmith-Thompson fit for single crystal samples \cite{sucksmith1954magnetic} and  Sucksmith-Thompson fit modified version proposed by Ram and Gaut for powder samples \cite{ram1983magnetic}. This last method was used by Dung and co-workers \cite{dung1988anomaly} and Kowalczyk\cite{kowalczyk1994local} to determine the anisotropic constants of YCo$_4$B, in order to determine the anisotropic energy by crystallographic site in this compound.

\begin{center}
\begin{figure}[h!]
\centering
\includegraphics[width=10cm]{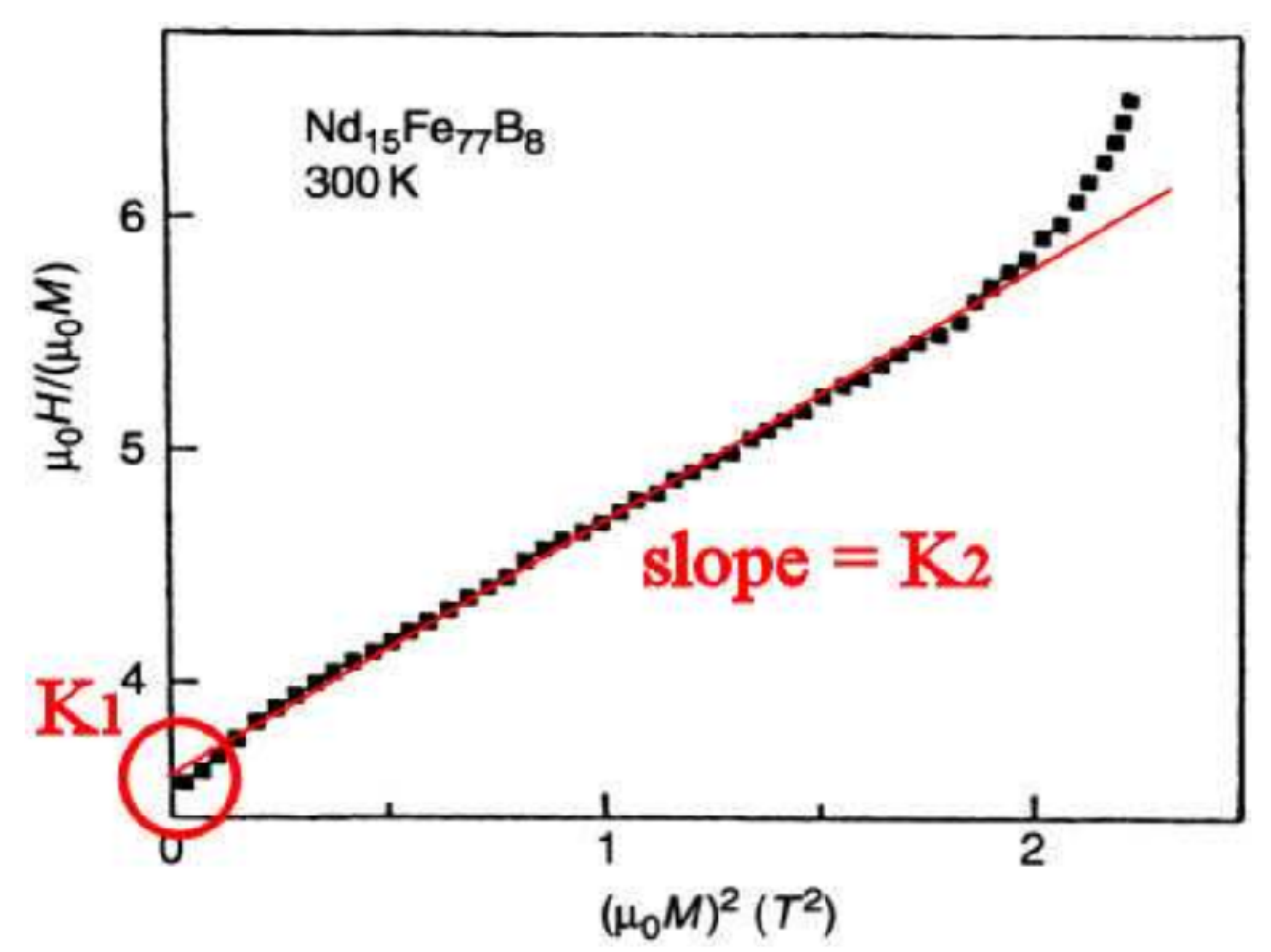}
\caption[Sucksmith-Thompson fit method]{Sucksmith-Thompson fit method for Nd$_{15}$Fe$_{77}$B$_8$, evidencing the extraction of the values $K_1$ and $K_2$ constants \cite{durst1986determination}}
\label{ultima}
\end{figure}
\end{center}

Sucksmith-Thompson fit modified consist in the construction of a graph of $H/ (\mu_0M)$ as a function of $(\mu_0M)^2$ from a powder sample oriented at small applied magnetic field, and easy direction. It is results in a linear relationship, where $K_1$ can be estimated by the vertical interception and $K_2$  by the slope. An example of this construction is shown in the Fig.\ref{ultima}.

	\chapter{Fundamentals of  magnetocaloric effect}\label{fundaMCE} 

In this chapter, we introduce basic concepts of the magnetocaloric effect (MCE). Here, the processes that define the MCE are described qualitatively and quantitatively. In addition, we will describe some aspects of how magnetic transitions affect the MCE.

\section{MCE phenomenology}

The MCE is an intrinsic thermodynamic property of magnetic materials and it was discovered in 1881 by Warburg \cite{warburg1881}, when he observed that iron absorbed  and emitted heat under the influence of an external magnetic field. However, Smith \cite{smith2013who} proposed that MCE was observed first by Weisse-Piccard \cite{pierre1917phenomene}. In simple term, MCE is the temperature change or heat exchange of a magnetic material with application of an external magnetic field. 

This effect can be observed in either an adiabatic or an isothermal process, due to a change of the applied magnetic field. We considering a magnetic material in an adiabatic process, and in the presence of a variable external magnetic field. The increase in the magnetic field causes magnetic dipoles to align themselves,  leading to a decrease in magnetic entropy. However, the total system  entropy should be constant, and in consequence, the lattice entropy increases causing an increase in the system temperature.

\begin{figure}
\begin{center}
\includegraphics[width=15cm]{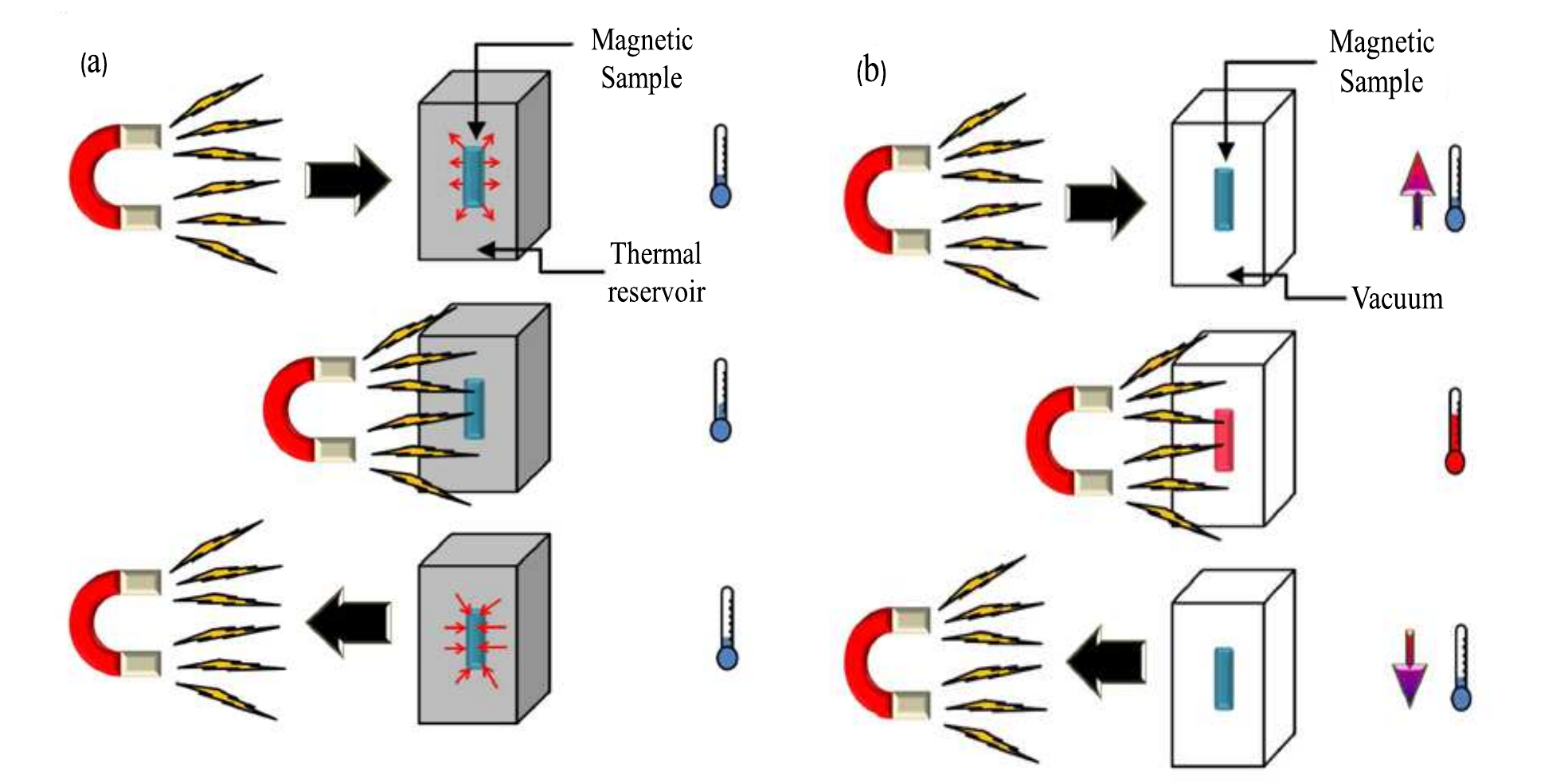}
\caption{MCE process: isothermal (a) and adiabatic (b).}
\label{proMCE}
\end{center}
\end{figure}

From an isothermal process, the material is in thermal equilibrium with a reservoir. Then, a variable external magnetic field is applied to align the magnetic dipoles and the magnetic entropy is changed. As it is an isothermal process, the internal energy of the system does not change, and the material must be in continuous heat exchange with the reservoir. To clarify these processes, see the Fig. \ref{proMCE}.

\section{General thermodynamic approach}

For the description of magnetothermal effects in magnetic materials, we considered the  Gibbs free energy $G$ as a function of the temperature ($T$), pressure ($p$) and magnetic field ($H$), we can write \cite{tishin}:

\begin{equation}
G= U-TS+pV-MH,
\label{EGibss}
\end{equation}

\noindent where $U$ is the internal energy and $S$ is the entropy. Considering an isobaric system, variations of $G$ are given by  \cite{marioIntro}:
\begin{equation}
dG= - SdT - MdH.
\label{difG}
\end{equation}

To find the internal parameters $S$ and $M$, we use the equations of state,

\begin{eqnarray}
S(T,H,p)= - \left(\frac{\partial G}{\partial T}\right)_{p,H}, \label{eqGS}\\
M(T,H,p)= - \left(\frac{\partial G}{\partial H}\right)_{T,p}\label{eqGM}.
\end{eqnarray}

From Eqs. \ref{eqGS} and \ref{eqGM}, we can find the so-called Maxwell equation,

\begin{equation}
\left(\frac{\partial S}{\partial H}\right) = \left(\frac{\partial M}{\partial T}\right). 
\label{eqMaxSM}
\end{equation}

Thus, we obtain the expression for the magnetic entropy change ($\Delta S$) from an initial magnetic field ($H_i$) to the final magnetic field ($H_f$),

\begin{equation}
\Delta S(T,\Delta H) = \int\limits_{H_i}^{H_f} \left( \frac{\partial M(T,H)}{\partial T}\right)_{H} dH.
\label{changeS}
\end{equation}

\begin{figure}
\begin{center}
\includegraphics[width=17cm]{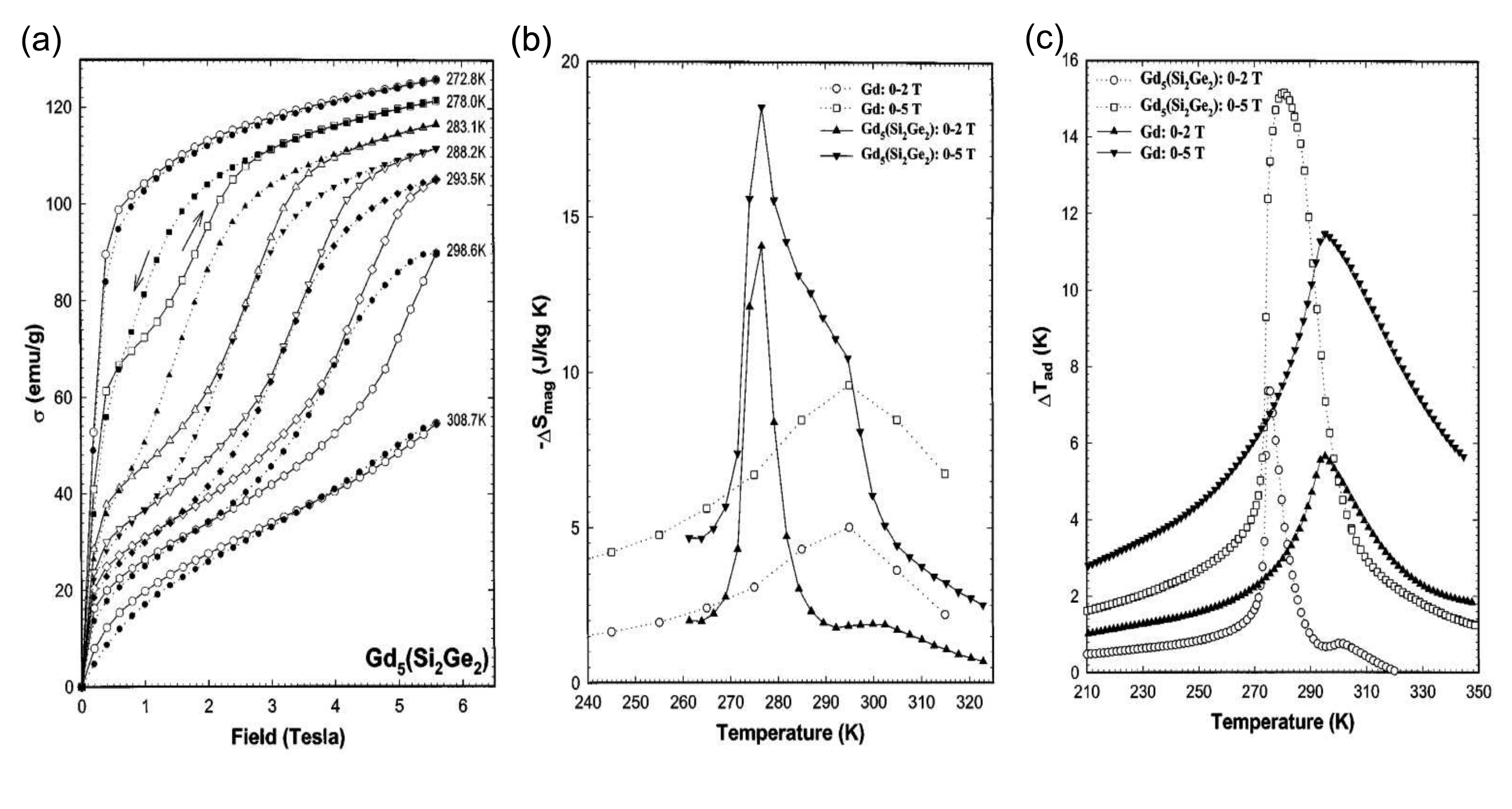}
\caption[Entropy calculated example]{Magnetic mapping of a  Gd$_5$(Si$_2$Ge$_2$) sample (a), for find the entropy changes (b). (c) Adiabatic temperature calculated from the magnetic and specific heat measurements in  Gd$_5$(Si$_2$Ge$_2$) and Gd samples for several $\Delta H$ \cite{pecharsky1997giant}.}
\label{exmplo}
\end{center}
\end{figure}

It can be easily observed that this quantity will be maximized around large variations in magnetization with temperature, as those that happen around the Curie temperature ($T_C$).  Experimentally is realized a mapping of the magnetization measurement as a function of the magnetic field around $T_C$, for then calculate of magnetic entropy changes. In  Fig.\ref{exmplo}.(a) and (b) is shown the magnetization mapping and magnetic entropy changes of Gd$_5$(Si$_2$Ge$_2$) from Ref. \cite{pecharsky1997giant}.

On the other hand, we can calculate the specific heat of a system with the second derivative of the Gibbs free energy and Eq. \ref{eqGS},

\begin{equation}
C_{H} =-T \left( \frac{\partial^2 G}{\partial T^2}\right)_{H,p} = T \left( \frac{\partial S}{\partial T}\right)_{H}.
\label{eqCapS}
\end{equation}

By considering the entropy as a function of the temperature and magnetic field, $S= S(T,H)$:

\begin{equation}
dS  = \left( \frac{\partial S}{\partial T}\right)_H dT + \left( \frac{\partial S}{\partial H}\right)_T dH.
\label{eqdS}
\end{equation} 

Then, in an adiabatic process ($dS=0$), using Eqs.\ref{eqCapS} and \ref{eqMaxSM}, we obtain:

\begin{equation}
\frac{C}{T} dT = - \left(\frac{\partial M}{\partial T} \right) dH.
\end{equation}

Consequently, we find the change in the adiabatic temperature,

$$\Delta T_{ad}= \int\limits_{H_i}^{H_f} \left( \frac{T}{C(T)_H}\right) \left( \frac{\partial M(T,H)}{\partial T}\right)_{H} dH,$$

\noindent and from the magnetic entropy \cite{pecharsky1999magnetocaloric},

\begin{equation}
\Delta T_{ad} (T) \approxeq  \frac{T}{C(T)_H} \Delta S(T)_{\Delta H}
\end{equation}

The calculation of $Delta T_{ad}$ requires the experimental measurements of the magnetization with the magnetic field and specific heat data. Thus, in the practice, it is mathematically and experimentally more difficult.  $\Delta T_{ad}$ results for Gd$_5$(Si$_2$Ge$_2$) sample compared with Gd sample in different $\Delta H$ is in Fig. \ref{exmplo}.(c).

\section{Magnetic-phase transitions and MCE}

Order phase transitions can be caused by either varying the temperature or applying a magnetic field, and they have been extensively studied in the context of magnetocaloric materials   \cite{pecharsky1999magnetocaloric,wada2003giant,liu2007determination,gama2004pressure}. For, a material undergoes the transition of a first-order, then the first-order derivatives of the thermodynamic potential change discontinuously, and values such as entropy, volume and magnetization display a {\it jump} at the point of transition and characterized by the existence of latent heat. In the second order transitions, the derivatives of thermodynamic potential are continuous, they are no associated latent heat, and the second derivatives are discontinuous.

As mentioned above, the MCE is maximized at the transition temperature in magnetic systems, for example $T_C$ in the ferro-paramagnetic transition. A way to approach this phenomena is by Landau theory \cite{tishin,marioIntro,pathria1986statistica}, which an analytical function, such as free energy ($F$), is taken and realized a potential expansion over the order parameter. The latter can be the magnetization for ferro-paramagnetic transitions, and then $F$ is expressed around the magnetic transition temperature as:

\begin{equation}
F= F_0 + \frac{1}{2} g_2(T) M^2 + \frac{1}{4}g_4(T) M^4,
\label{TL1a}
\end{equation}

\noindent where $g_2(T) = \alpha(T-T_C)$ is a null parameter at $T=T_C$ and $g_4$ is constant \cite{marioIntro,tishin}. Then, from the minimization of $F$, we can determine the equilibrium condition to obtain:

\begin{equation}
M=0, \hspace{2cm}\mbox{and} \hspace{2cm} M_{\pm} = \pm \left( \frac{\alpha(T-T_C)}{2g_4} \right)^{1/2},
\end{equation}

For $M=0$, the system is at $T> T_C$ dominant disorder state, while $M_{\pm}$ is taken when $T,T_C $ represents the minimum values of $F$ and in the ferromagnetic magnetization is proportional to $(T_C - T)^{1/2}$.

When a magnetic field ($H$) is applied, an additional term is added.

\begin{equation}
F= F_0 + \frac{1}{2} g_2(T) M^2 + \frac{1}{4}g_4(T) M^4 - HM.
\end{equation}

From minimized $F$ for $T=T_C$, this result can be reached,

$$\frac{H}{M}= \beta M^2,$$ 

\noindent where $ \beta = 4g_4$ is the parameter that determines the magnetic transition type. From experimental construction of $H/M$ $vs$ $M^2$ graph, it is possible found $\beta$, which is $\beta > 0$ in a second-order transition, while that $\beta <0$ is first-order transition. This, is known as the Banerjee criterion \cite{banerjee1964}.

	\chapter{Sample preparations}\label{capitulo4}

The development of many experimental works is based on the quality of the samples studied, which allows us to study efficiently their physical  properties. For this reason, sample preparation plays an important role in the development of this Thesis. Thus, the goal of this  chapter is to describe the techniques used for sample synthesis, which were implemented in the {\it Laborat\'orio de prepara\c c\~ao de amostras} of the {\it Instituto de F\'isica da Universidade Federal Fluminense (UFF)}.

\section{Intermetallic synthesis by arc melting furnace}

\subsection{Arc melting furnace}

\begin{center}
\begin{figure}[h!]
\centering
\includegraphics[width=15cm]{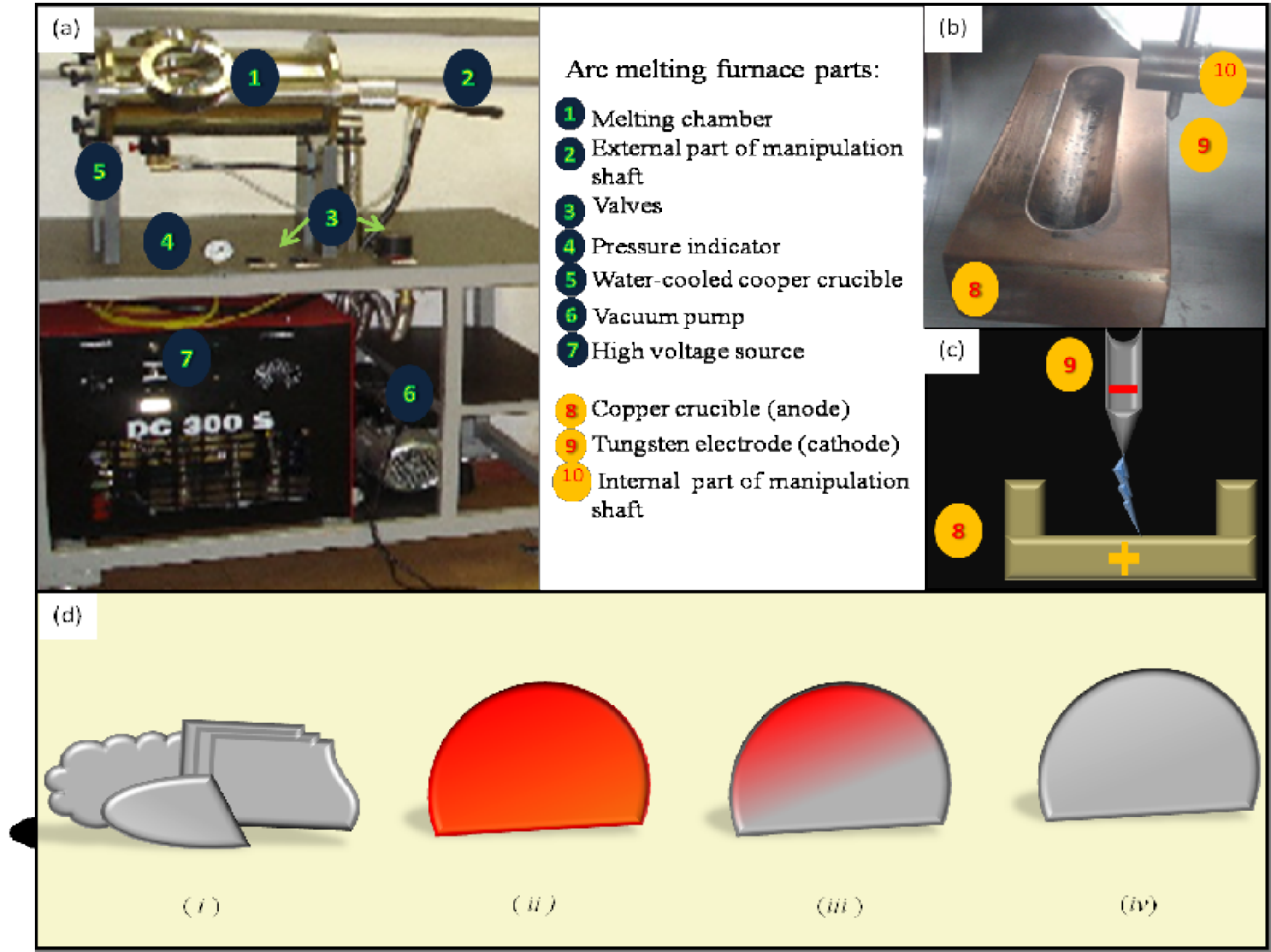}
\caption[Components on arc melting furnace]{Arc melting furnace parts: external (a) and internal (b). Schematic process of arc formation (c). Fusion process is schematized in (d): (i) separate element, (ii) the molten mass is red because of a high temperature, (iii) the mass formed begins its cooling process, (iv) the new sample is formed.}
\label{Horno}
\end{figure}
\end{center}

Arc melting furnace is extensively used both in industries and in research laboratories as a mean of producing samples of intermetallic compounds. The furnace used by us was fabricated at {\it Universidade de Campinas} and consists of a vacuum chamber, incorporated internally by tungsten electrode and a copper crucible cooled with water. Additionally, it has externally pressure valves, a vacuum pump, a high voltage source and other components that are shown in Figs. \ref{Horno} (a) and (b).

The fusion of the elements is carried out by a voltage source, which generates a potential difference between the tungsten electrode (cathode) and the copper crucible cooled with water (anode), where is achieved temperatures from 2000 $^{\circ}$C to 3000 $^{\circ}$C, Fig. \ref{Horno}(c) illustrated the process scheme. The process begins when the compounds are placed into the copper crucible, for later, close the chamber and we realize the vacuum and subsequent purges. Then, these materials are melted by the potential difference, and a red mass (live red) is obtained due to the high temperature achieved. This mass is cooled by a water flux, and a new sample is formed, as can be seen in Fig.\ref{Horno}(d).

\subsection{Compounds used}

For the preparation of intermetallic samples by the arc furnace, the compounds used  should be treated carefully for improving the efficiency of the fusion process and to obtain samples of good quality. Here, we enunciate some aspects to consider:

\begin{itemize}

\item[$\Diamond$] {\it Purity}:  the purity of the components used is very important for obtaining samples of high quality. However, some of the compounds have not the same purity. In this case, the quality of the sample will be evaluated from the component with minor purity. Actually, to our knowledge, the most widely used method to check that the samples crystallized in a single phase is  the X-rays diffraction. The purity of the components used in the preparations of the compounds is shown in Table \ref{compo}.

\item[$\Diamond$] {\it Melting and boiling point}: the research on elements to be used in an arc furnace shall consider their melting and boiling points, the temperature achieved in the furnace. An example of this is phosphorus (P), which has a boiling temperature of 550 K (277 $^{\circ}$C) \cite{hayne2011melting}. This is much lower than the temperatures of the arc furnace, and might cause total evaporation of this material. As a result it is virtually  impossible to synthesize materials with phosphorus in an arc furnace. The melting and boiling points of elements used in the synthesis of the intermetallic samples in this {\bf Thesis} are presented in Table \ref{compo}.

\begin{table}[h!]
\centering
\begin{tabular}{|c|c|c|c|}
\hline
Element & Purity(\%) & Melting point(K)& Boiling point (K)\\
\hline
Y	& 99,9	& 1799 & 3609 \\
Mn	& 99,98	& 1519 & 2334 \\
Fe	& 99,9  & 1808 & 3023 \\
Co	& 99,98 & 1768 & 3200 \\
Ni	& 99,99 & 1728 & 2730 \\
Si	& 99,99 & 1687 & 3173 \\
Ga	& 99,99 &  303 & 2477 \\
B	& 99,9  & 2349 & 4200 \\
\hline
\end{tabular}
\caption[Elements used for manufacturing our samples]{Elements used for manufacturing our samples. The melting and boiling points are taken from Ref. \cite{hayne2011melting}.}
\label{compo}
\end{table}

\item[$\Diamond$] {\it Additional mass}: it is important to add a suitable amount of mass of some of the elements for obtaining the desired stoichiometry. However, it is necessary to add additional mass due to evaporation of the elements. This is caused by the  temperature achieved by the arc, which is higher than the boiling point of the material. To avoid the loss of stoichiometry, we recommend to find the additional mass  before fusion. For example in our work \cite{vivas2014competing} on YNi$_{4-x}$Co$_x$B, we measured the yttrium mass before and after the melting  and found that the required  additional mass of yttrium was 7\%. While that for Fe$_2$MnSi$_{1-x}$Ga$_x$ \cite{vivas2016experimental}  3\% of manganese was added. 

\end{itemize}

\subsection{Sample annealing}\label{Trata}

The next step after the fusion is sample annealing. The annealing of a sample at high temperatures is important to obtain homogeneous samples without spurious phases. The main variables for carrying out annealing of samples are: ($i$) the temperature used, and ($ii$) the annealing time. Therefore, the information obtained from the literature before of sample annealing is important. Temperature and time of annealing can be seen in Table \ref{trato}.

\begin{table}[h!]
\centering
\begin{tabular}{|c|c|c|}
\hline
Compounds & Temperature (K) & time (days) \\
\hline
{\bf YNi$_{4-x}$Co$_x$B}& 	&  \\
$x$ =0, 1, 2, 3 and 4 & 1323 & 10 \\
\hline 
{\bf Fe$_2$MnSi$_{1-x}$Ga$_x$} & & \\
$x$ =0, 0.02, 0.04, 0.09& &\\ 
0.12, 0.20, 0.30 and 0.50 & 1323 & 3 \\
\hline 
{\bf Co$_2$FeSi} & & \\
0d & 1323 & 0\\
3d & 1323 & 3\\
6d & 1323 & 6\\
15d & 1323 & 15\\
\hline
\end{tabular}
\caption[Temperature and time of annealing]{Temperature and time employed for annealing of all the intermetallic samples manufactured in this work. Here, Co$_2$FeSi samples were annealed during 3 (3d), 6 (6d) and 15 (15d) days, while one was no annealed (0d).}
\label{trato}
\end{table}

\begin{figure}[h!]
\centering
\includegraphics[width=8cm,height=6cm]{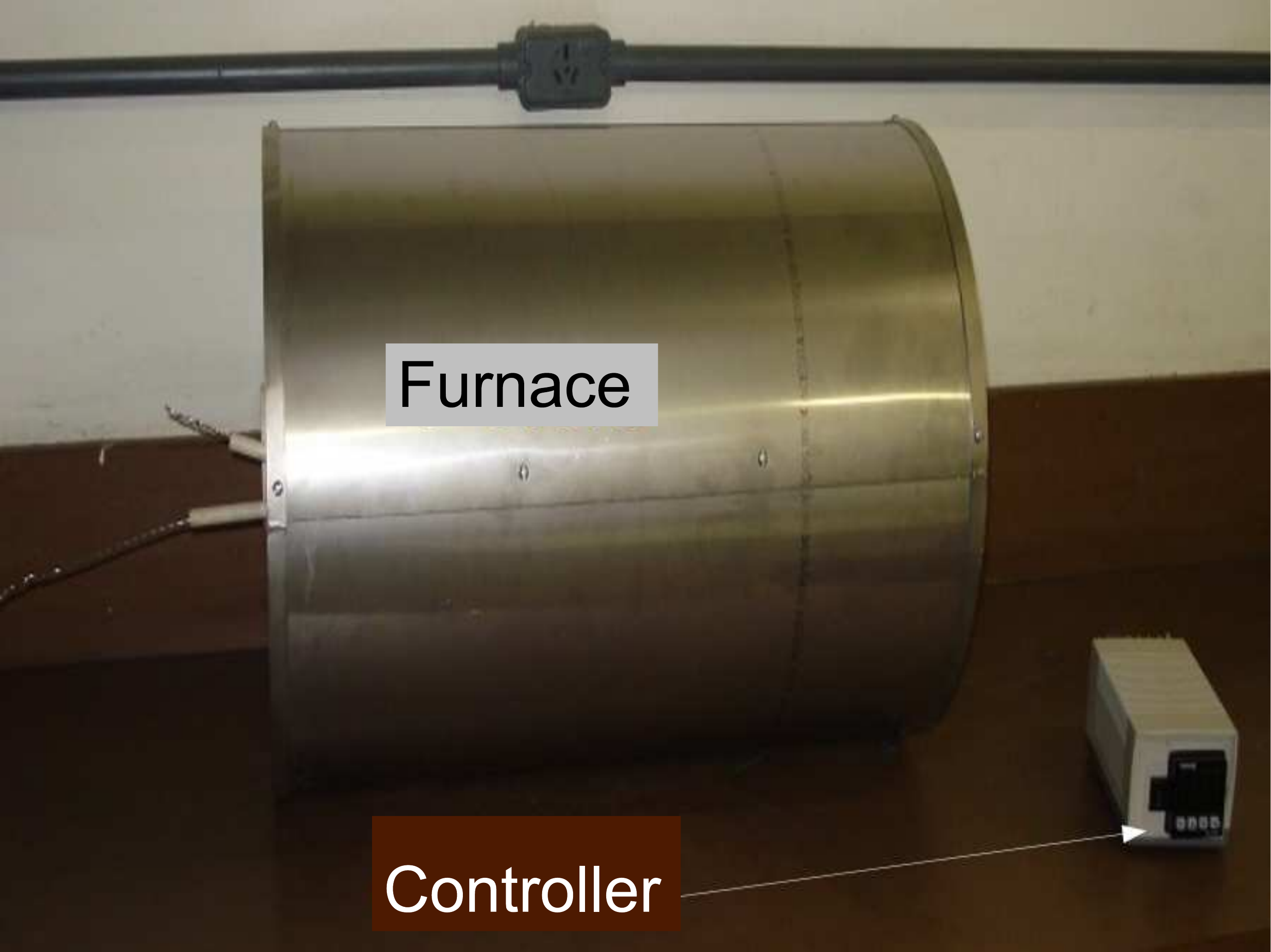}
\caption[Annealing furnace]{Annealing furnace}
\label{fotohorno}
\end{figure}

To carry out the annealing process first, we need to wrap the samples in tantalum (Ta) foils in order to prevent contact between them. Ta foils are used since their melt temperature is  3290 K \cite{hayne2011melting} and do not contaminate the samples. Then, the samples are encapsulated within a quartz tube filled with argon (Ar) to avoid oxidation and promote uniformity of the annealing temperature. Finally, the samples are placed inside the annealing furnace (see the Fig. \ref{fotohorno}), and subsequently, they are subjected to quenching in water for maintaining the crystalline structure obtained at high temperatures.

\section{Pechini method (sol-gel)}

The experimental technique used for sample preparation of perovskites (manganites and cobaltites) was the Pechini method,  also known as the {\it sol-gel method}. This method was developed by Maggio P. Pechini in 1967 \cite{sprague1967method}, mainly motivated by difficulties in obtaining oxide compounds of high quality by other methods, such as solid state reaction or mechanically-ground mixture.

\begin{figure}[h!]
\centering
\includegraphics[width=8cm,height=8cm]{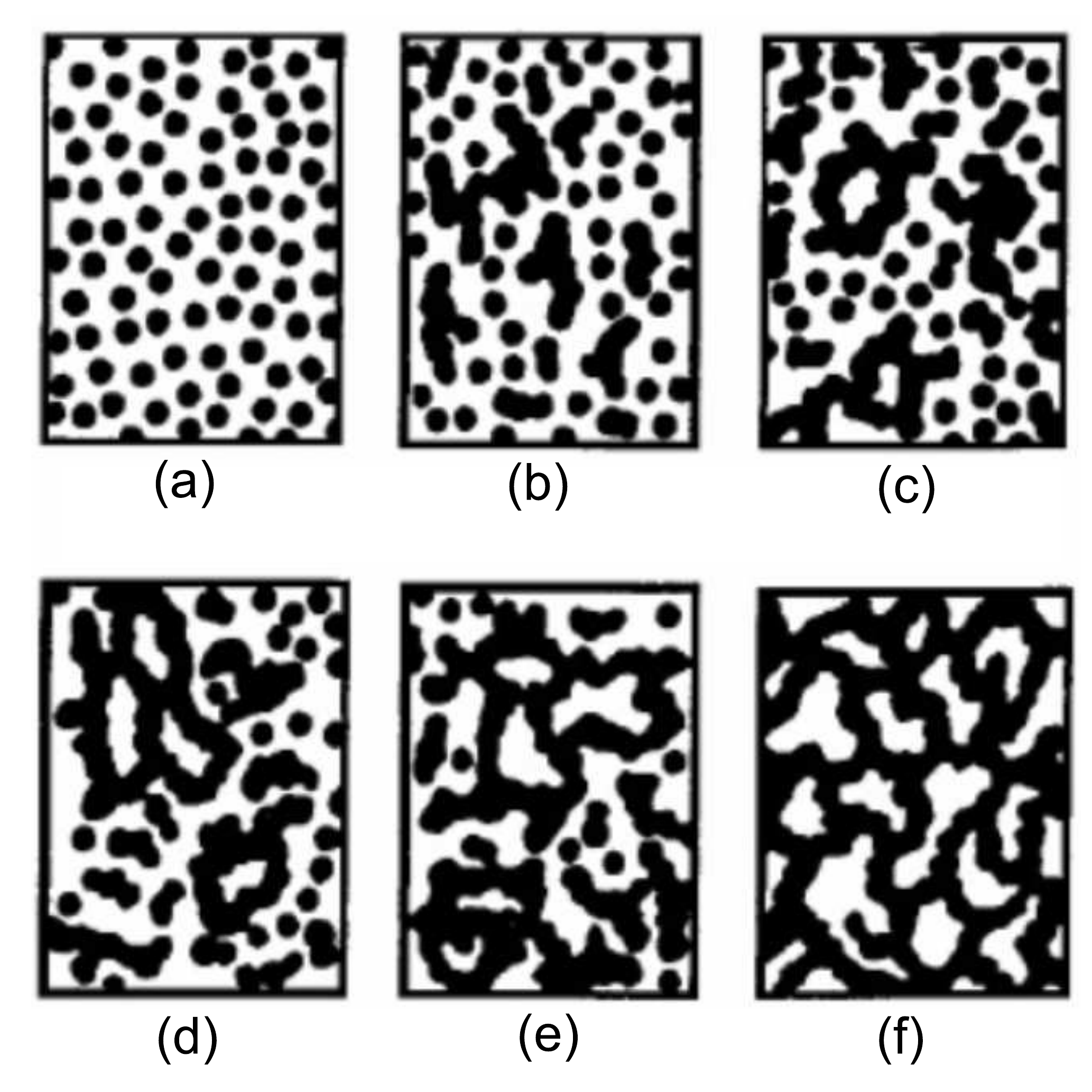}
\caption[Schematic representation of sol-gel process]{Schematic representation of sol-gel process. (a) Metal dissociation in an acidic solution form colloidal particles. Particle chains are formed generating microgel regions (b) with the same density and without particles decant (c). (d) The gel point is achieved, and (e-f) the chains occupy the entire volume forming the gel.}
\label{esqsolgel}
\end{figure}

The Pechini method uses a chemical route for the production of compounds, and consists of separating oxygen from their cations using an acidic solution. From this solution, and with a polymerizing agent, a gel ({\it sol-gel}) with the desired stoichiometry is obtain. This process is schematically illustrated in Fig. \ref{esqsolgel}.

\subsection{Synthesis of the nanoparticles }

For the synthesis of perovskite nanoparticles, we have used the reagents listed in Tab. \ref{muestras2016}, and taking the following steps:

\begin{table}[h!]
\centering
\begin{tabular}{|c|c|c|c|}
\hline
Compounds & Reagents & Acid for dissociation & Polymerize agents\\
\hline
La$_{0.6}$Sr$_{0.4}$MnO$_3$ & La$_2$O$_3$& nitric & polyethylene glicol\\
	& SrCO$_3$& + & + \\
	& Mn$_2$O$_3$ & citric & ethylene diamine\\
\hline
Nd$_{0.5}$Sr$_{0.5}$CoO$_3$ & Nd(NO$_3$)$_3 \times$ 6H$_2$O &citric & ethylene diamine\\
	& Sr(NO$_3$)$_2$ & & \\
	&  Co(NO$_3$)$_3 \times$ 6H$_2$O& & \\
\hline
\end{tabular}
\caption{Compounds used for the  synthesis of perovskites oxides}
\label{muestras2016}
\end{table}

\begin{itemize}

\item[$\bullet$] The quantities of reagents (oxides or nitrates) used were calculated by their molecular weights and percentages of each compound to ensure the desired stoichiometry.

\item[$\bullet$] {\it Dissociation of metal}: when the samples are synthesized from {\it nitrate reagents}, we dissolved these reagents in  a solution of citric acid (or another organic acid) and deionized water (alcohol can also be used) at room temperature, with high acid pH ($\approx 1$). Cases when the reagents are dissociate individually or together happens, but all have to be placed in the same solution.

When the samples are synthesized from {\it oxide reagents} the process is more complicated since the dissociation can take several hours. In this case, the reagents are dissolved separately in the solutions with 100 ml of deionized water, nitric acid, and citric acid to obtain homogeneous solutions. Then, these solutions are mixed, and from here, the following two processes are similar.

\begin{center}
\begin{figure}[t!]
\centering
\includegraphics[width=16cm]{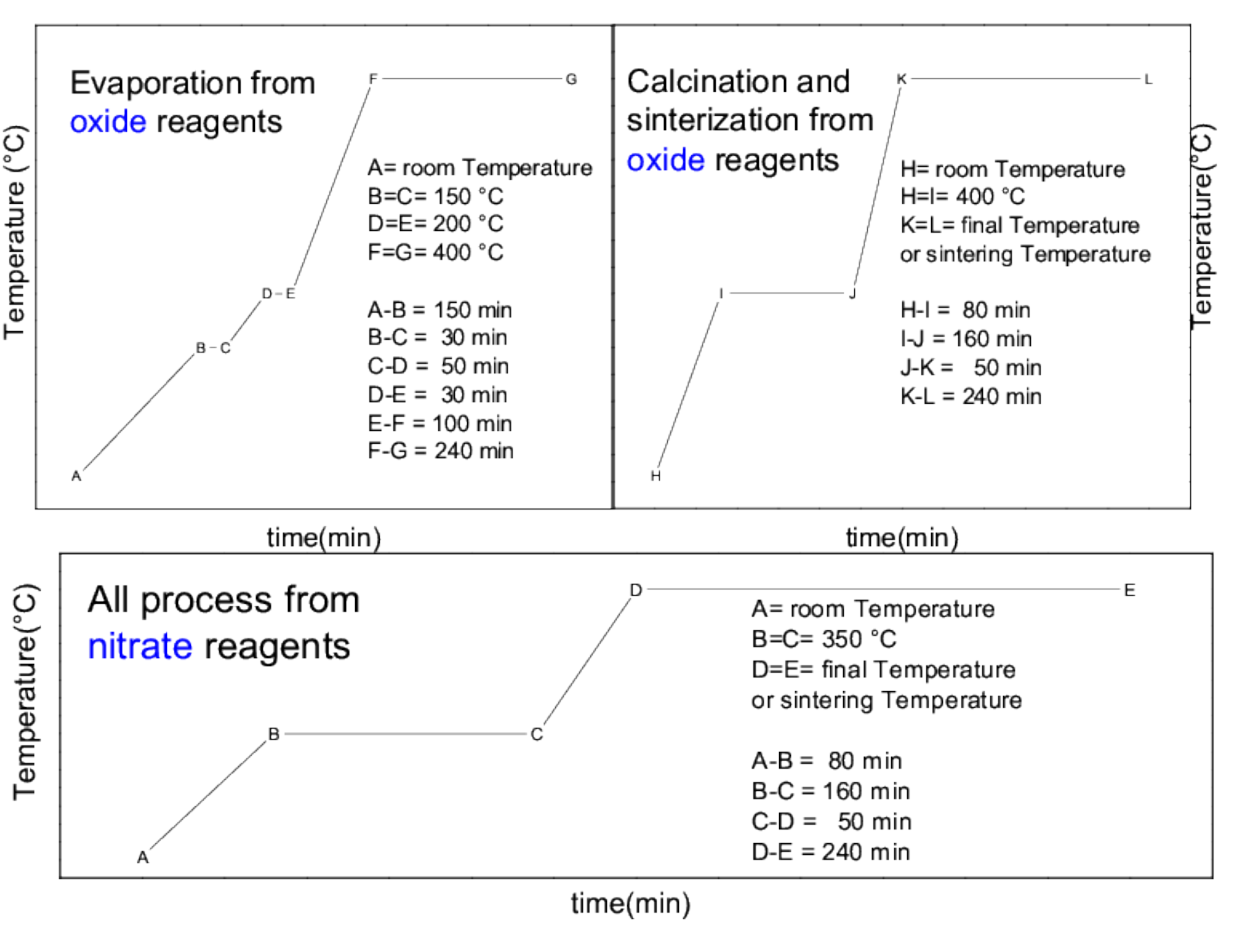}
\caption[Schemes of evaporation and annealing process for oxides and nitrates reagents]{Schemes of evaporation and annealing process for oxides(top) and nitrates (bottom) reagents.}
\label{tratados}
\end{figure}
\end{center}

\item[$\bullet$] {\it Polymerization }: in this process a polymerizing agent such as ethylene glycol, which produces organic chains, is added, thus capturing the dissociated metals from the previous step. For the pH control ($\approx$ 5), diamine ethylene should be added. This solution is stirred at 70 $^{\circ}$C for approximately 4 h, to finally obtain a gel.

\item [$\bullet$] {\it Evaporation and annealing}: because of the various compounds that can be synthesized by this method, a large number of evaporation and annealing processes are required to obtain the materials. Our case depends on the reagents used in the synthesis: oxides (see Fig. \ref{tratados}-top ) or nitrates (see Fig. \ref{tratados}-bottom).

- From {\it oxide} reagents for synthesis of La$_{0.6}$Sr$_{0.4}$MnO$_3$ the evaporation process is as follows: The samples are heated for 150 min from the room temperature to 150 $^{\circ}$C. After 30 minutes at 150$^{\circ}$C, it is heated for 50 min up to 200 $^{\circ}$C, and waiting for 30 min at 200 $^{\circ}$C. Then, the sample is heated for 100 min up to 400 $^{\circ}$C and the temperature is maintained for 240 min. This process is for the evaporation of acid and other liquid compounds. The final product of this evaporation is a brown-red powder, which it is not the final compound of interest. To obtain the manganite phase, the powder is treated at 400 $^{\circ}$C for 4 hs. Then, the powder is separated in portions for annealing at 700 $^{\circ}$C (973 K), 800 $^{\circ}$C (1073 K), 900 $^{\circ}$C (1173 K) and 1000 $^{\circ}$C (1273 K). 

- In the case of {\it nitrates} used from precursor reagents, only one process that includes both two evaporation and annealing treatments is necessary. The gel is placed in an oven at room temperature and is heated at a rate of 5 $^{\circ}$C/min to 360 $^{\circ}$C, where it remains during 4 hs to promote the evaporation. Next, it is warmed up to the annealing temperature, and finally remains at this temperature for 6 hs (depending on the sample the temperature and the annealing time are different).

\end{itemize}

\begin{center}
\begin{figure}[h!]
\centering
\includegraphics[width=16cm]{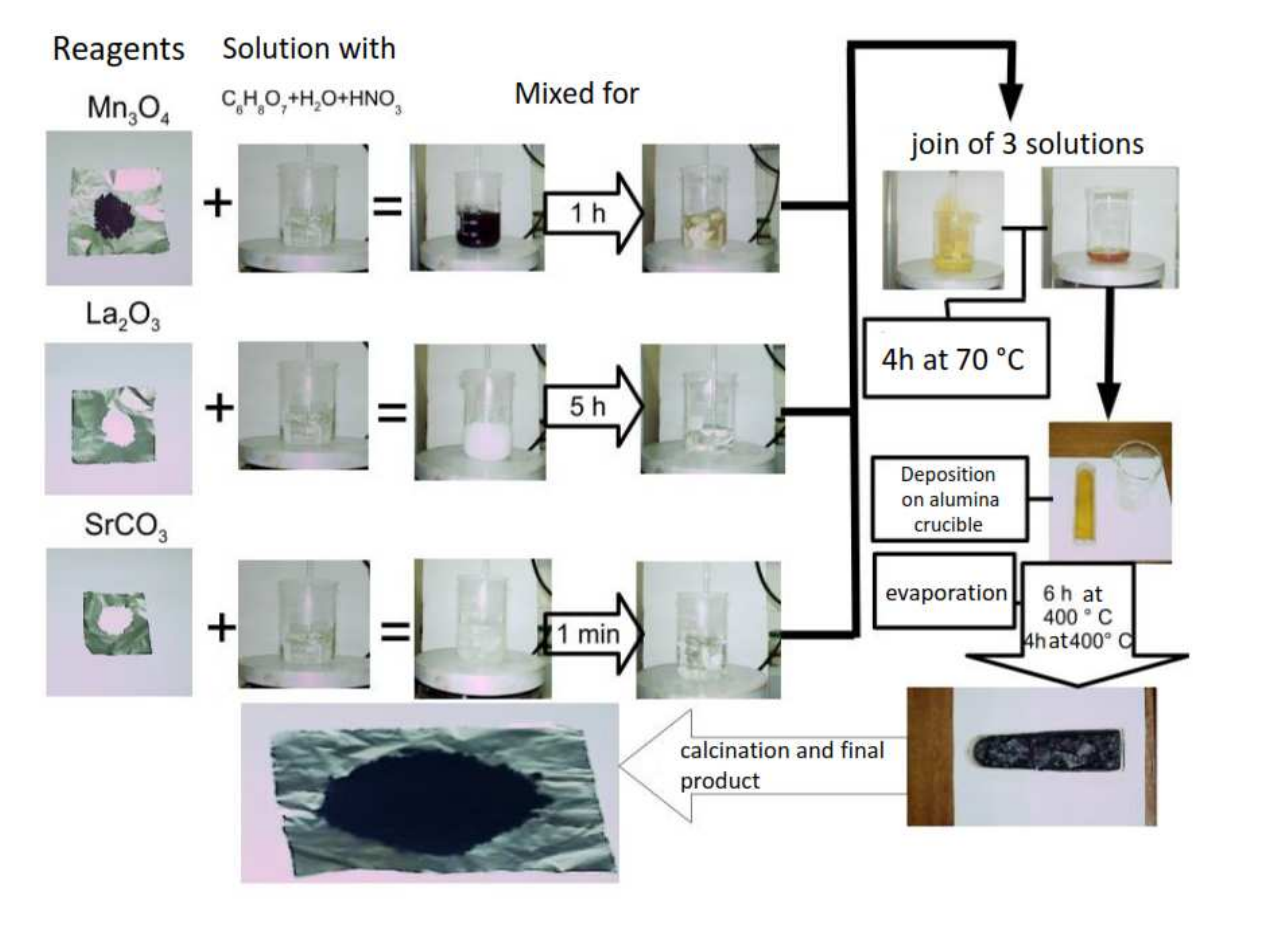}
\caption[Nanoparticles preparations from oxides]{Nanoparticle preparations for La$_{0.6}$Sr$_{0.4}$MnO$_3$ nanoparticle oxides.}
\label{Npsoxides}
\end{figure}
\end{center}

\begin{center}
\begin{figure}[t!]
\centering
\includegraphics[width=12cm]{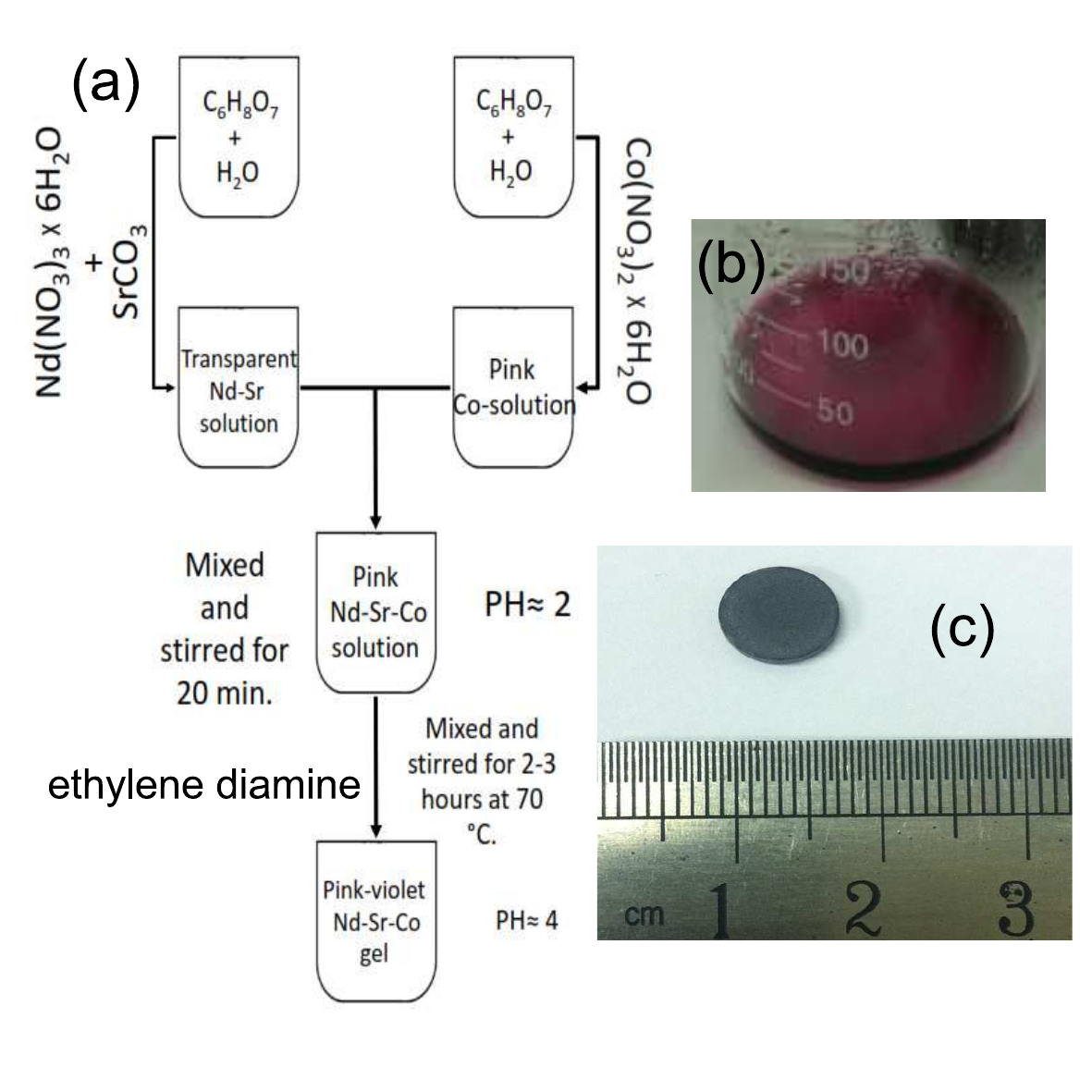}
\caption[Sol-gel preparation from nitrates]{Sol-gel synthesis from nitrates of Nd$_{0.5}$Sr$_{0.5}$CoO$_3$. (a) Schematic synthesis, (b) resulted gel and (c) bulk sample after sintering annealing.}
\label{Npsnitrates}
\end{figure}
\end{center}

The scheme in Fig. \ref{Npsoxides} shows the process for oxide reagents (here for La$_{0.6}$Sr$_{0.4}$MnO$_3$ samples), while that of nitrate reagents is shown in Fig. \ref{Npsnitrates} (here is Nd$_{0.5}$Sr$_{0.5}$CoO$_3$ sample). Both processes result in little particles of the order of nanometers or hundreds of nanometers. In the case of manganite samples, we obtain nanoparticles with a diameter between 5 nm and 100 nm of high quality. They will be explored in the following chapters. For Nd$_{0.5}$Sr$_{0.5}$CoO$_3$ cobaltite, we obtain a powder of paticles with diameters ranging between 500 nm and 1000 nm, which is pilled for getting a bulk sample shown in Fig. \ref{Npsnitrates} (c).

\section{Sample characterization by X-ray diffraction}

X-rays diffraction (XRD) is the most common technique used to (among other purposes) characterization of the samples, because their wavelength $\lambda$ is typically the same order of magnitude (1-100 \AA) as the spacing d between planes in the crystal.  Physicist W.L. Bragg considered a family of parallel planes separated by a distance $ d $. The path difference between the rays reflected by neighboring planes is $2d \sin \theta $, where $ \theta $ is the angle of incidence. The rays reflected by the different planes interfere constructively, when the path difference is equal to an integer $n$ of wavelengths $ \lambda $, that is, when \cite{kittel},

\begin{equation}
2d\sin \theta =  n \lambda .
\end{equation}

\begin{figure}[h!]
\centering
\includegraphics[width=10cm,height=7cm]{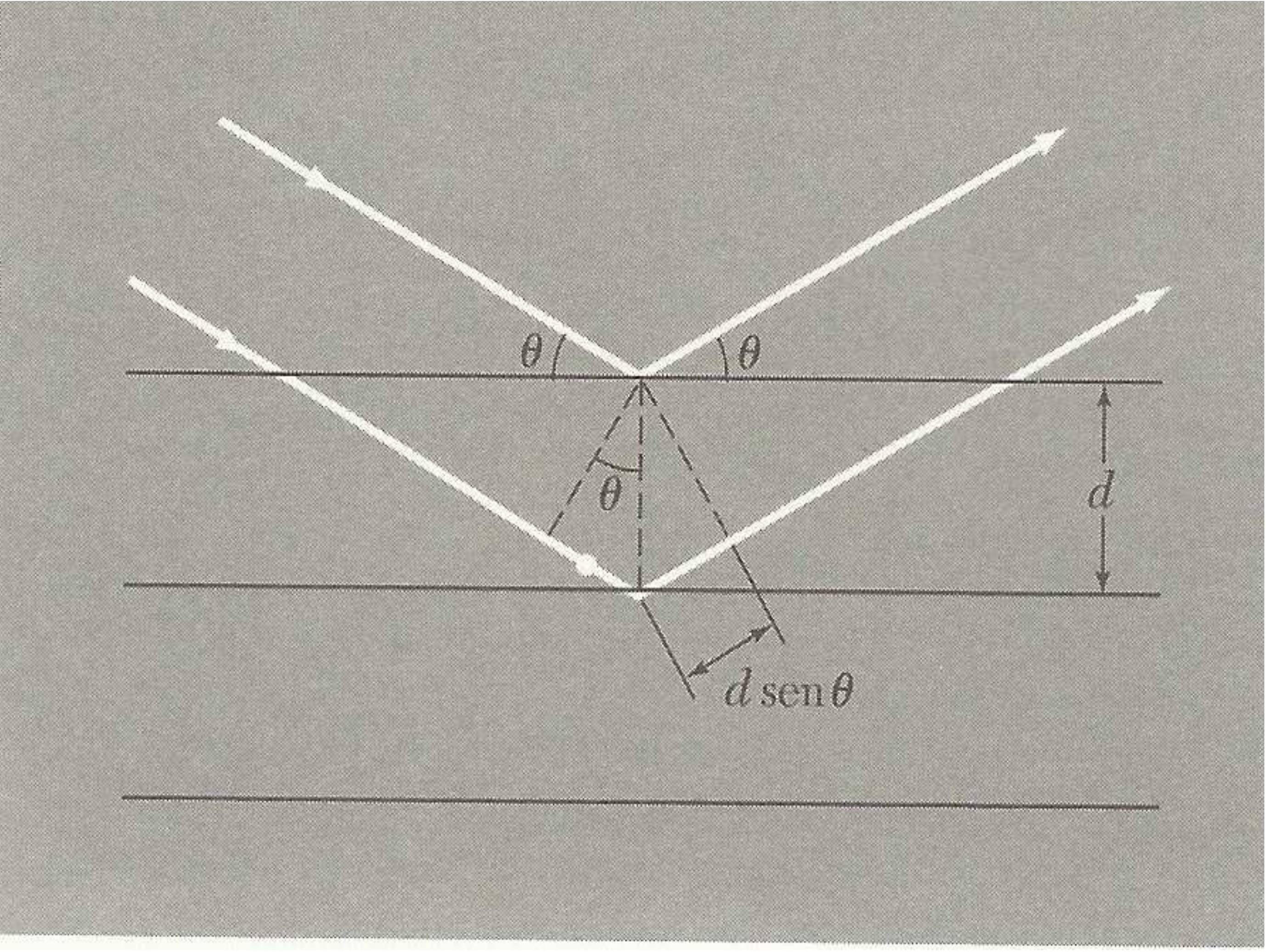}
\caption[Bragg law]{Bragg's law Ref.\cite{kittel}. The incoming beam (coming from upper left) causes each scatterer, which are arranged symmetrically with a separation $d$, these waves will be in sync (add constructively) only in $2d$ $\sin \theta = n \lambda$. In that case, part of the incoming beam is deflected by an angle $2\theta$, producing a reflection spot in the diffraction pattern.}
\label{Lbragg}
\end{figure}

This is Bragg's law \cite{bragg1929diffraction}, illustrated in Fig. \ref{Lbragg}. Among other parameters, the intensity of the reflections in the diffraction patterns depends on the structure factor ($ F_{hkl}$), which depends on how the radiation in the crystallographic planes of the material is scattered. This quantity depends of the atomic scattering factor, which is defined as
\begin{equation}
f = f_0(\theta)+f^{\prime}(E)+if^{\prime\prime}(E),
\end{equation}

\noindent where $f_0$ is known as the normal scattering factor, while $f^{\prime}$ and $ f^{\prime\prime}$ are dispersive and absorption terms, respectively. The last two terms are not taken into account in conventional XRD, however, they are very important in the anomalous X-ray diffraction (AXRD) and their values may be found in the {\it International Tables for Crystallography}   \cite{tablaC}.

XRD conventional  data of the polycristal samples  were obtained at {\it Laborat\'orio de Raios X at Universidade Federal Fluminense} and at room temperature, using a Bruker AXS D8 Advance diffractometer with CuK$_\alpha$ radiation ($\lambda$ = 1.54056 \AA), 40 kV and 40 mA. 
 
Synchrotron radiation was used for obtain AXRD. It is made of X-ray beams generated by the accelerated electrons confined in a  circular loop using magnetic fields. AXRD patterns were obtained at room temperature using a MYTHEN 24K system, from Dectris$^\circledR$ at XRD1 beamline at {\it Laborat\'orio Nacional de Luz S\'incrotron}. In addition, a NIST SRM640d standard Si powder was used to determine the X-ray wavelengths with precision. 

\begin{figure}[t!]
\centering
\fbox{
\includegraphics[width=10cm]{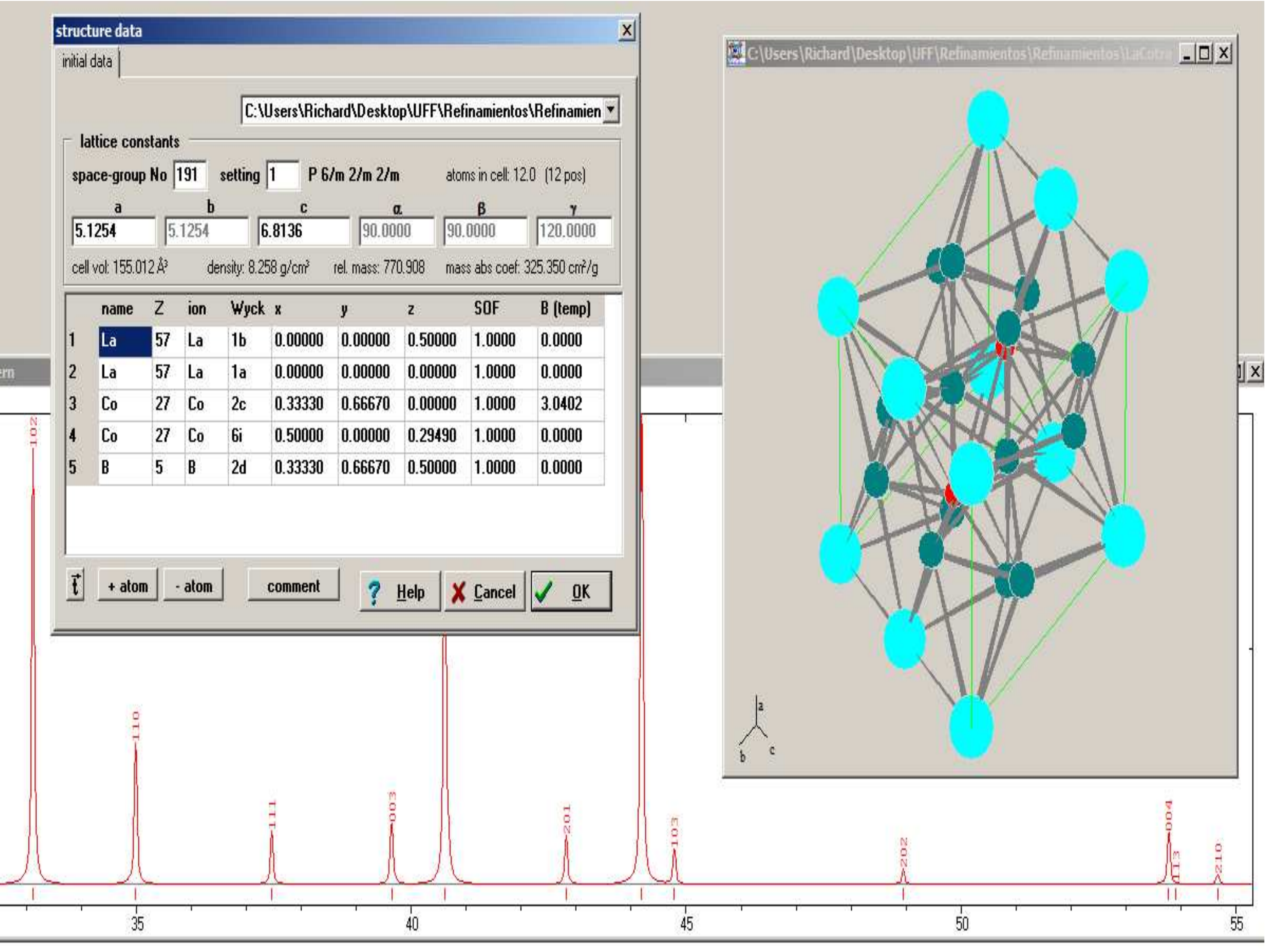}}
\caption[Example of the window provided by the Powder Cell program]{Example of the window provided by the Powder Cell program during refining. In this figure one can see the theoretical diffractogram, a design of the structure and the structural data that are introduced in the program.}
\label{PowCell}
\end{figure}

For analysis of XRD, the Rietveld {\bf Powder Cell} \cite{Pow} refinement program was used. The experimental diffractograms are adjusted by calculated model generated by the program from the crystallographic data of the compound. In Fig. \ref{PowCell} we see an example of how the data fits and provides us with a theoretical model to characterize. The AXRD data was analyzed by  using FULLPROF suite software \cite{fullprof} in the same way; however, it takes the correction factors for the data fit.  

\section{Magnetic measurement}

The magnetic measurements as a function of magnetic field and temperature were carried out in order of explore the magnetic properties of all samples. The equipment used were a commercial vibrating sample magnetometer (VSM) and a commercial superconducting
quantum interference device (SQUID) at {\it Universidade Estadual de Campinas} and {\it Quantum Condensed Matter Division, Oak Ridge National Laboratory}, in the temperature range between 4 K to 320 K and magnetic field between 0 and 70 kOe. In general, these measurement equipment operate based on the  magnetic flux change of the sample inside a detector coil. 

The most commonly used magnetization measurement protocols are zero field cooling (ZFC), field cooling (FC) and warming field cooling (WFC). ZFC is a measurement type of magnetization as a function of the temperature, which consists in  cooling the sample without an external magnetic field applied, to lowest measurement temperature. Then, the magnetic field is switched on, and the data is acquired during the temperature rise process. FC measurement is realized during the cooling process of the sample with applied magnetic field. WFC differs from FC because it is performed when the sample is in the heating process after being cooled with applied external magnetic field.

From these magnetic data, we can obtain the susceptibility $\chi$ as a function of the temperature (as in Chapter \ref{funda}), and  inverse susceptibility as a function of the temperature. From lineal fit at paramagnetic regime and Curie-Weiss law, we can obtain parameters such as  the effective magnetic moment, the Curie-Weiss constant and the Curie  temperature.


\part{Intermetallic alloys} \label{parte2}

	\chapter[Intermetallics with B: YNi$_{4-x}$Co$_x$B compounds ]{Intermetallics with B: competing anisotropies on $3d$ sub-lattice of YNi$_{4-x}$Co$_x$B compounds}\label{boratos}

In this chapter, we focus on competing anisotropies on $3d$ sub-lattice of YNi$_{4-x}$Co$_x$B alloys, which plays an important role on the overall magnetic properties of hard magnets. Intermetallic alloys with boron (for instance, $R$-Co/Ni-B) belong to those hard magnets family and are useful objects to help to understand the magnetic behavior of $3d$ sub-lattice, specially when the rare earth ions $R$ are not magnetic, for instance YCo$_4$B. Interestingly, YNi$_4$B is a paramagnetic material and the Ni ions do not contribute to the magnetic anisotropy. Here, we focused our attention on YNi$_{4-x}$Co$_x$B series, with $x=0, 1, 2, 3$ and $4$. For this purpose, we development two model for Co occupation into the crystallographic sites in the CeCo$_4$B type hexagonal structure.   
X-ray powder diffraction data were obtained at UFF and at room temperature, the magnetic measurements were carried out using a commercial vibrating sample magnetometer (VSM) and a commercial superconducting quantum interference device (SQUID) at Unicamp, and in order to determine composition and topology of the samples, we carried out scanning electron microscopy (SEM) measurements at IF Sudeste MG.

\section{Brief survey of $R_{n+1}$Co$_{3n+5}$B$_{2n}$ family alloys}

Since the 70's intermetallic alloys with boron, like Nd$_2$Fe$_{14}$B \cite{kirchmayr1996permanent} and SmCo$_4$B \cite{ido1994new}, have been very much studied due to their permanent magnets properties. Some of these materials were inspired by SmCo$_5$ \cite{Zhao1991}, from the substitution of Co by B into the $R_{n+1}$Co$_{3n+5}$B$_{2n}$ family ($R$= rare earth), with $n=1, 2, 3$ and $\infty$  \cite{Oda, Burzo, Zlotea}. It is well known that the magnetic anisotropy is an important property that rules the magnetic hardness of the material, specially the anisotropy from the $3d$ sub-lattice.

The aim of this effort is to provide further knowledge about the $3d$ magnetic anisotropy of intermetallic alloys with boron. To this purpose, we consider a non-magnetic rare-earth (yttrium), in order to be sure that the magnetic contributions are coming only from the $3d$ sub-lattice. In addition, we considered two transition metals: Ni and Co. From one side, Ni ions do not contribute to the magnetic anisotropy \cite{Bartolome200711}, while Co ions are extremely anisotropic \cite{Zhao1991L231}.

Thus, YNi$_{4-x}$Co$_x$B alloys were synthesized  in an arc furnace under argon atmosphere with appropriate amounts of cobalt, nickel, boron, and yttrium , with $x=0, 1, 2, 3$ and $4$. YNi$_4$B ($x=0$) is a paramagnetic material and does not present signatures of anisotropy \cite{mazumbar}, while YCo$_4$B ($x=4$) has its Curie temperature at 380 K and spin reorientation (due to a strong anisotropy competition), at 150 K \cite{kowalczyk1994local,cadogan,Thang1}. Therefore, it is  clear that the anisotropy of YNi$_{4-x}$Co$_x$B alloys depends on the Co content $x$. 

To explore these features, we develop a statistical model of Co occupation among the crystallographic (Wyckoff) sites ($2c$ with axial anisotropy and, $6i$ with planar anisotropy \cite{cadogan}), in which predicts a strong competing anisotropy among these two sites and spin reorientation for all samples of this series. On the other hand, a preferential model, in which Co ions go into a preferential position into Wyckoff sites, is developed and predicts that only $x=2$ and $x=4$ samples would have strong competing anisotropies with spin reorientation. Experimental measurements of magnetization on those samples verify that this last model successfully describes the nature of $3d$ magnetic anisotropy of this family. This preferential occupation of Co into $3d$ sites has a simple physical meaning: maximization of Co-Co distances. Indeed, this kind of approach was already experimentally verified with neutron diffraction measurements in other samples, like  PrNi$_{5-x}$Co$_x$ \cite{rocco} and YCo$_{4-x}$Fe$_x$B \cite{chacon2000652}. 

\section{Crystallography of YNi$_{4-x}$Co$_x$B alloys}\label{cristalografia}

\begin{figure}[h!]
\begin{center}
\includegraphics[width=10cm]{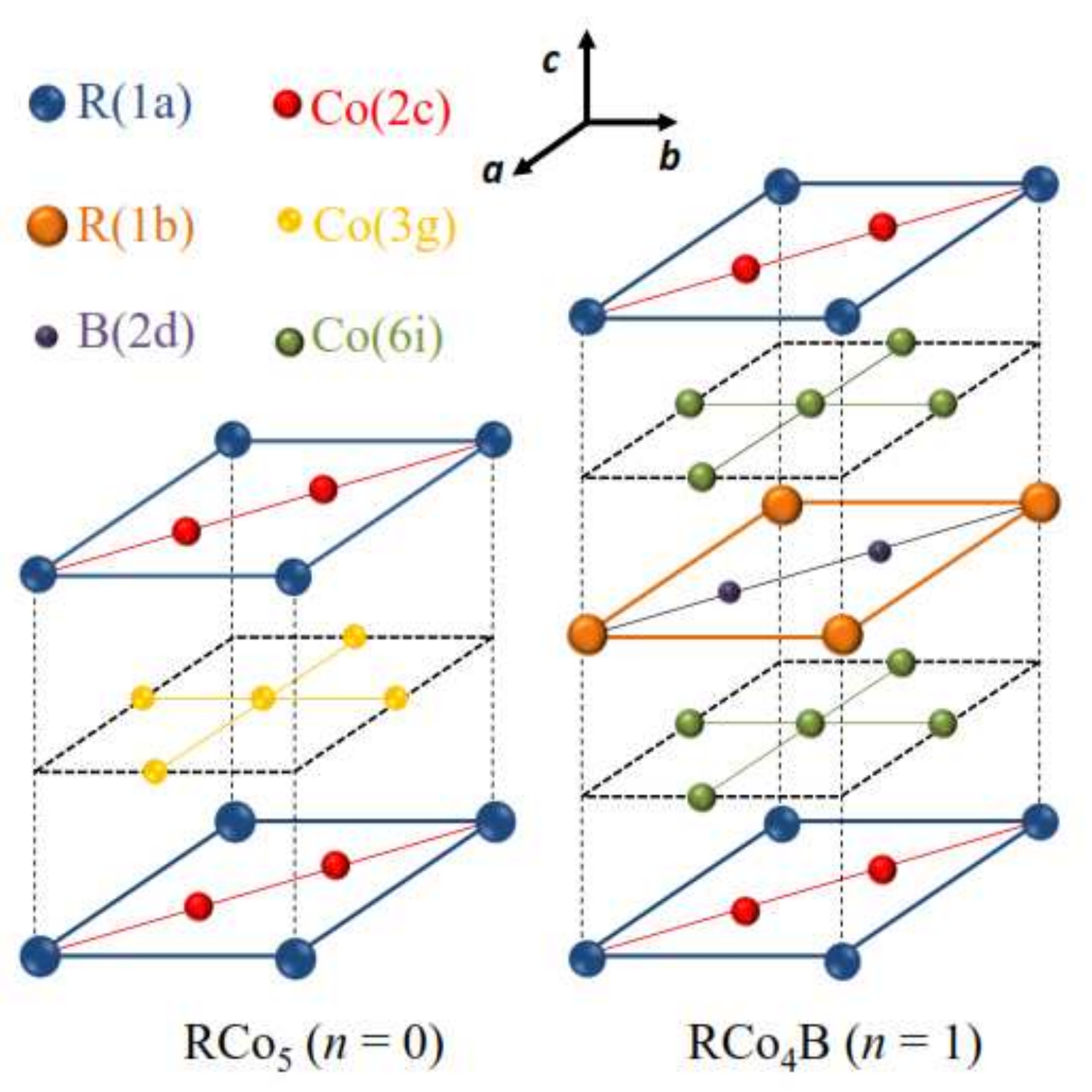}
\caption[Crystal structures by  R$_{n+1}$Co$_{3n+5}$B$_{2n}$ family]{ Crystal structure of  RCo$_5$ ($n=0$) followed by RCo$_4$B ($n=1$). Note that the crystal structure on the left has one unit formula (RCo$_5$), while the right structure has two unit formulas (R$_2$Co$_8$B$_2$).}\label{structure}
\end{center}
\end{figure}

$R_{n+1}$Co$_{3n+5}$B$_{2n}$ structures with $n = 1,2,3$ and $\infty $  are possible due to the replacement of Co by B in every second layer of $R$Co$_5$ ($n=0$) \cite{Oda, Burzo, Zlotea}, as illustrated in Fig. \ref{structure}. More precisely, $R$Co$_4$B ($n=1$) compound consist of two crystallographic sites for rare-earth: $1a$ and $1b$; two crystallographic sites for Co (or $3d$ ions): $2c$ and $6i$; and only one site for B ions: $2d$ \cite{systemynib}. This can be seen in the figure \ref{structure}, where the RCo$_5$ ($n=0$) case is also shown for comparison.

These compounds have the CeCo$_4$B type structure, with space group P6/mmm (ISCD n$^{\circ}$191) \cite{spada}. The first YNi$_4$B alloy was reported by Niihara \cite{niihara} with the same structure as above. Later, Kuz'ma and Khaburskaya \cite{systemynib} reported a superstructure with lattice constant $a=3a_0$ and $c=c_0$, where $a_0$ and $c_0$ are the lattice constant of the original structure found by Nihara. This superstructure was also found on YNi$_{4-x}$Co$_x$B series by Isnard and co-workers \cite{chaconf, chacon2, isnard1}.

\begin{figure}
\center
(a)\\\includegraphics[width=10cm]{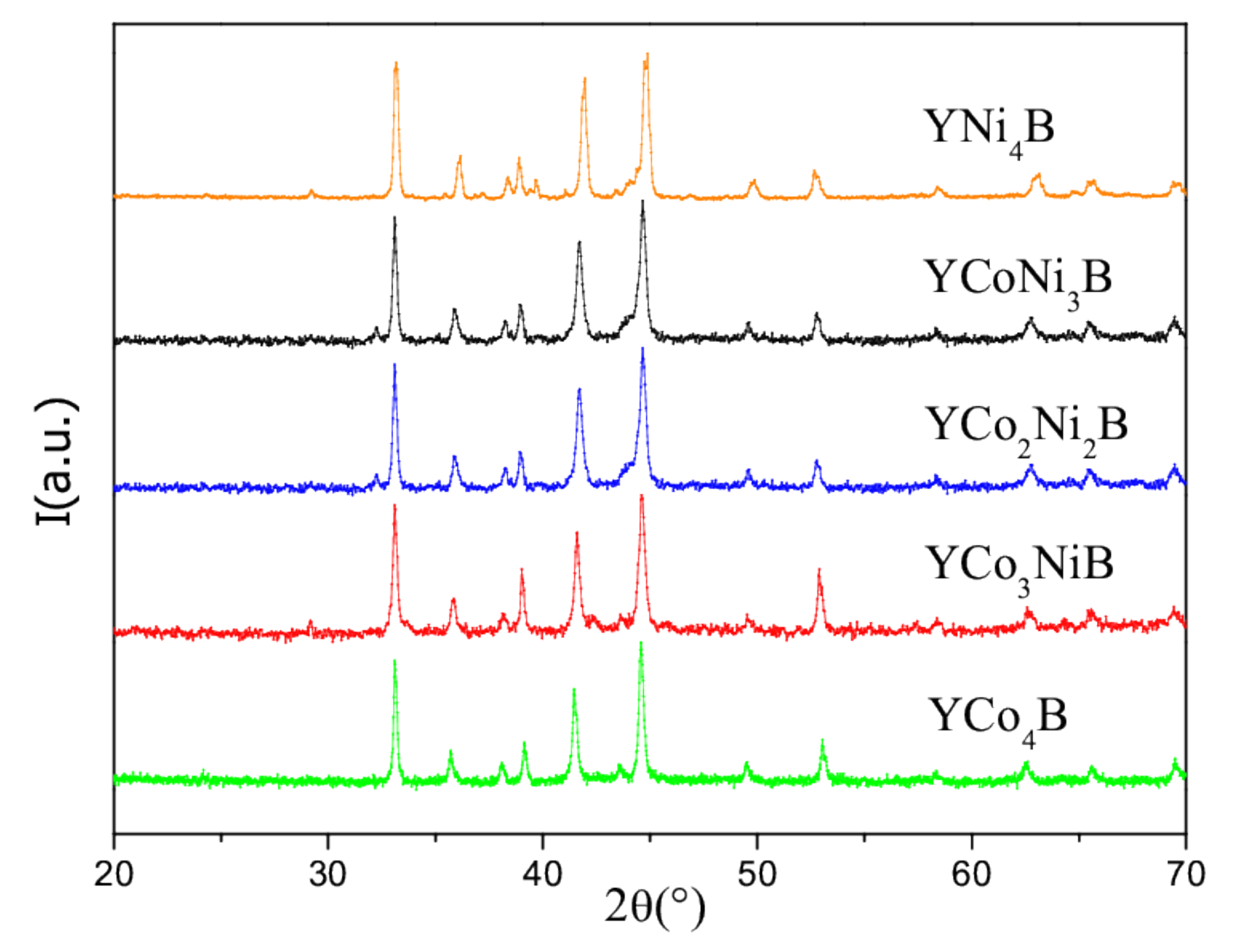}\\
(b)\\\includegraphics[width=8cm]{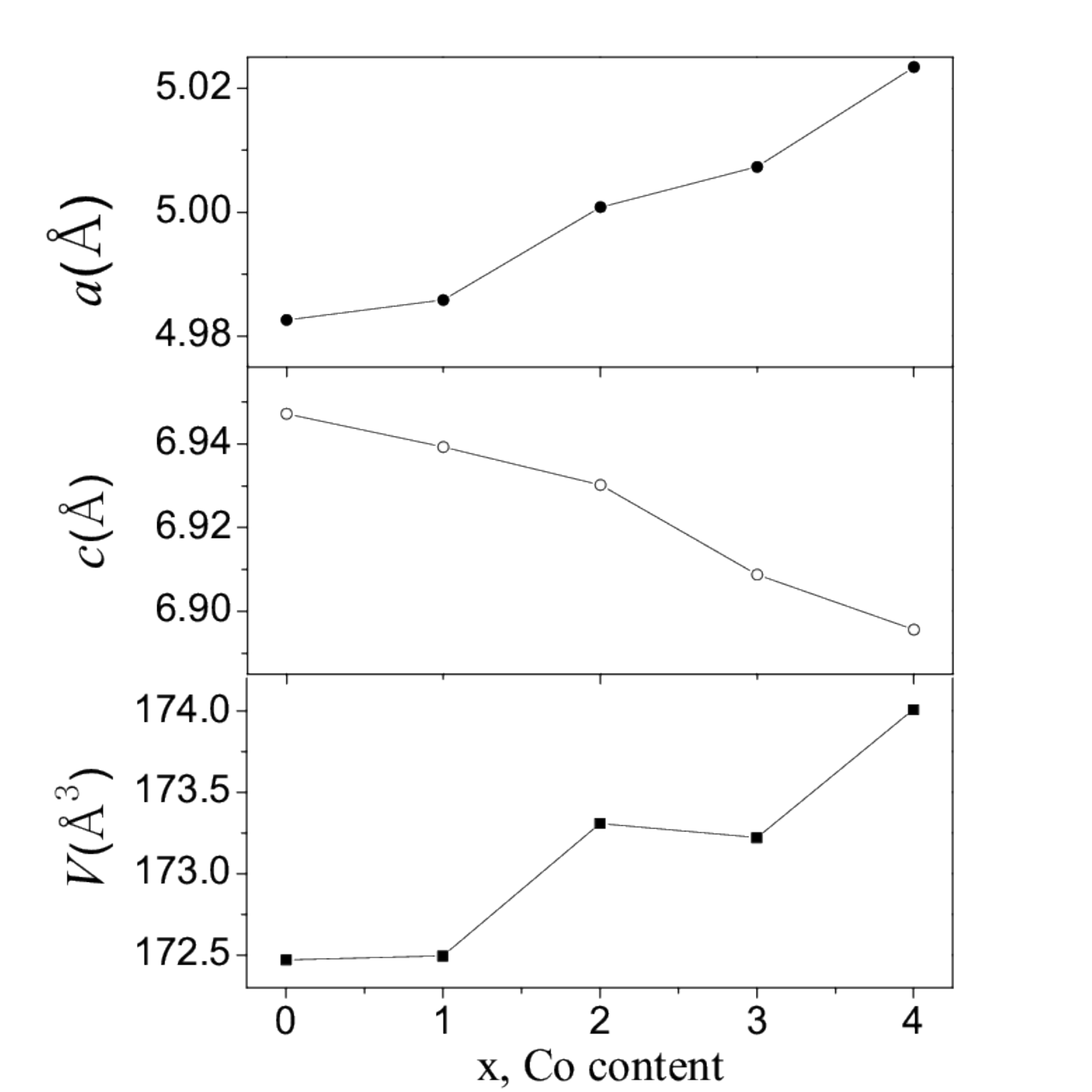}
\caption[X-ray patterns and cell parameters of YNi$_{4-x}$Co$x$B alloys]{Powder XRD pattern for YNi$_{4-x}$Co$_x$B alloys (a). Lattice constant and volume of the cell as a function of Co content (b). This behavior is similar to other R-M-B compounds  \cite{agil,chacon2}.}\label{acvol}
\end{figure}

\begin{figure}
\center
\includegraphics[width=15cm]{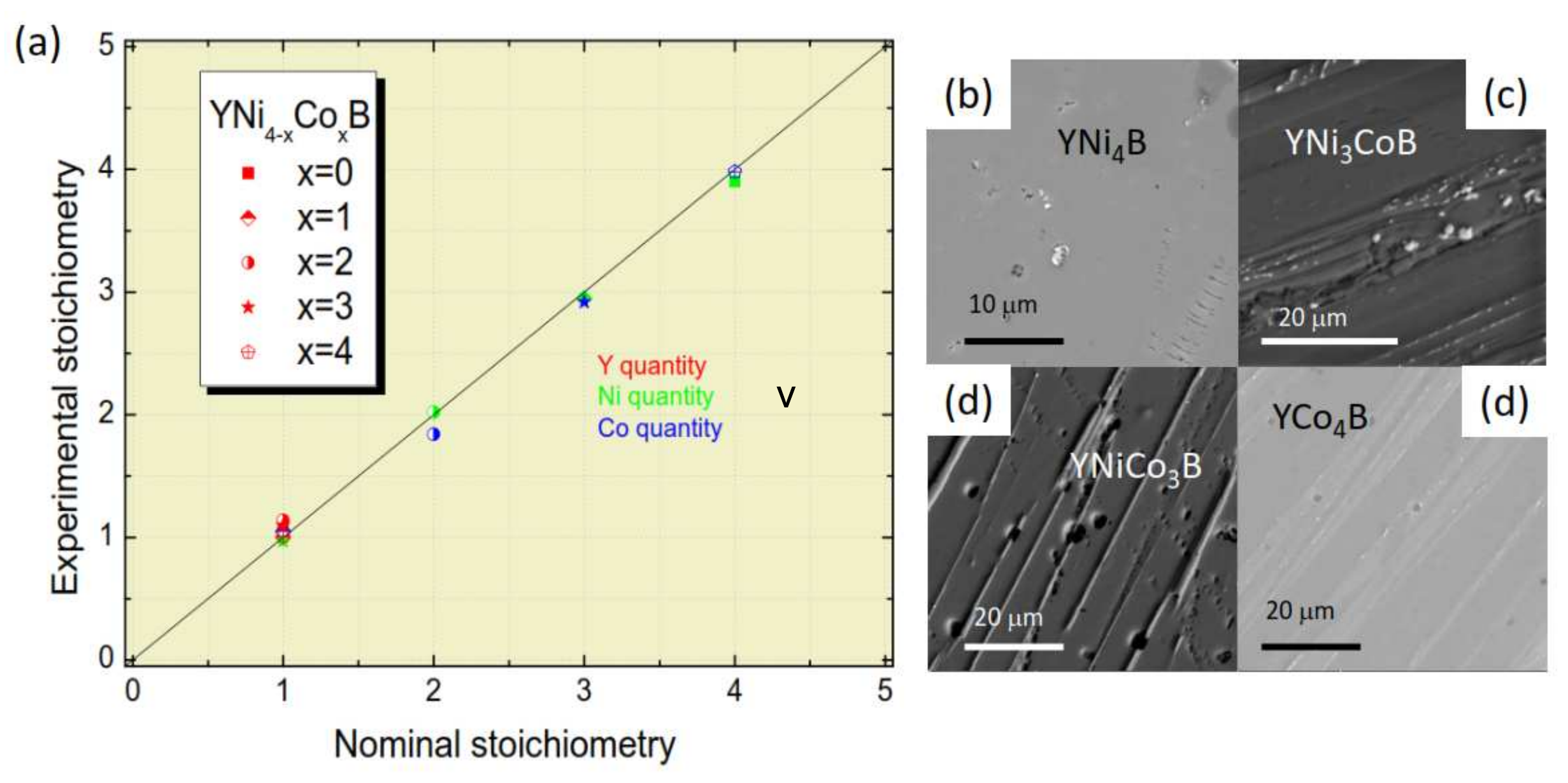}
\caption[SEM measuremnts of YNi$_{4-x}$Co$x$B alloys]{(a) Experimental and nominal stochiometry of the samples. SEM imagens realized at IF Sudeste MG of the samples from (b) to (e).}\label{MevYCoNi}
\end{figure}

X-ray diffraction on our samples show that all those crystallize in a single phase, similar to CeCo$_4$B structure (see figure \ref{acvol}-a), without extra picks in the diffractograms associated to the superstructure reported by Isnard et al.  \cite{isnard1}. The lattice parameters $a$ and $c$ were determined using the standard pattern matching method of the Powder Cell software \cite{Pow}; and these  change almost linearly as a function of Co content, as can be seen in figure \ref{acvol}-b. A similar behavior was found by Chacon on YNi$_{4-x}$Co$_x$B \cite{chacon2} and A$\check{\mbox{g}}$il et al. on PrNi$_{4-x}$Co$_x$B \cite{agil}. SEM measurements also show that the stoichiometry of the experimental composition of the samples are according with nominal composition (see Fig. \ref{MevYCoNi}).

\section{Competing anisotropies in 3d sub-lattice}\label{anisotropia}

The magnetic anisotropy of YNi$_{4-x}$Co$_x$B compounds is due to the presence of Co ions, since Ni ions do not contribute to the anisotropy \cite{mazumbar}. To understand the anisotropy of these compounds is necessary know from the crystallographic point of view, the mechanism of Ni/Co substitution. Let us first consider the YCo$_4$B compound,  where $2c$ and $6i$ sites are fully filled of Co. The anisotropy/ion, at 0 K, are known \cite{kowalczyk1994local} $\epsilon_{2c}=47.6514 \times 10^{-23}$ J/ion in $2c$ site  and $\epsilon_{6i}=-15.9804 \times 10^{-23}$ J/ion in $6i$ site; and therefore the total anisotropy for each site reads as: $E_{2c}=1\epsilon_{2c}= 47.6514 \times 10^{-23}$ J and $E_{6i}=3\epsilon_{6i}=-47.9412 \times 10^{-23}$ J \cite{kowalczyk1994local}. Note the pre-factor are the corresponding occupation factor of the Wyckoff sites (3/4 for $6i$ and 1/4 for $2c$), and we are considering only one formula unit, i.e., YCo$_4$B. On the other hand, $2c$ site has its magnetic moment with axial anisotropy, while $6i$ site has planar anisotropy (result from M\"ossbauer measurement \cite{cadogan}). These facts lead therefore to a strong competition of anisotropies, since the magnitude of those two contributions are the same, but the directions are different. The consequence is simple: a minor energy addition to the system (either thermal or magnetic, for instance), is able to unbalance this fragile equilibrium; and indeed it occurs: a spin reorientation from the plane to the axis happens at 150 K. Note these competing anisotropies lead to an almost vanishing overall anisotropy energy  $E_a = E_{2c}+ E_{6i}$.

To understand the magnetic anisotropy of the proposed series, we need analyses the mechanism of Ni/Co substitution. Thus, for a given compound of the YNi$_{4-x}$Co$_x$B series let us consider
\begin{equation}\label{bin}
P_{k} (x)= \frac{x!}{k!(x-k)!} p^k (1-p)^{x-k},
\end{equation}

\noindent as the probability of finding $k$ Co ions in the $6i$ site, for a given Co content $x$. It simply considers $6i$ site has a weight of $p=3/4$, due to its bigger size. Based on this distribution probability, let us focus on two different models: one with statistical distribution, in which all probabilities distribution are taken into account; and a second model, in which only the most probable distribution is considered. This latter represents a preferential site occupancy for the Ni/Co substitution, and its hypothesis has been verified in PrNi$_{5-x}$Co$_x$ \cite{rocco} and YCo$_{4-x}$Fe$_x$B \cite{chacon2000652}.

\begin{figure}[h!]
\center
\includegraphics[width=12cm]{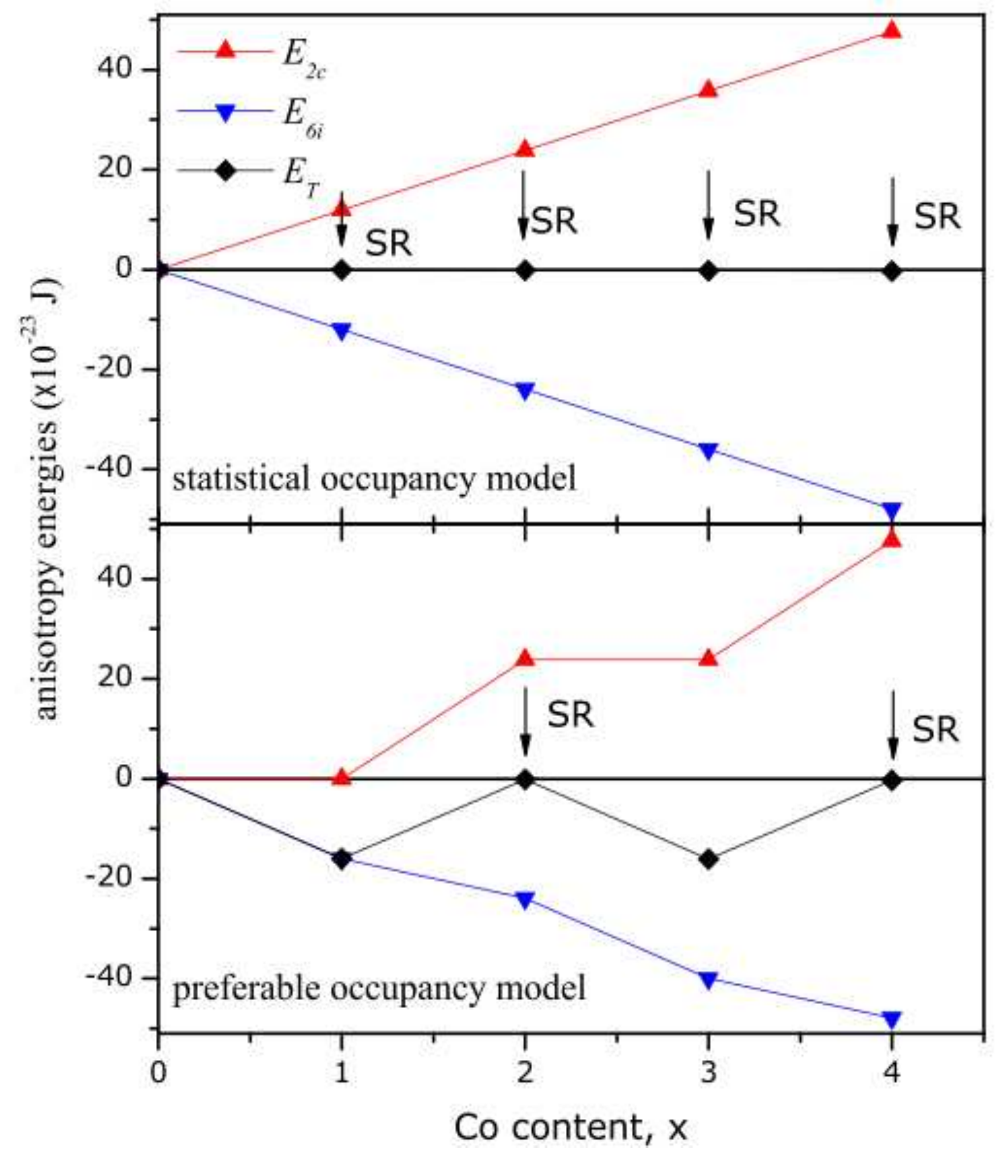}
\caption[Anisotropy energies models]{Anisotropy energies for the two considered models: statistical (top) and preferential (bottom) occupancy. SR means Spin Reorientation. Note the model on the top panel predicts SR for all concentrations with Co, while the model on the bottom panel predicts SR only for $x=2$ and 4, in accordance with the experimental magnetization data (see section \ref{magnetizacion}). \label{cuentas}}
\end{figure}

The first model considers all possibilities of occupancy to obtain the anisotropy energy for each site. Thus, it is straightforward to write:
\begin{align}
E_{6i}&= \sum^{x}_{k=0} P_k(x) k\, \epsilon_{6i},\\
E_{2c}&= \sum^{x}_{k=0} P_k(x) (x-k)\, \epsilon_{2c}.
\end{align}
Evaluation of the above energies leads to the result shown in top panel of Fig. \ref{cuentas}. This model predicts that the anisotropy energy of $2c$ and $6i$ sites are the same in magnitude for all samples, and therefore a strong anisotropy competition would be observed with further spin reorientation on all of them.

In a different fashion as before, the preferential occupation model considers that the Co ions are distributed among those two Wyckoff sites in a preferential fashion. To evaluate this idea, we considered only the most probable element of the set $\{P_{k} (x)\}$ and save its corresponding $k$ value, named as $k_{max}$, i.e.: $P_{k_{\mbox{{\tiny max}}}}= \mbox{max}\{ P_k(x)\}$. Thus, the anisotropy energy of each site can be written as 
\begin{equation}
E_{6i}= k_{\mbox{{\tiny max}}}\epsilon_{6i},
\end{equation}
and
\begin{equation}
E_{2c}= (x-k_{\mbox{{\tiny max}}})\epsilon_{2c}.
\end{equation}

The most probable distributions, for each value of Co content $x$, are shown on Table \ref{tablaFig}. Note the physical meaning of this preferential occupation model: Co ions try to keep the maximum distance of each other. 

The bottom panel of Fig. \ref{cuentas} summarizes the results of the of latter model. For $x=1$, it predicts that the anisotropy energy of the $2c$ site is zero (there are no Co ions in this site for such Co concentration), while the anisotropy energy of $6i$ site is finite. As a consequence, there is not a competing anisotropy and the magnetic moment of the $6i$ site stays in the basal plane. Obviously, without anisotropy competition there is no spin reorientation. This analysis is similar to  associated with the case  $x=3$, i.e., there is neither anisotropy competition no spin reorientation. The scenario is different for the samples with $x=2$ and $x=4$. For these two samples, the anisotropy energy of each site is comparable, leading to a strong anisotropy competition and, therefore, to a spin reorientation. Summarizing, the present model considers a preferential occupation of the Wyckoff sites, given by the most probable value of the distribution considered in equation \ref{bin}. The physical roots of this model, interestingly, is to maximize the Co-Co distances. As a consequence, we found magnetic anisotropies for all samples of the series. However, competing anisotropies with a consequent spin reorientation is found only for samples with $x=2$ and $x=4$. It is important take account that this kind of model was experimentally verified previously in similar materials: PrNi$_{5-x}$Co$_x$ \cite{rocco} and YCo$_{4-x}$Fe$_x$B \cite{chacon2000652}. 

\begin{table}[h!]
\centering
\begin{tabular}{|c|c|c|c|c|c|c|}
\hline
    & Sub-lattice 3d &$E_{2c}$($\times 10^{-23}$ J) & $E_{6i}$($\times 10^{-23}$ J)& T$_{SR}$(K) & T$_c$(K)\\
\hline
YNi$_4$B& \includegraphics[width=2.5cm]{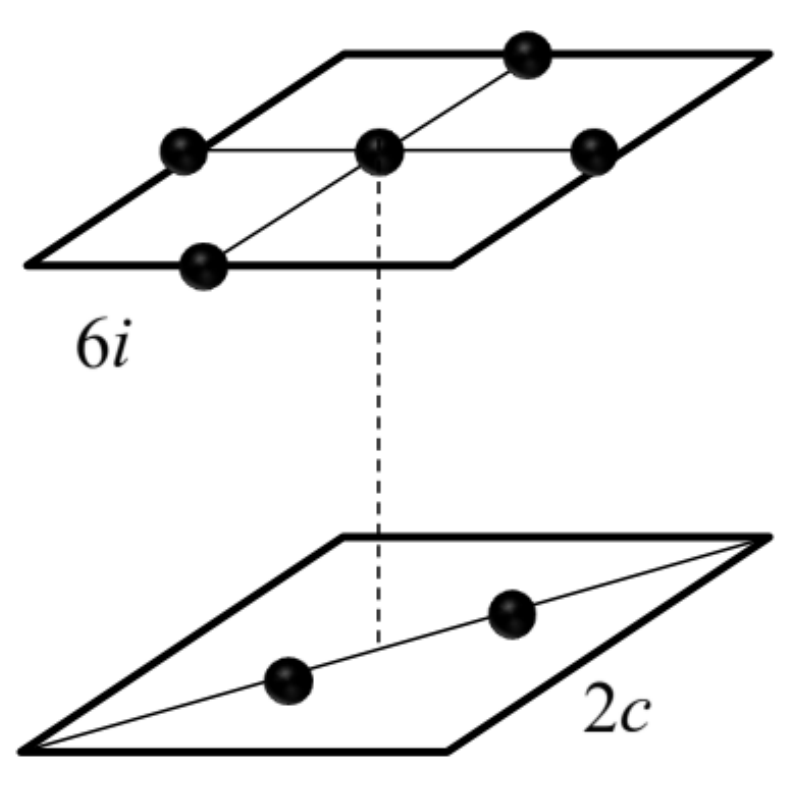}& 0& 0  & no & no\\
\hline
YNi$_3$CoB & \includegraphics[width=2.5cm]{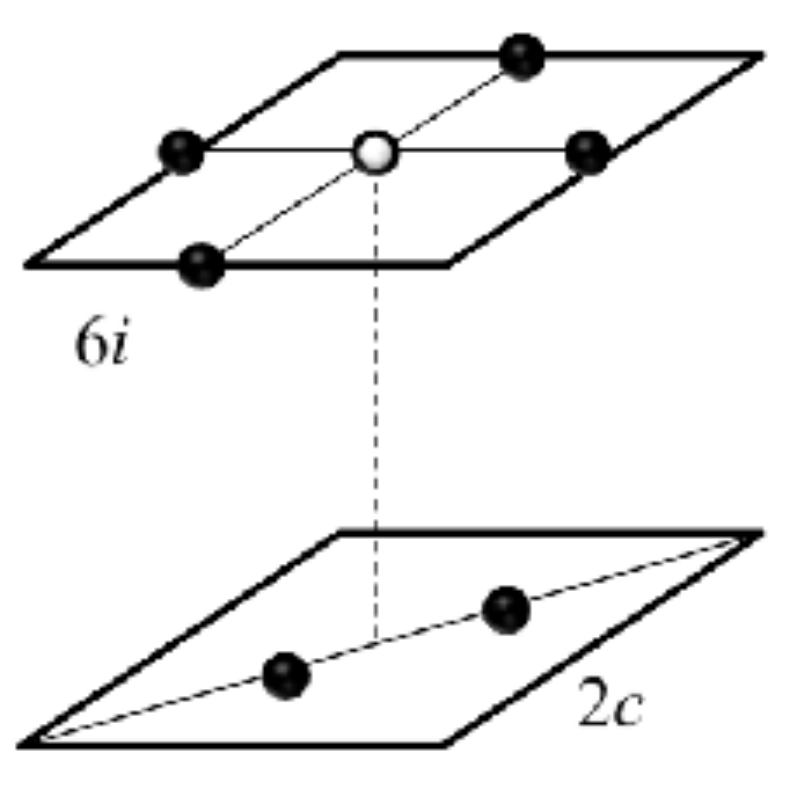}&  0 & -15.98 &  no & 180
\\
\hline
YNi$_2$Co$_2$B& \includegraphics[width=2.5cm]{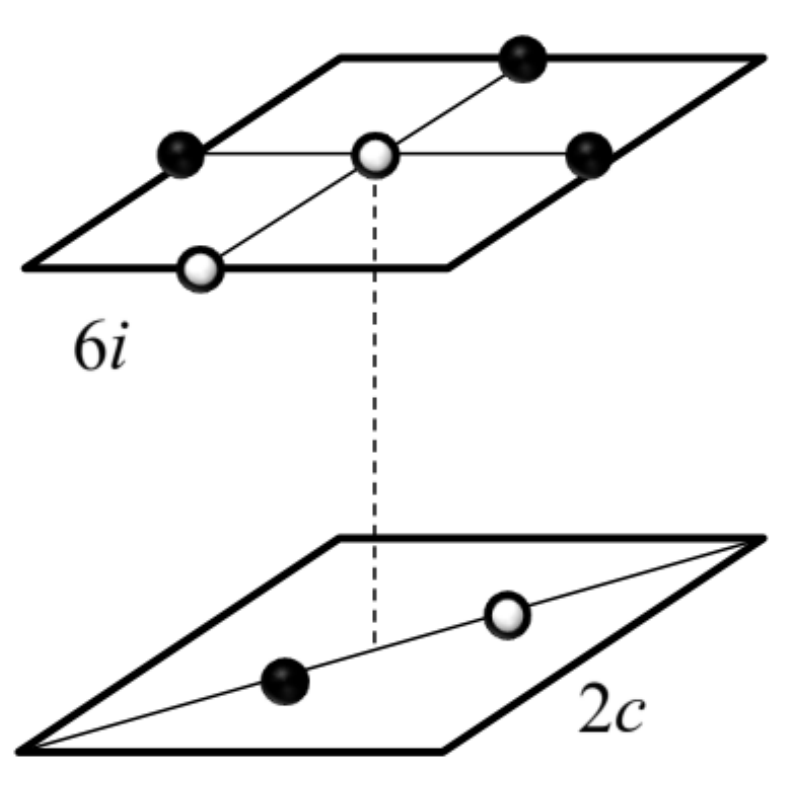}& 23.83 & -23.97 &  150& 307\\
\hline
YNiCo$_3$B& \includegraphics[width=2.5cm]{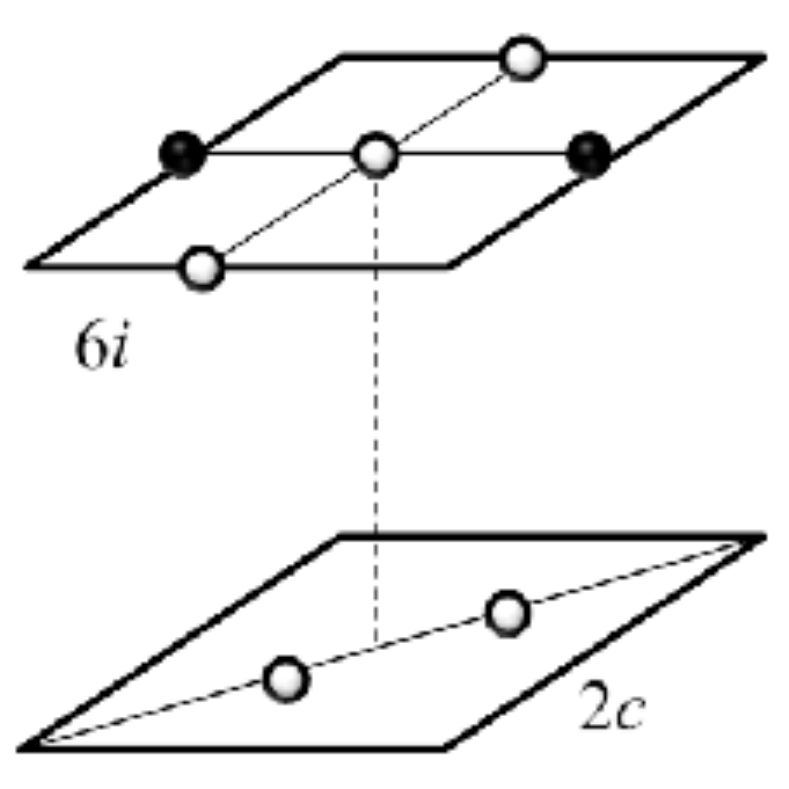}&  23.83& -39.95&  no & 314\\
\hline
YCo$_4$B& \includegraphics[width=2.5cm]{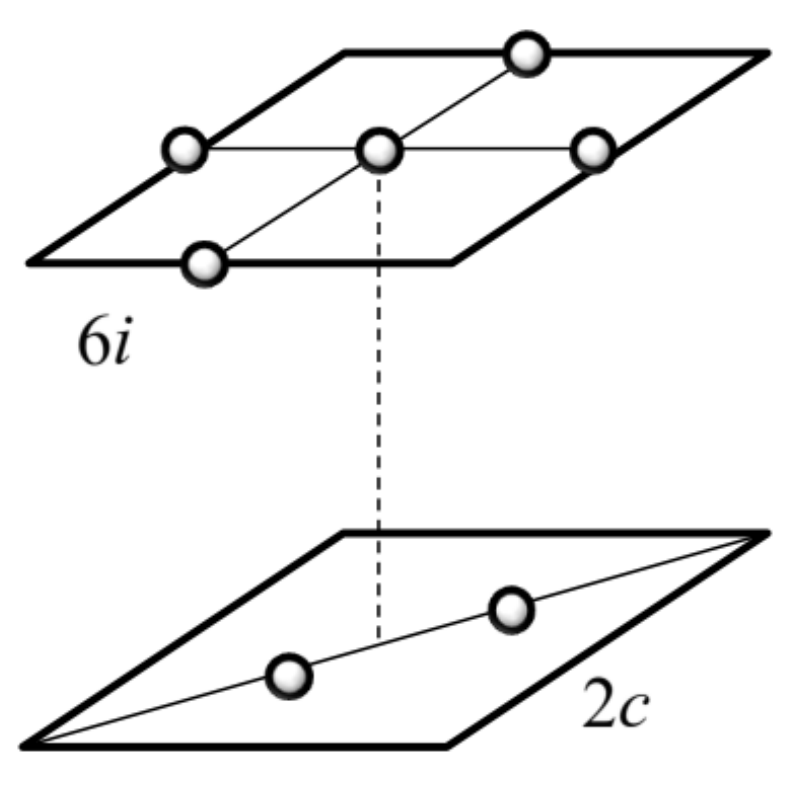}&  47.65 &-47.94 & 150& 380\\
\hline
\end{tabular}
\caption[YNi$_{4-x}$Co$_x$B alloys and 3d competing]{YNi$_{4-x}$Co$_x$B compounds and the corresponding $3d$ site anisotropy energies and crystallographic preferantial distribution among the Wyckoff sites.  $\bullet$ represents Ni  and  $\circ$ represents Co. Note $E_{2c}$ and $E_{6i}$ are theoretical values, while T$_{SR}$ and T$_c$ are experimental.} \label{tablaFig}
\end{table}

\section{Magnetism in YNi$_{4-x}$Co$_x$B alloys}\label{magnetizacion}

Nagarajan et al. \cite{Mazumdar10} studied the YNi$_4$B compound and observed that it behaves as a paramagnet from room temperature down to 12 K. Below this threshold temperature, they found a superconducting behavior. Later \cite{Andrzejewski, Nagarajan1, Hong}, this superconducting behavior was ascribed to be from an additional phase containing carbon. On the other hand, YCo$_4$B is a ferromagnetic material with  $T_c=380$ K, exhibiting spin reorientation at 150 K due to the competition between the two crystallographic sites of Co \cite{Thang1}. The magnetic properties of the RNi$_{4-x}$Co$_x$B compounds were studied for R = Sm \cite{mazumbar}, Pr \cite{agil}, Nd \cite{Soulie} and La \cite{Ito1}. In all these systems the saturation magnetization ($M_s$) and the Curie temperature ($T_c$) increase monotonically with Co content.
	
\begin{figure}[h!]
\begin{center}
(a)\\
\includegraphics[width=12cm]{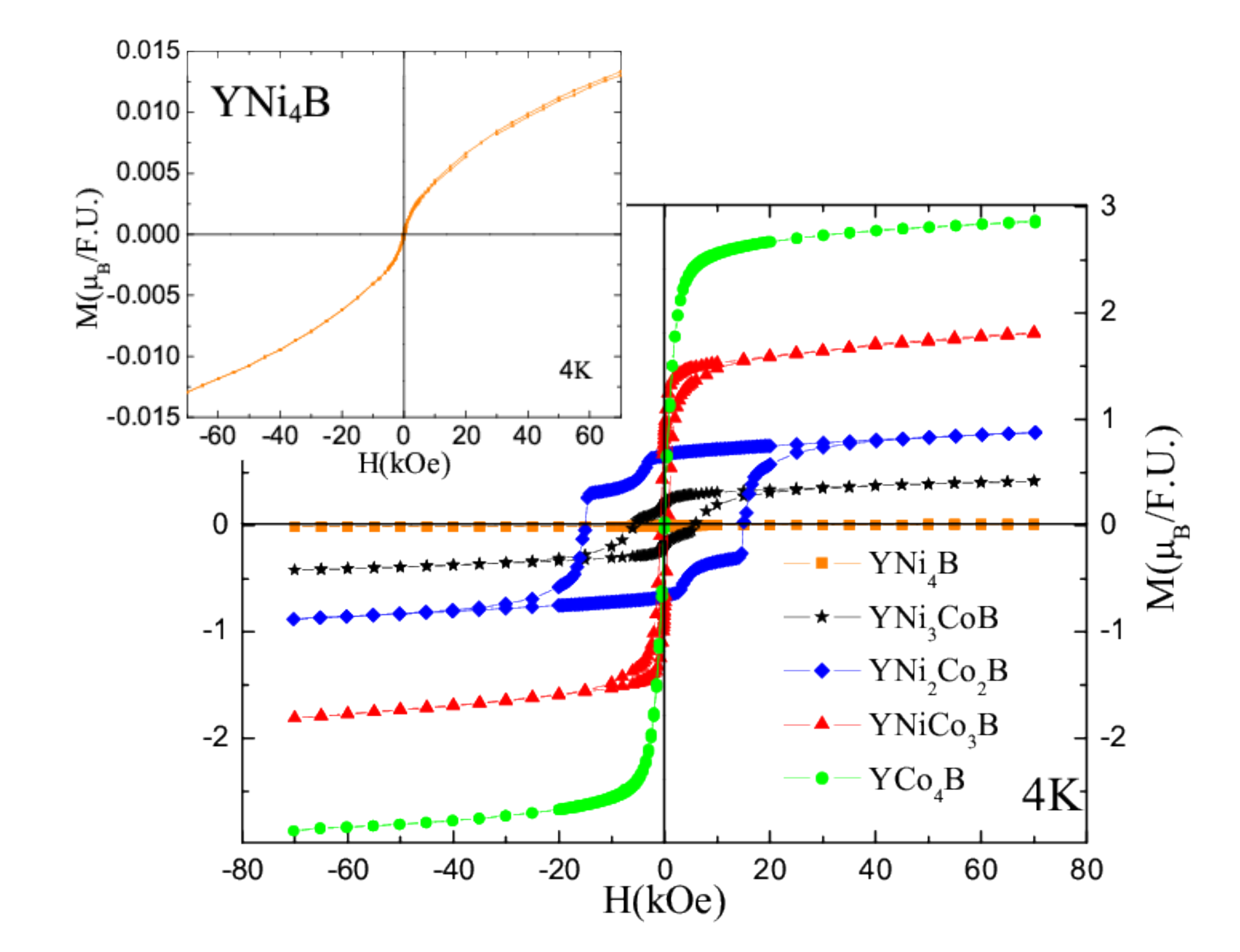}\\
(b)\\
\includegraphics[width=12cm]{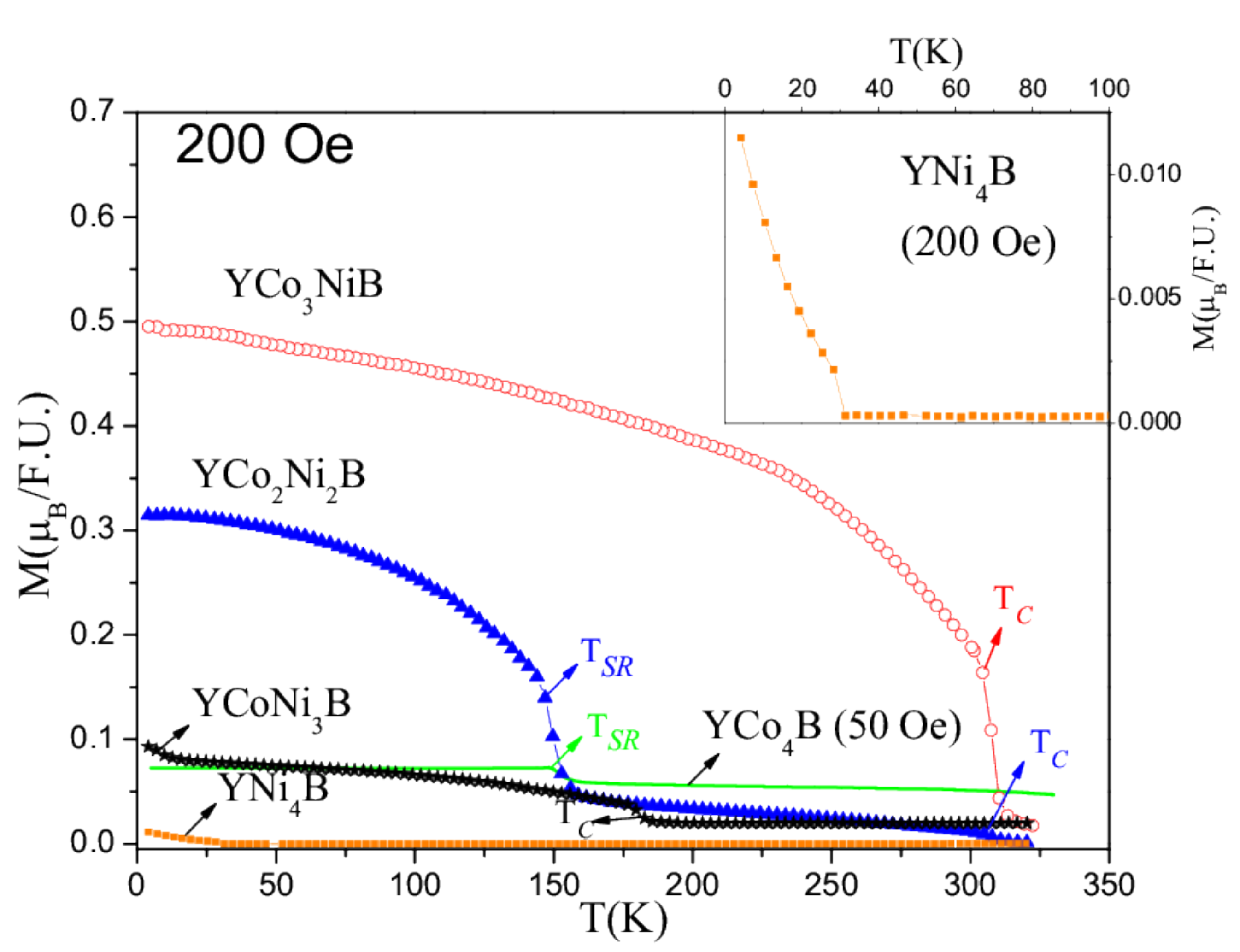}
\end{center}
\caption[Magnetims in YNi$_{4-x}$Co$x$B alloys]{Magnetic moment per formula unitary  as a function of (a) external magnetic field and (b) temperature. The inset magnifies the YNi$_4$B result, since it was hidden when presenting the result of all samples together. \label{M_H_T}}
\end{figure}

In our samples, we measured magnetization as a function of magnetic field at 4K (see Fig. \ref{M_H_T}-a). The YNi$_4$B sample shows no hysteresis, and has a quite small value of magnetization. Increasing the Co content, the hysteresis width becomes larger, with the maximum width occurring for $x=2$. This fact is in accordance with both models, since the anisotropy energy of each Wyckoff site promotes this hysteresis. Note also that the saturation value of magnetization for these samples increases with the Co content.

The temperature dependence of the magnetization was also measured, and the results exrepessed as the magnetic moment per formula unitary are presented in  Fig. \ref{M_H_T}-b. YNi$_4$B is indeed paramagnetic with a possible superconducting behavior below 20 K, in accordance with references  \cite{Andrzejewski, Nagarajan1, Hong}. By adding Co, the compound become ferromagnet, for example, Isnard \cite{isnard1} showed that the sample with $x=1$ has a ferro-paramagnetic phase transition at T$_c=180$ K, \emph{without} a spin reorientation phenomena. For  $x=2$, we observed strong drop of magnetization at 150 K as a spin reorientation,  and a much higher Curie temperature of 310 K. This series is able to receive more Co ions, for $x=3$ sample, we observe the Curie temperature at 307 K, in accordance with reference \cite{isnard1}. Finally, the last sample of our series exhibits a spin reorientation at 150 K and the para-ferromagnetic Curie temperature at 380 K, in agreement with Refs. \cite{kowalczyk1994local,cadogan,Thang1}. These remarkable temperatures are exhibits in Table \ref{tablaFig}.

It is worth to note that our experimental result are in excellent  agreement with the model of preferencial occupancy, which predicts spin reorientation for  the samples with $x=2$ and $x=4$ only (see Fig \ref{cuentas} and Table \ref{tablaFig}).

\section{Concluding remarks on YNi$_{4-x}$Co$_x$B alloys}\label{conclusiones}

Two possible occupation models for Co ions in the $3d$ sub-lattice of the YNi$_{4-x}$Co$_x$B samples has been analyzed. One  considers a statistical distribution of Co/Ni ions, while the other considers a preferential occupation for Co ions. The former predicts strong anisotropy competition among the two $3d$ possible Wyckoff sites ($2c$ and $6i$) with spin reorientation for all Co contents. In contrast, the second model predicts that both sites have strong anisotropies, however, the competition ($|E_{2c}|=|E_{6i}|$) and spin reorientation arises only for $x=2$ and $x=4$. Our experimental data of magnetization as a function  of magnetic field and temperature show that only $x=2$ and $x=4$ compositions exhibits spin reorientation in agreement with the preferential occupation model. Similar results have been previously obtain for other compounds \cite{rocco,chacon2000652}. From the physical point view, this preferential occupation model interestingly mimics the case in which Co-Co distances are maximized.


	\chapter{Heusler alloys: general consideration}\label{capitulo6}

The Heusler alloys family have attracted great scientific and technological interest mainly in the fields of spintronics and magnetocaloric effect, among others. In this chapter, we review some general properties of Heusler alloys, which will help us to  the following chapters that deal with these compounds.

\section{Introducing Heusler alloys}

\begin{figure}[t]\begin{center}
\includegraphics[width=15cm]{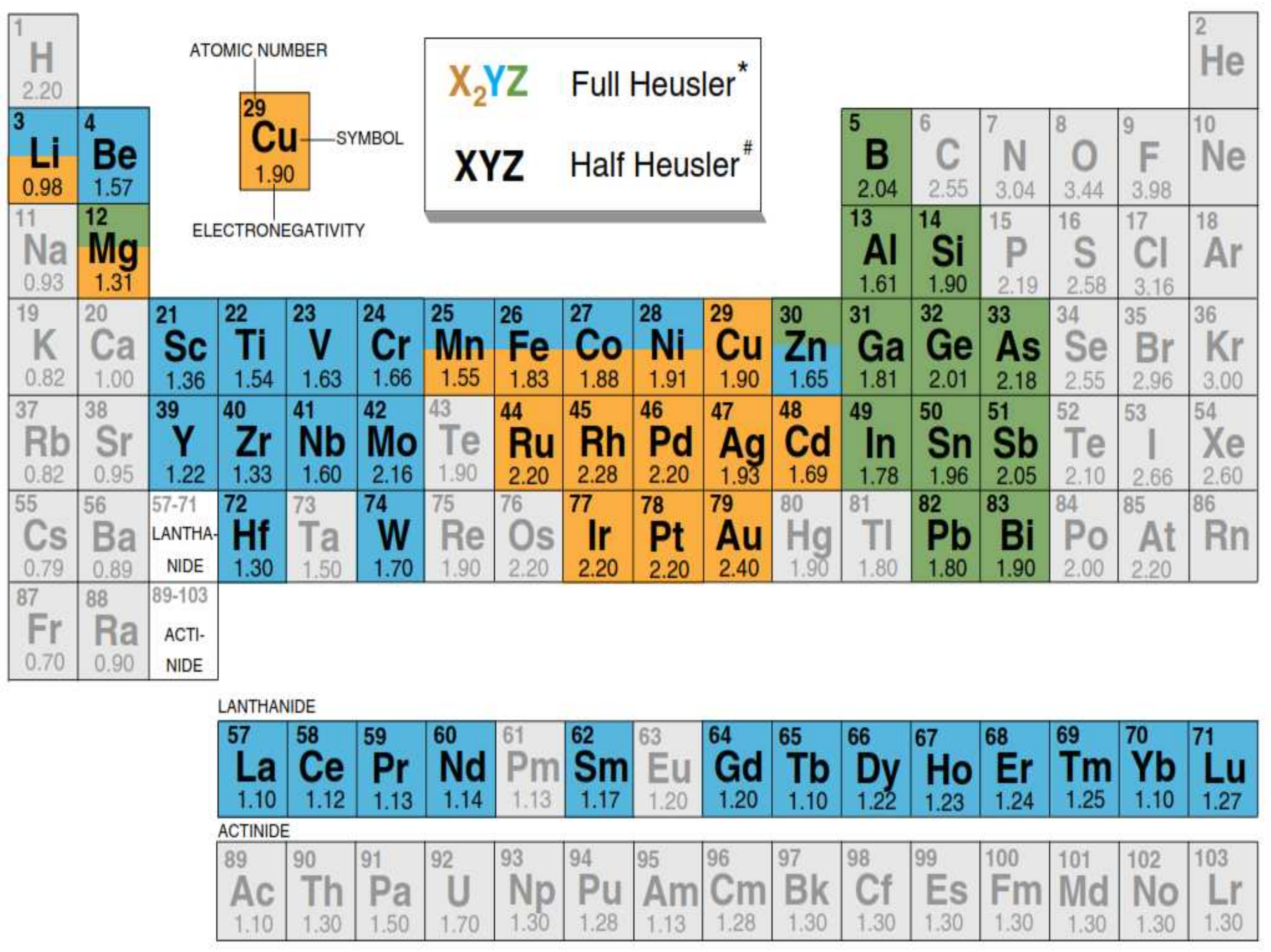}
\caption[Periodic table of Heusler alloys]{Periodic table of Heusler alloys. The huge number of full-Heusler compounds can be formed by combining different elements according to the color scheme. For half-Heusler compounds XYZ, the elements can be ordered according to their electronegativity. This table is according to IUPAC and is found in Ref. \cite{bai2012data}.}
\label{TPH}\end{center}
\end{figure}

In 1903 Fritz Heusler discovered that it was possible to make ferromagnetic alloys entirely from non ferromagnetic elements such as copper-manganese bronze alloyed with tin, aluminum, arsenic, antimony, bismuth, or boron \cite{heusler1904,webster1969heusler}. Then, the constituent elements of the Heusler compounds cover almost the whole periodic table, as shown in Fig. \ref{TPH}.

We can subdivide the Heusler alloys into the following categories, according with the stoichiometric function, nomenclature, and ion positions:

\begin{itemize}

\item[$\Box$] {\it Half-Heusler alloys} (or semi-Heusler alloys) have the form XYZ, where X, Y and Z are mainly transition metals or are replaced by rare earth metals, metalloids, or other elements like Li, Be, and Mg. Thus, the nomenclature order in the literature varies greatly, ranging from sorting the elements alphabetically, according to their electronegativity, or randomly, and then all three possible permutations can be found according to the convenience of the authors. This semi-Heusler type crystallize in the $C1_b$ cubic structure.

\item[$\Box$] {\it Full-Heusler alloys} (or simply Heusler alloys) are intermetallics with the X$_2$YZ form, where X and Y are transition metals, and metalloids are mainly used for Z. Thus, the nomenclature of these compounds generally has two parts of the first metal followed by other metal element of minor quantity and finally by a metalloid, e.g., Co$_2$FeSi, Fe$_2$MnGa, and many others. The first Heusler alloys studied crystallized in the $L2_1$ structure, which consists of 4 interpenetrating $fcc$ sublattices in an $A$, $B$, $C$ and $D$ order. Here, the sequence of the atoms is X-Y-X-Z, in a  Cu$_2$MnAl-type structure  \cite{bradley1934crystal}. In this thesis, we focus on the study of this Heusler alloys type.

\item[$\Box$] {\it Inverse Heusler alloys} are compounds that also have the chemical formula of X$_2$YZ, but in their case, the valence of an X transition metal atom is smaller than the valence of an Y transition metal atom. As a consequence, inverse Heusler compounds crystallize in the so-called $XA$ or $X_{\alpha}$ structure, where the sequence of atoms is X-X-Y-Z and the prototype is Hg$_2$TiCu-type \cite{ozdog2009first}. This is energetically preferred to the $L2_1$ structure of the usual full-Heusler compounds. Several inverse Heusler alloys have been studied using first-principles electronic structure calculations \cite{ozdog2009first,kervan2012first,bayar2011half}. Inverse Heusler alloys are interesting for applications since they combine coherent growth of films on semiconductors with large Curie temperatures (e.g. Cr$_2$CoGa \cite{galanakis2011high}). 

\item[$\Box$] {\it Quaternary Heusler alloys} are compounds with the chemical formula of XX$^{\prime}$YZ, where X, X$^{\prime}$, and Y are transition metal atoms. The valence of X$^{\prime}$ is lower than the valence of X, and the valence of the Y element is lower than the valences of both X and X$^{\prime}$. The sequence of the atoms in the $fcc$ structure is X-Y-X$^{\prime}$-Z which is energetically the most stable \cite{alijani2011quaternary}. A large series of such compounds has recently been studied \cite{xu2013new,ozdougan2013slater}.

\end{itemize}

\begin{figure}\begin{center}
\includegraphics[width=13cm]{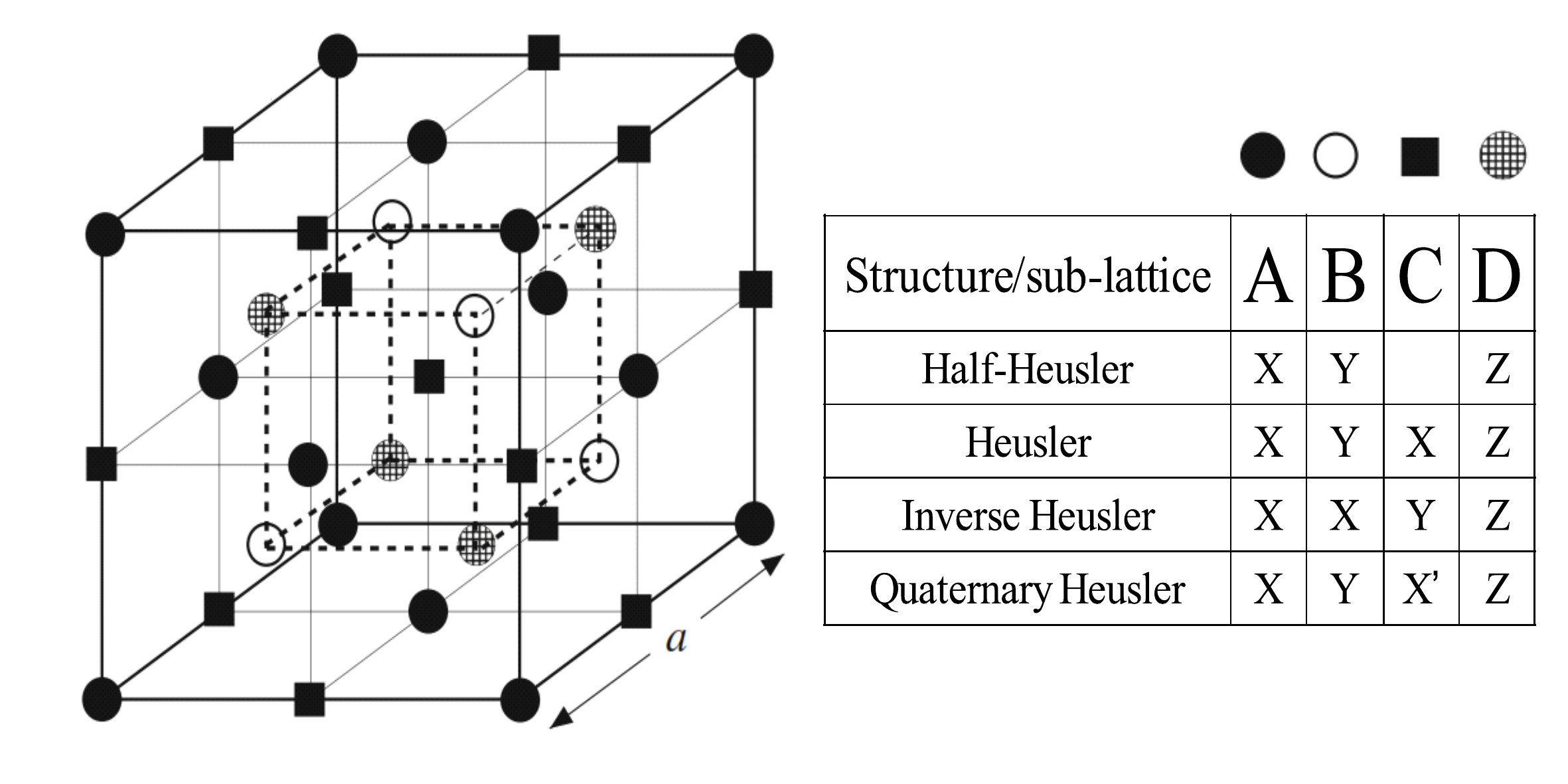}
\caption[Schematic representation of Heusler alloys]{Schematic representation of various structures of semi- and full-Heusler compounds. In all cases, the lattice consists of four interpenetrating $fcc$ lattices. If all atoms were identical, the lattice would simply be $bcc$ \cite{book2015heusler}.}
\label{eH}\end{center}
\end{figure}

Fig. \ref{eH} shows the positions of the constituent atoms in each Heusler alloy type. In all cases, the lattice consists of four interpenetrating $fcc$ lattices, except for half-Heusler alloys, where the $C$ sublattice  is not occupied.

\section{Crystal structure of Heusler alloys}\label{secCrystalHeusler}

The family of full-Heusler compounds X$_2$YZ crystallize in the cubic space group $Fm\bar{3}m$ (space group no. 225 ICSD) in a Cu$_2$MnAl ($L2_1$) structure type \cite{bradley1934crystal}. X atoms occupy Wyckoff position 8c (1/4,1/4, 1/4), Y and Z atoms are located at 4a (0, 0, 0) and 4b (1/2,1/2, 1/2), respectively. This structure consists of four interpenetrating $fcc$ sublattices, two of which are equally occupied by X and the remaining atoms occupy the other two sublattices, as illustrated in the Fig. \ref{eH}.

However, various variants of the $L2_1$ structure can be formed, if X and/or Y atoms are intermixed at the respective crystallographic positions, leading to different (local) symmetries and structure types \cite{bacon1971chemical}. We describe the most common types of structures in what follows.

\begin{figure}[t]
\begin{center}
\includegraphics[width=15cm]{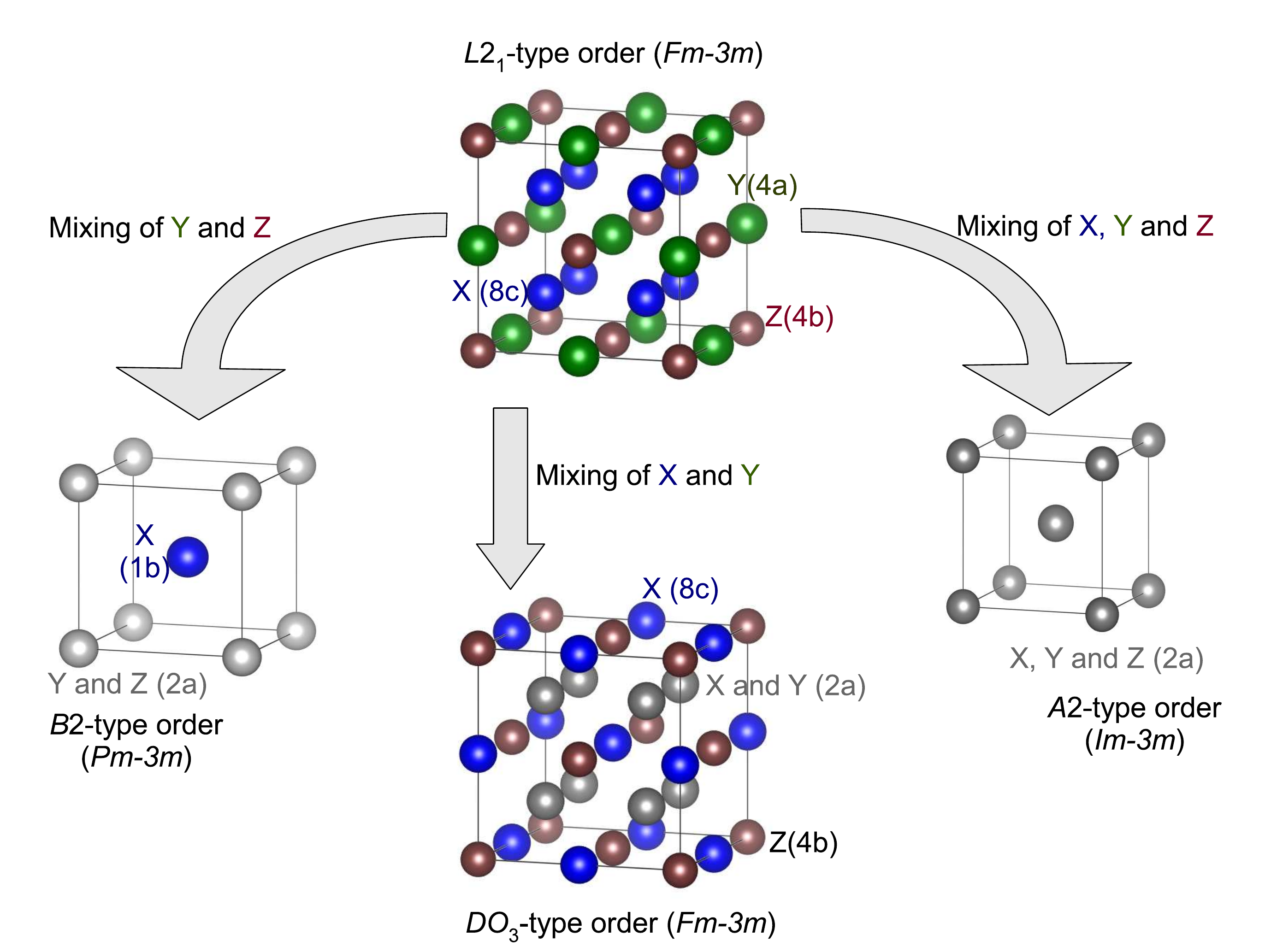}
\caption[Structure types for X$_2$YZ Heusler compound]{Structure types for X$_2$YZ Heusler compounds \cite{book2015heusler}.}
\label{typesHA}\end{center}
\end{figure}

\begin{itemize}

\item[$\bullet$] {\bf $B2$-type structure}

If Y and Z atoms are randomly intermixed at their crystallographic positions, a $B2$-type structure is obtained, in which Y and Z sites become equivalent. This structure may also be described using a CsCl lattice, and as a result of this intermixing, the CsCl lattice with X at the center of the cube randomly surrounded by Y and Z atoms is obtained (see Fig. \ref{typesHA}). The symmetry is reduced, and the resulting space group is $Pm\bar{3}m$. All X atoms are at the $1b$ Wykhoff position, Z and Y atoms are randomly distributed at the $1a$ position.

\item[$\bullet$] {\bf $A2$-type structure}

A completely random intermixing at the Wykhoff $2a$ position in X$_2$YZ Heusler compounds between all sites results in the $A2$-type structure  $Im\bar{3}m$ with reduced symmetry. The X, Y, and Z sites become equivalent leading to a body-centered cubic lattice, also known as the tungsten (W) structure-type (see Fig. \ref{typesHA}).

\newpage
\item[$\bullet$] {\bf $DO_3$-type structure}

The space group $Fm\bar{3}m$ is kept, but if X and Y atoms are mixed at their crystallographic positions, a $DO_3$-type structure is obtained; the corresponding ICSD notation is BiF$_3$ structure type (see Fig. \ref{typesHA}).

\end{itemize}

The different structure types as described above will lead to the formation of different local environments for each atom. This is known as the atomic disorder and causes modifications in the intrinsic properties of these materials. In many cases, X-ray diffraction with a Cu$K_{\alpha}$ source is not enough for experimental determination of the disorder, and  special measurement setups such as anomalous X-ray diffraction (AXRD), Nuclear Magnetic Resonance Spectroscopy or M\"ossbabuer spectroscopy are required.  In Chapter \ref{cap7CFS}, this problem  is discussed in detail.

\section{Half-metallic ferromagnetism Heusler alloys}\label{SPHA}

The first attractive property of Heusler compounds comes from their magnetic characteristics.  F. Heusler found that the Cu$_2$MnSn alloy is ferromagnetic although it is comprised of nonferromagnetic elements at room temperature \cite{heusler1904}. Despite this, the impact of these compounds in the scientific community was not high for several decades. Until the eighties, the electronic structure of several Heusler compounds were investigated, and an unexpected result was found: depending on the spin direction, certain Heusler materials showed metallic as well as insulating properties at the same time; a feature called half-metallic ferromagnetism \cite{de1984half,kubler1983formation}, which can be found in other materials such as half-metallic oxides (Sr$_2$FeMoO$_6$, Fe$_3$O$_4$, La$_{0.7}$Sr$_{0.3}$MnO$_3$) \cite{coey2003half,saloaro2016towards}, diluted magnetic semiconductors \cite{Dietl}, for instance.

Half-metallic ferromagnets are metals with 100 \% spin-polarized electrons at the Fermi energy, i.e., electron with one spin direction behave as an insulator o semiconductor, while those with opposite spin are metallic. These compounds can be used for spintronic devices such as spin filters \cite{Victora}, and tunnel junctions \cite{Tanaka, Caballero}, among other. Formally, the complete spin polarization of charge carriers in half-metallic ferromagnets is only reached in the limiting case of zero temperature and vanishing spin-orbit interactions. Since most of the Heusler compounds containing only $3d$ elements do not show significant spin-orbit coupling, they are ideal candidates to exhibit half-metallic ferromagnetism \cite{graf2011simple}. 

\begin{figure}\begin{center}
\includegraphics[width=12cm]{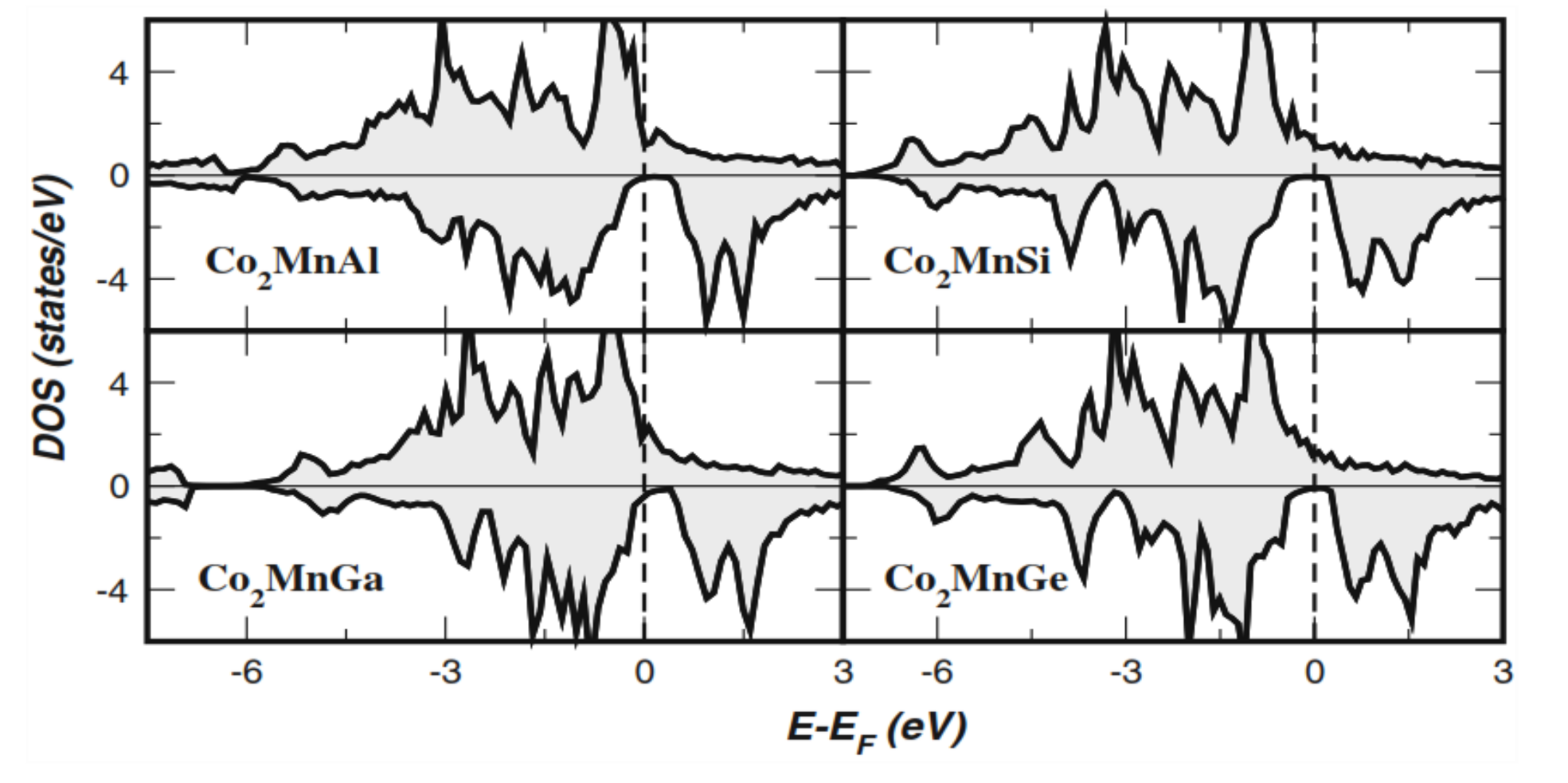}
\includegraphics[width=12cm]{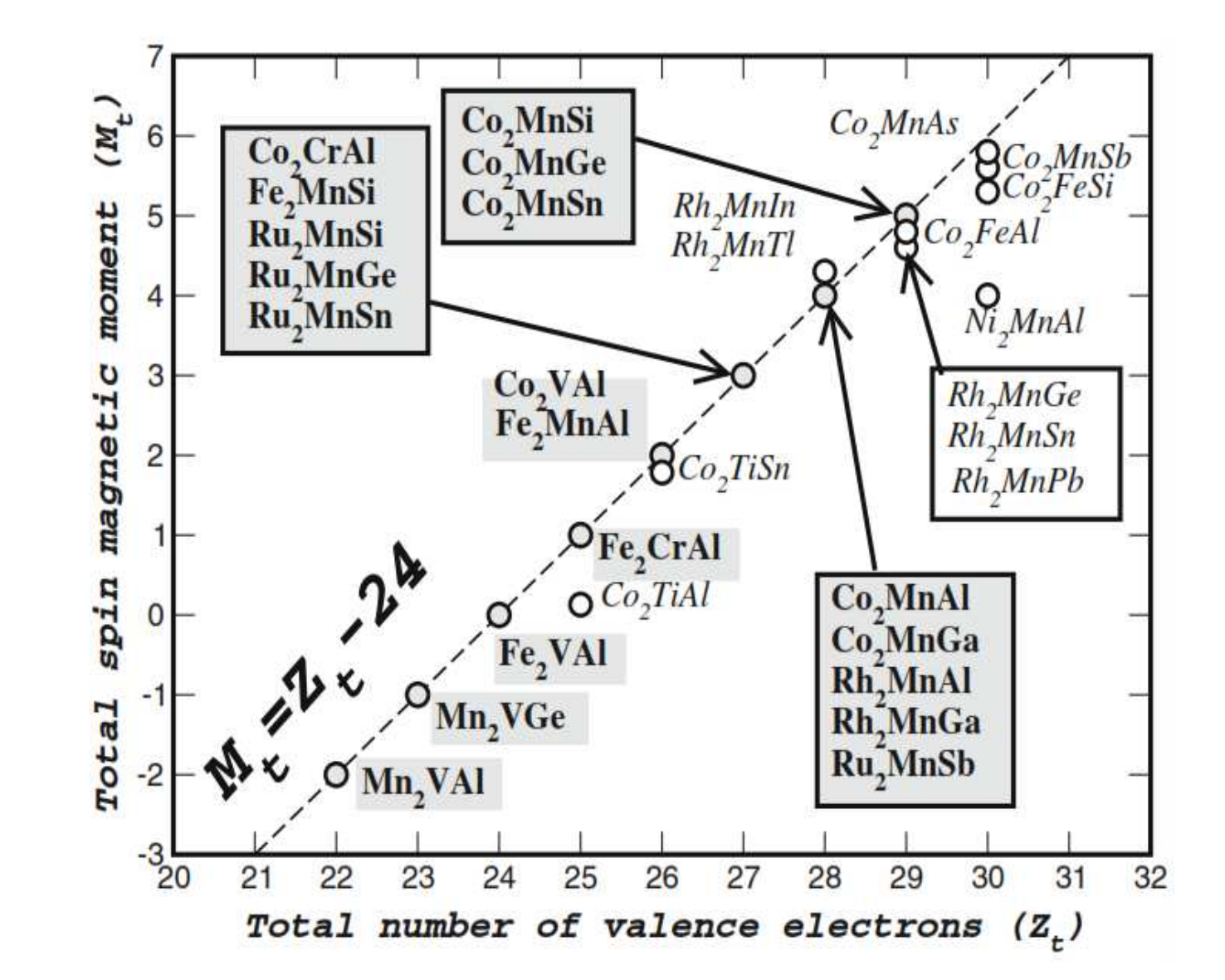}
\caption[Calculated total spin moments for full Heusler alloys]{(Top)Total DOS example for the Co$_2$MnZ compounds with Z = Al, Si, Ge, Sn from Ref. \cite{brown1998photoelectron}. (bottom) Total spin moments (give in $\mu_B$) for full-Heusler alloys. The dashed line represents the Slater-Pauling behavior. \cite{book2015heusler}}
\label{DOSfig}\end{center}
\end{figure}

The half-metallic properties of Heusler alloys can be verified, for example,  by calculations of density of states (DOS) at the Fermi level, using first-principle methods (see Fig. \ref{DOSfig}-top). The other method is by the total magnetic moment of the compound, which has to obey the generalized Slater-Pauling rule.

Slater and Pauling discovered that the magnetic moment $m$ of $3d$ elements and their binary alloys can be estimated based on the average valence electron number ($N_V$) per atom. The materials are divided into two groups depending on $m(N_V)$ \cite{slater1936ferromagnetism,pauling1938nature}. The first group has low valence electron concentrations ($N_V \leq 8$) and localized magnetism. Here, mostly bcc and bcc-related structures are found. The second group has high valence electron concentrations ($N_V \geq 8$) and itinerant magnetism. Here, systems with closed packed structures ($fcc$ and $hcp$) are found. Iron is located at the border line between localized and itinerant magnetism. The magnetic moment, measured in unit of the magnetons ($\mu_B$) is given by \cite{book2015heusler}:
\begin{center}
\begin{equation}
m = N_V - 2n_\downarrow,
\end{equation}
\end{center}

\noindent where $2n_\downarrow$ denotes the number of electrons in the minority states. In the case of X$_2$YZ Heusler material, the magnetic moment per formula unit can be written as \cite{book2015heusler}:
\begin{center}
\begin{equation}
m = N_V - 24.
\end{equation}
\end{center}

The bottom panel in Fig. \ref{DOSfig} shows the magnetization as a function of the number of valence electrons per formula unit.

\section{Magnetocaloric effect in Heusler alloys}

\begin{figure}\begin{center}
\includegraphics[width=16cm]{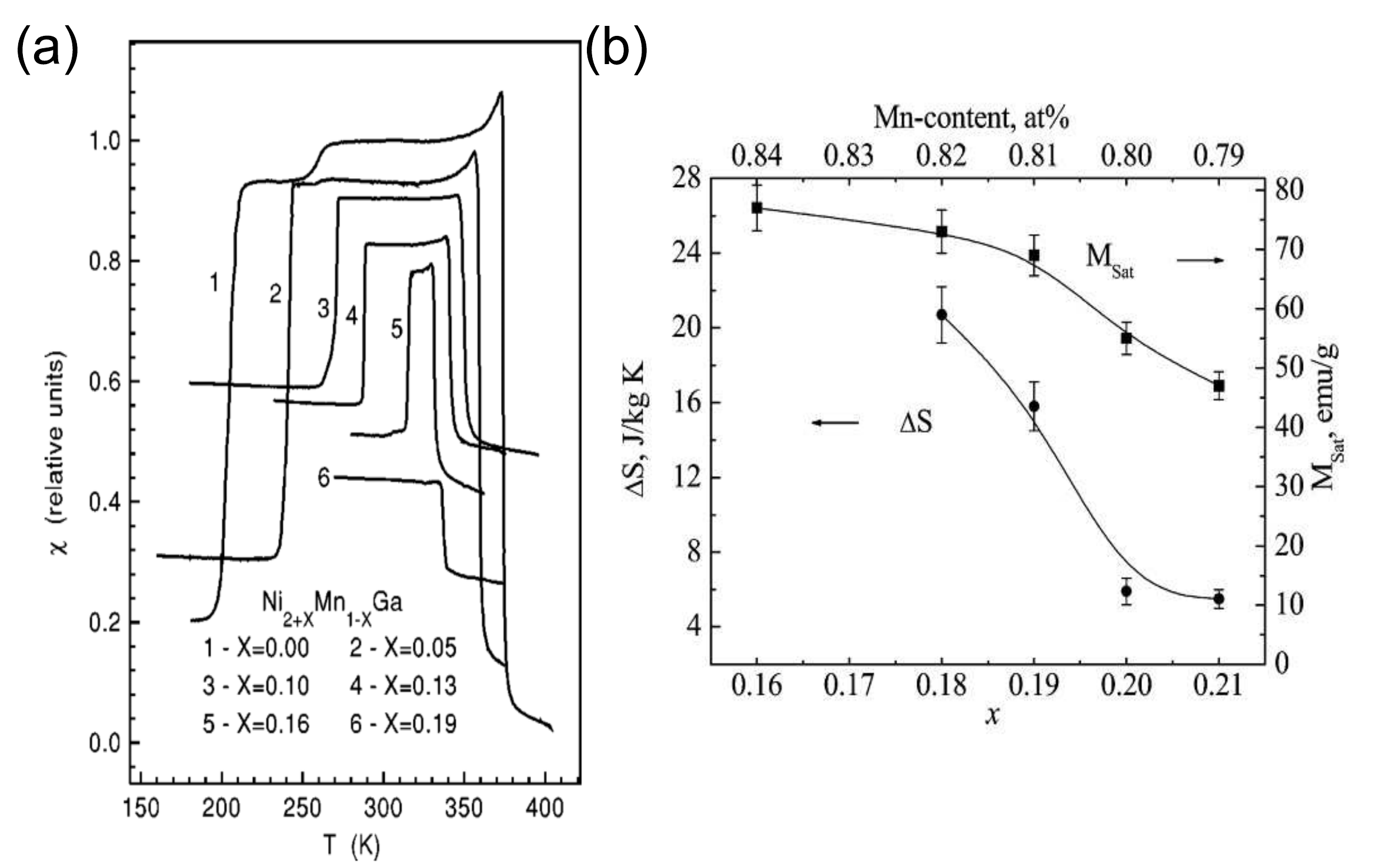}
\caption[MCE on Ni-Mn-Ga alloys]{Magnetic susceptibility as a function of the temperature for Ni-Mn-Ga alloys \cite{vasil1999structural}, where it shows the changes of the structural and magnetic transition based on the amount of Ni (a). (b) Entropy changes and saturation magnetization as a function of Mn content for Ni-Mn-Ga alloys \cite{cherechukin2004magnetocaloric}}
\label{SPfig}\end{center}
\end{figure}

The magnetocaloric effect is present in Heusler alloys. In addition, they may exhibit structural transitions under the influence of the temperature or external magnetic field, and  some of these materials are shape-memory alloys \cite{graf2011simple}. The most famous compounds of this family with these characteristics are those based on Ni-Mn-Ga  \cite{vasil1999structural,planes2009magnetocaloric}, because of the occurrence of coupled  magneto-structural transition near the room temperature. This has motivated several scientists to optimize these alloys for applications in magnetic refrigeration. Fig. \ref{SPfig}-a shows the low-field ac magnetic susceptibility as a function the temperature of several samples of Ni-Mn-Ga Heusler alloys, where both the structural and magnetic transition changes can be observed, depending on the Mn concentration, from Ref. \cite{vasil1999structural}. The entropy changes as a function of Mn concentration are shown in Fig. \ref{SPfig}-b \cite{cherechukin2004magnetocaloric}.

In Chapter \ref{capFMS}, we provide more information on the magnetocaloric effect in Heusler alloys, and discuss how it is affected by the increase of the valence electron number.


	\chapter{Intermixed disorder effect in Co$_2$FeSi Heusler alloy}\label{cap7CFS}

Co$_2$-based Heusler alloys as half-metallic compounds present promising properties such as $T_C$ above 1000 K and large saturation magnetization. However, atomic disorder can affect the half-metallic properties of these materials and decreases their potential use in spintronic devices. Here, our aim is to evaluate the role played by atomic disorder in Co$_2$FeSi, which has a Curie temperature at 1100 K and magnetic moment of 6 $\mu_B$ according to the Slater-Pauli rule \cite{inomata2006structural}. We have synthesized samples of Co$_2$FeSi following the process described in the Chapter \ref{capitulo4}, and annealed, then for 0, 3, 6 and 15 days at 1323 K with subsequent quenching in water. Here, the samples will be called 0d, 3d, 6d, and 15d based on the annealing times. Anomalous X-ray diffraction were obtained at room temperature at {\it Laborat\'orio Nacional de Luz S\'incrotron}. The M\"ossbabuer spectroscopy measurements at room temperature were performed at {\it Laborat\'orio de M\"ossbabuer} at UFF. Density functional theory (DFT) calculations were carried out in collaboration with the University of Aveiro in Portugal, using the SPR-KKR (spin polarized relativistic Korringa-Kohn-Rostoker) package \cite{ebert2012munich,ebert2011calculating}, which implements the KKR-Green's function formalism. It is known from previous studies \cite{wurmehl2005geometric} that the local density approximation (LDA)  fails to reproduce the half-metallic behavior of Co$_2$FeSi; therefore, the LDA+U approximation (where U is the calculation of the Hubbard) \cite{anisimov1991band} was used. The U values (U$_{\mbox{\scriptsize{Fe}}}$=3.8, U$_{\mbox{\scriptsize{Co}}}$=3.75 ) were chosen to obtain the expected behavior of the ordered system and were then fixed for the disordered system calculations. To simulate the atomic disorder, the Coherent Potential Approximation (CPA) \cite{soven1967coherent} was used and the lattice parameters were kept fixed at the experimental values for Co$_2$FeSi. The angular momentum cut-off was set at $l=3$, and 2119 $k$-vectors in the irreducible Brillouin zone was used in all calculations, which were carried out in a relativistic approach.

\section{Co$_2$FeSi atomic disorder}

\begin{figure}[h!]
\center
\includegraphics[width=14cm]{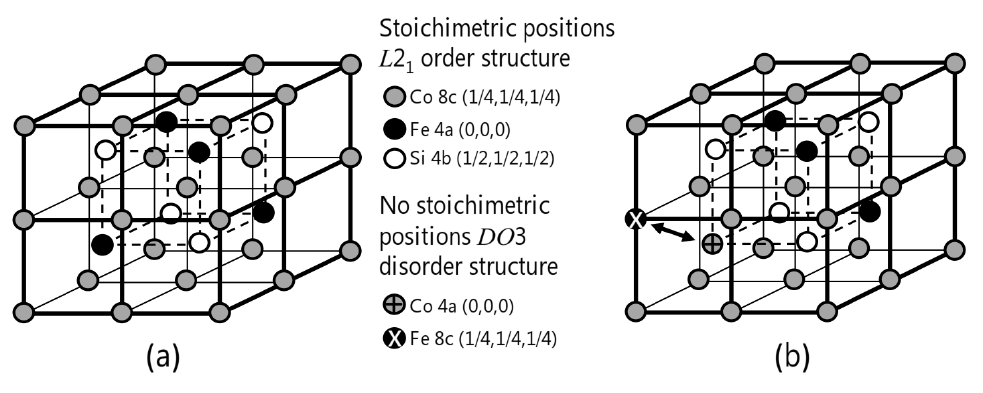}
\caption[Co$_2$FeSi  stoichiometric structure in $L2_1$ and  $DO3$ type]{Co$_2$FeSi  stoichiometric structure of $L2_1$ order type (a) and $DO3$ disorder structure (b).}
\label{estruturafig1}
\end{figure}

Full-ordered Co$_2$FeSi crystallizes in the Cu$_2$MnAl-type structure, where atoms are localized in  $8c$ site (Co), $4a$ site (Fe) and $4b$ site (Si)(see Fig.\ref{estruturafig1}.(a). However, it is possible to occur interchanges between Co-Fe atoms in Co$_2$FeSi (see Fig. \ref{estruturafig1}.(b), and observed the variations in the magnetic and thermodynamic properties that affect the half-metallic behavior of this system. We show that the potential of Co$_2$FeSi for spintronics decreases with the increase in atomic disorder, and new ways for optimizing the production process of these materials are required to improve their performance.

\begin{figure*}
\centering
\includegraphics[width=17cm]{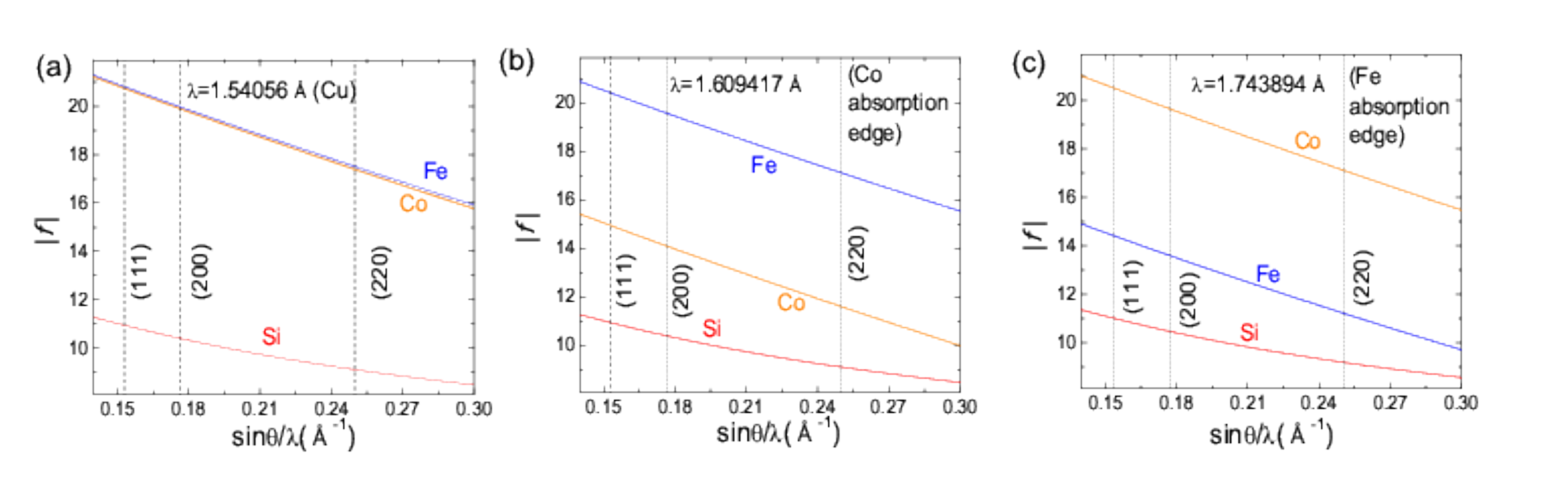}
\caption[Atomic scattering factors]{Atomic scattering factors. (a) Cu radiation, (b) Co absorption edge and (c) Fe absorption edge. Note the differences between the atomic scattering factors of Cu K$_\alpha$ radiation and  Co and Fe absorption edges, which makes them the ideal parameters for the determination of the chemical disorder of the samples.}
\label{borda}
\end{figure*}

Co$_2$FeSi crystallized in the full-ordered $L2_1$ structure, the Co atoms occupy the $8c$ Wyckoff position at (1/4,1/4,1/4), the Fe atoms are located at the $4a$ sites (0,0,0) and the Si atoms occupy the $4b$ site (1/2,1/2,1/2). Nevertheless, the Co and Fe atoms interchanges their positions, leading to what is known as $DO3$ disorder.  X-ray diffraction is commonly used to evaluate chemical disorder in Heusler alloys \cite{basit2009heusler,takamura2009,takamura2010}; however, there are difficulties using this technique  with a Cu$K_{\alpha}$  source. The main problem of  X-ray diffraction with a Cu$K_{\alpha}$ source is the fact that the atomic scattering factors ($f_{hkl}$) of Co and Fe ions have very close values (see Fig. \ref{borda}.a ). Making this technique inefficient for detection of the atomic disorder in Co$_2$FeSi. Other measurements setups, such as AXRD and M\"ossbabuer spectroscopy are required to evaluate the atomic disorder these systems.

\subsection{Anomalous X-ray difraction}

For Co$_2$FeSi,  the AXRD data were collected close to the absorption edge energies of Co (7709 eV) and Fe (7112 eV), where the atomic scattering factors of Co and Fe are well distinguishable( see Figs.\ref{borda}.b and Fig.\ref{borda}.c ), allowing us to evaluate the  $DO3$ disorder. 

\begin{figure}[h!]
\center
\includegraphics[width=8cm]{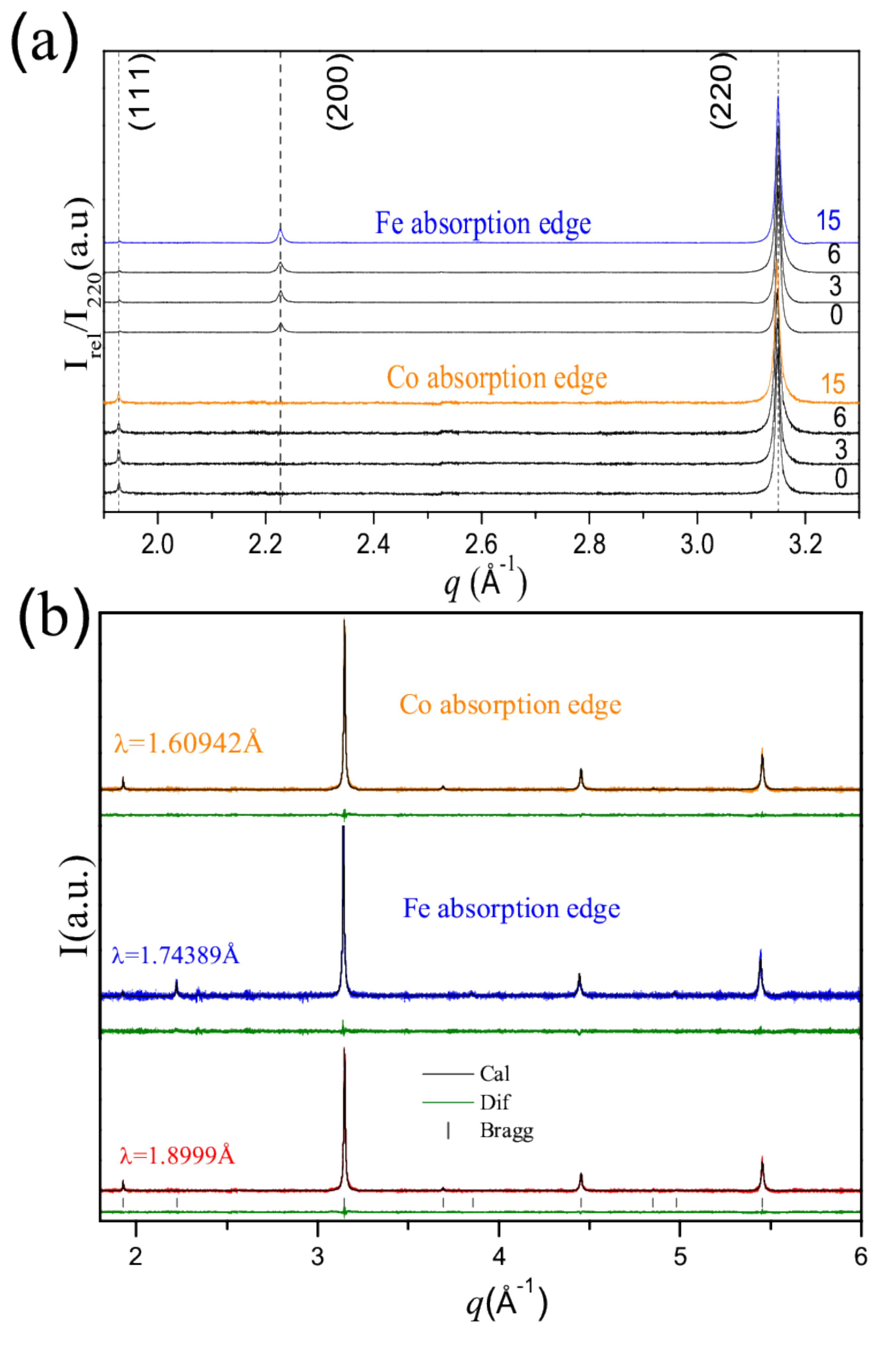}
\caption[Anomalous X-rays diffraction]{(a) AXRD in the samples as a function of the time of annealing. (b) Refinement of the data using Full Prof software  for determination of structural disorder for 15d sample with radiation with wavelengths of $\lambda$ = 1.6092 \AA (Co absorption edge), 1.7438 \AA (Fe absorption edge), and 1.8999 \AA. }
\label{axrd}
\end{figure}

AXRD of all the annealed samples of Co$_2$FeSi are presented in Fig.\ref{axrd}.a. A suitable $q$-range was chosen to show details of the first three peaks, which are the most sensitive to change in the energy of the focused beam. The differences in the intensities  are due to the structure factors of each crystallographic plane: $F_{111} \propto f_{\mbox{\scriptsize{Fe}}}- f_{\mbox{\scriptsize{Si}}}$ for (1,1,1) plane and $F_{200} \propto 2f_{\mbox{\scriptsize{Co}}} - f_{\mbox{\scriptsize{Fe}}}+ f_{\mbox{\scriptsize{Si}}}$ for (2,0,0) plane, where $f_{\mbox{\scriptsize{Co}}}$, $f_{\mbox{\scriptsize{Fe}}}$ and $f_{\mbox{\scriptsize{Si}}}$ are the atomic scattering factors of Co, Fe and Si respectively.  The full $q$-range for the 15d sample is shown in Fig \ref{axrd}.b  for three wavelengths: 1.60942 \AA$ $ (Co edge absorption), 1.74389 \AA$ $ (Fe edge absorption) and 1.8999 \AA. The data analysis was realized using the FULLPROF software \cite{fullprof}, and was conducted simultaneously for three diffraction patterns of each sample. We found that all samples crystallize in the single phase Cu$_2$MnAl structure type, and the lattice parameters $a$= 5.645 \AA$ $  does not change. However, the occupation of Co and Fe ions are different in all sample; and this is due to the influence of the annealing time. Thus, we have Co$_{2-x}$Fe$_x$Fe$_{1-x}$Co$_x$Si samples, where $x$ is the disorder degree, for $x=1$ the sample is  completely ordered, while, $x=1$ represents the  completely disordered sample, i.e., the $4a$ site is occupied completely by Co ions. The 0d sample (non-annealed) presents the larger disorder degree in comparison with other samples (3d, 6d and 15d), which have disorder degree between $ x \approx 0.12$ and 0.18. The Fig. \ref{desordem} shows these results in comparison with M\"ossbauer spectroscopic results. 

\begin{figure}[h!]
\center
(a)\\
\includegraphics[width=10cm]{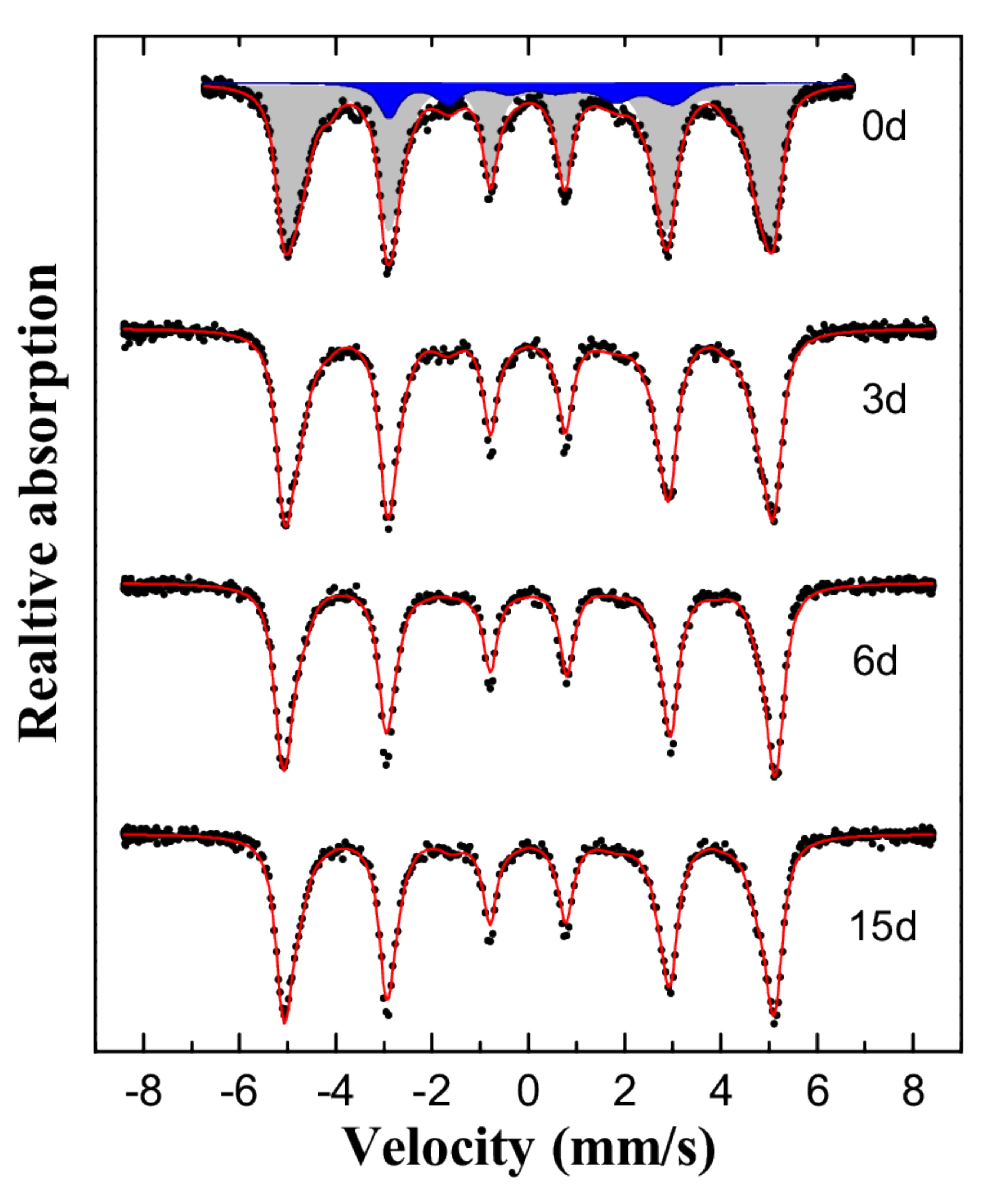}\\
(b)\\
\includegraphics[width=10cm]{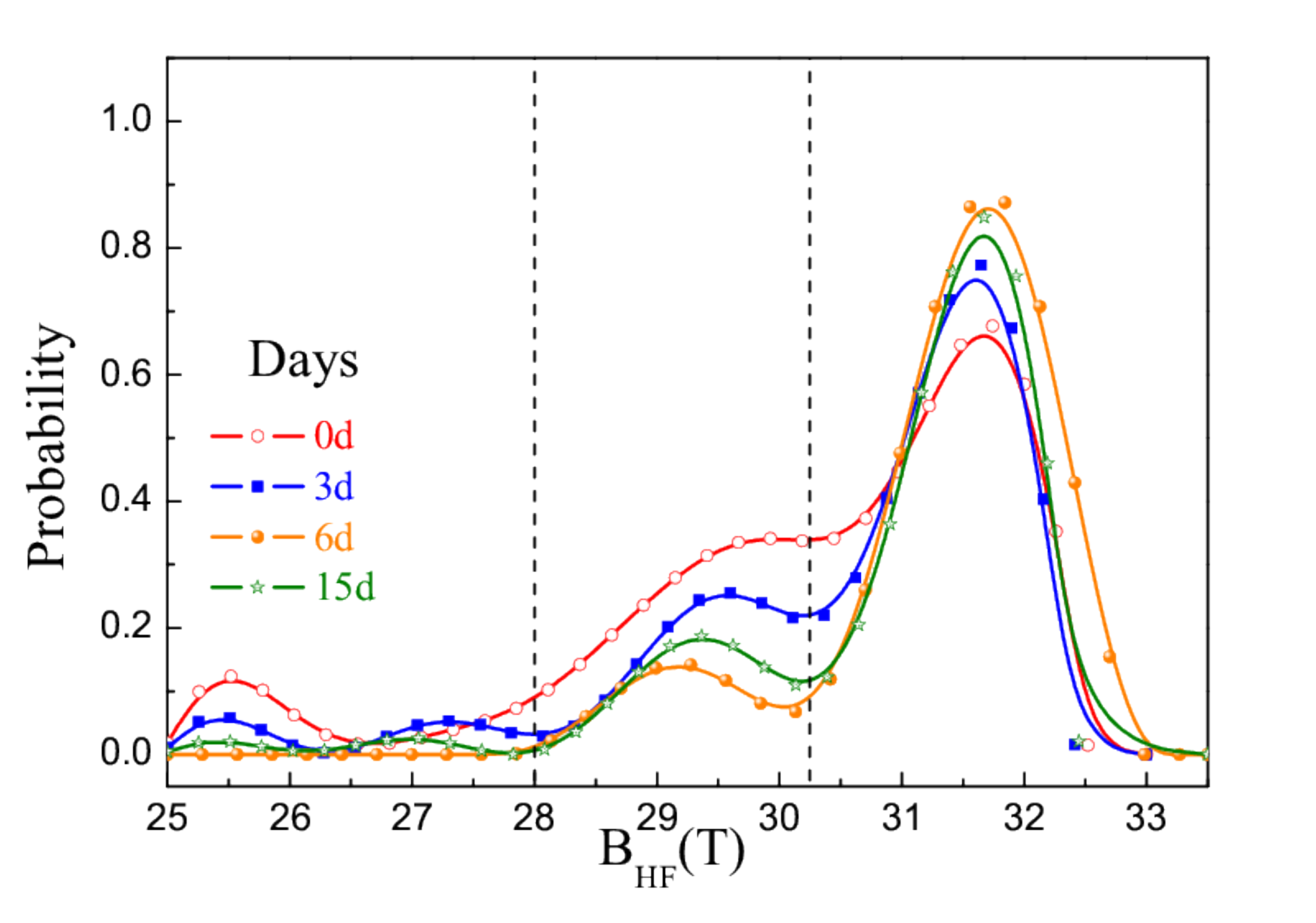}
\caption[M\"ossbauer spectra at room temperature]{(a) M\"ossbauer spectra at room temperature  of Co$_2$FeSi samples as a function of the  annealing time. Note the example shown over the 0d sample data, the gray area is Fe ions at the stoichiometric $4a$ sites, while the blue spectrum correspond to Fe ions in the $8c$ sites disordered position, and the red line is the sum of the two spectra. (b) Probability of find Fe ions with a specific hyperfine magnetic field. The dashed lines separate the ranges of hyperfine magnetic field characteristic of  $4a$ sites ($B_{\mbox{\scriptsize{HF}}}$ = 30.1 to 33.1 T), and  $8c$ sites  ($B_{\mbox{\scriptsize{HF}}}$ =28 T to 31.1 T).}\label{moss}
\end{figure}

\subsection{Surrounding of the Fe ions by M\"ossbauer spectroscopy}

Fe$^{57}$ M\"ossbauer spectroscopy was used by us, for investigate the local environment of the Fe ions in the samples. This technique based on the M\"ossbauer effect discovered by Rudolf M\"ossbauer in 1958 \cite{mossbauer1958kernresonanzfluoreszenz}, and consists in energy level transitions of the nuclei, which can be associated with the emission or absorption of a $\gamma$-ray.  These changes in the energy levels can provide information about local environment of an Fe-ions within  of the material \cite{de2007structure}. Therefore, we performance Mossbauer spectroscopy measurements and determine the probability of the Fe ions has hyperfine magnetic field, which are characteristic of the Fe sites (stoichiometric and disordered) in the samples.
\begin{figure}[h!]
\center
\includegraphics[width=10cm]{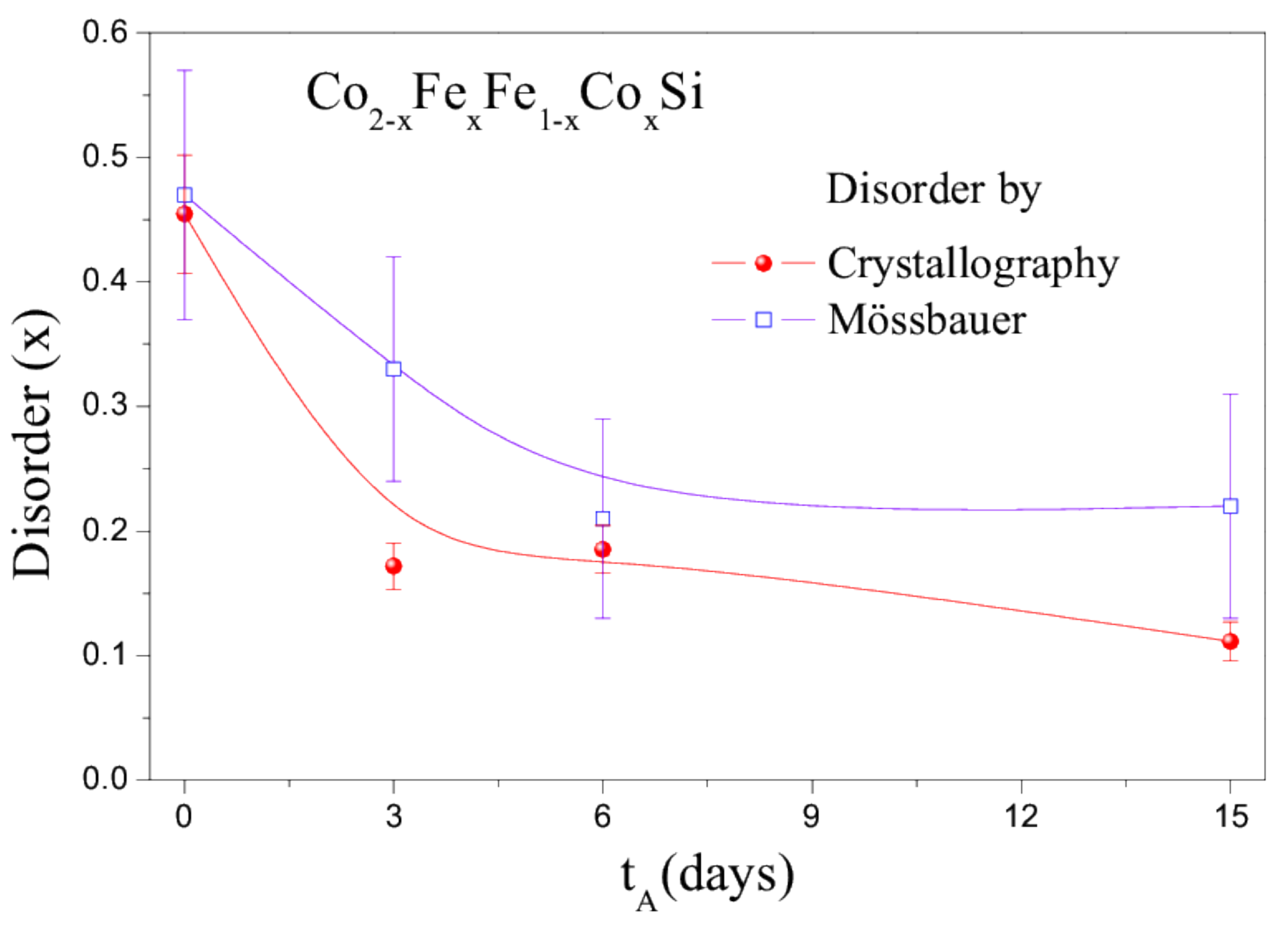}
\caption[Disorder evolution in function of time annealing]{Disorder evolution as a function of annealing time in all the samples from the crystallographic and M\"ossbauer results.}
\label{desordem}
\end{figure}

For Co$_2$FeSi full ordered  $L2_1$ structure, the M\"ossbauer spectrum at room temperature should show a single sextet, corresponding to the Fe ions occuping the $4a$ sites, while an additional sextets is observed in a disordered structure when the Fe ions occupy the Co Co position ($8c$ sites) \cite{srinivas2015effect}. The M\"ossbauer spectra at room temperature of all samples are shown in Fig.\ref{moss}-left. We found two sextets for each sample, indicating the presence of atomic disorder. The main sextet due to Fe ions in stoichiometric $4a$ site being surrounded by eight magnetic Co ions \cite{Jung2008Fe}, and therefore the hyperfine magnetic field of these is between $B_{\mbox{\tiny{HF}}}^{\mbox{\tiny{order}}}$= 31.3 and 32.2 T \cite{Jung2008Fe}. The additional sextet is due to Fe ions occupying the disordered $8c$ site, is surrounded by four magnetic Fe ions (at $4a$ sites) and four non-magnetic Si ions (at $4b$ sites ) as the nearest neighbors \cite{galanakis2006,Srinivas2013,srinivas2015effect}. In this case, the hyperfine magnetic field from disordered ions is in the range $B_{\mbox{\tiny{HF}}}^{\mbox{\tiny{disorder}}}$ = 28.0 and 31.1 T. 
From this information, we obtain the probability that a Fe atom has a hyperfine field between the characteristic ranges of each crystallographic site. The Fig. \ref{moss} shown these results, it is possible observed  that the samples have larger probability of possessing high value of hyperfine magnetic field, which indicate that they have greater portions of Fe in the stoichiometric site.  

Fig.\ref{desordem} shows the $DO3$ disorder parameter ($x$) of  Co$_{2-x}$Fe$_x$Fe$_{1-x}$Co$_x$Si samples from the measurements M\"ossbauer spectroscopy in comparison with AXRD results (discuss above). Using both techniques, we found that the 0d sample is more atomically disordered with $x \approx 0.46$, which  is expected since this sample was not subject to annealing. M\"ossbauer spectroscopy observes Fe ions directly, and is able to identify their positions with this technique, we obtain disorder coefficients between $ x \approx 0.21$ and 0.33. In contrast with AXRD technique, we found that the disorder changes more that 10 \% with the time annealing. This result can be explained by the fact the elements required more time the annealing for arrangement in their stoichiometric position.

\section{Half metallic properties of disordered Co$_2$FeSi }

\begin{figure}
\centering
\includegraphics[width=14cm]{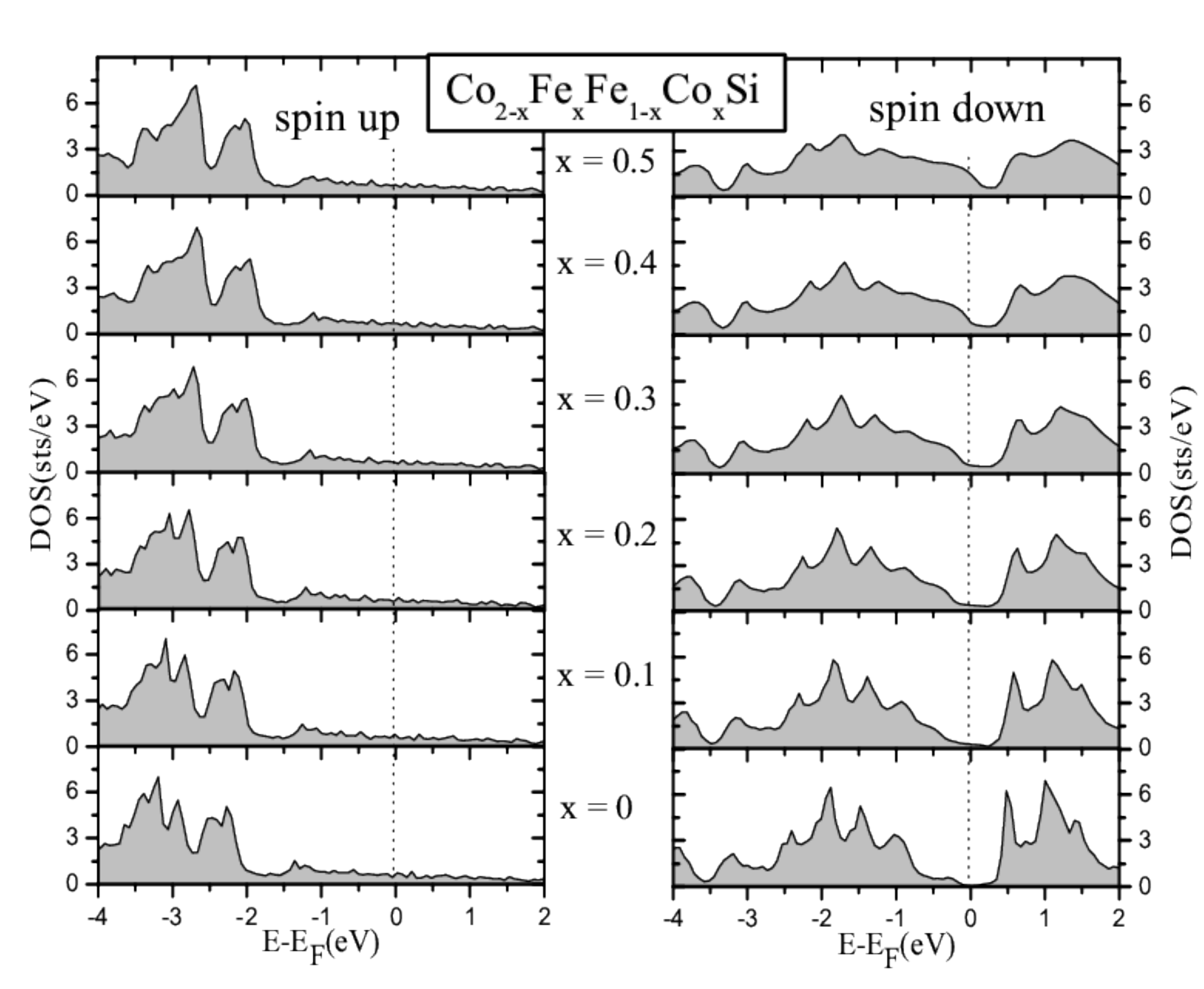}
\caption[Density of states (DOS)]{DOS obtained by the DFT calculations for several values of the disorder ($x$) for Co$_{2-x}$Fe$_x$Fe$_{1-x}$Co$_x$Si the spin-up (left) and spin-down (right) bands.}
\label{dospt}
\end{figure}

\begin{figure}[t!]
\center
\includegraphics[width=8cm]{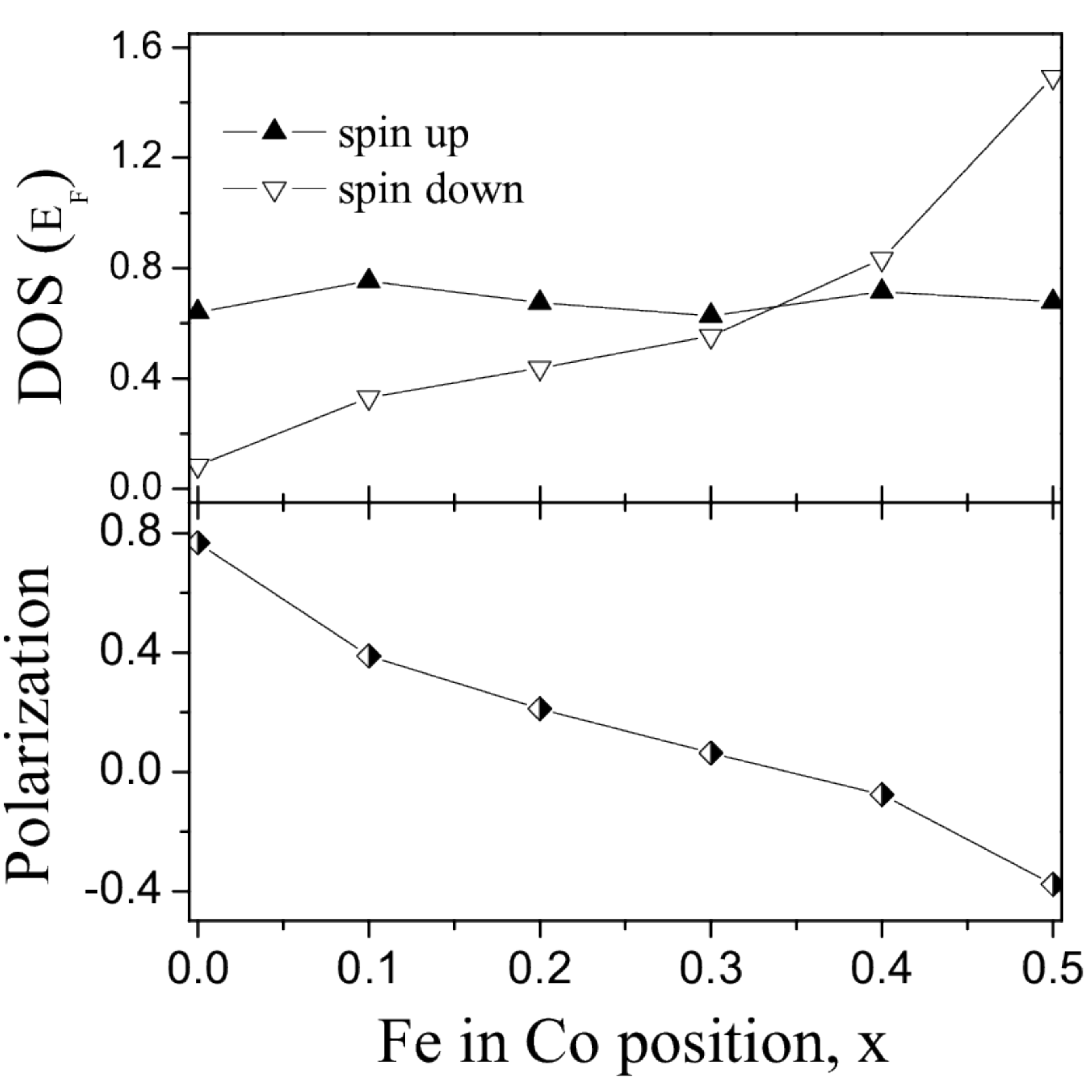}
\caption[Half-metallic properties in function of disorder parameter]{Half-metallic properties as function of the disorder parameter ($x$) for Co$_{2-x}$Fe$_x$Fe$_{1-x}$Co$_x$S. (a) DOS at Fermi level for the spin-up and spin-down band. (b) Polarization at Fermi level. (c) Saturation magnetization from the  experimental results and DFT calculations.}
\label{para_diso}
\end{figure}

DOS for Co$_{2-x}$Fe$_x$Fe$_{1-x}$Co$_{x}$Si system were obtain by DFT calculation for atomic disorder from $x=0$ to 0.5, is shown in the Fig. \ref{dospt}. The number of states at the Fermi energy ($E_F$) for the spin-up and spin-down band are plotted as a function of the disorder in Fig.\ref{para_diso}.a. The increase in disorder augments the number of states in the spin-down band, while the spin-up band remains almost constant. This leads to a decreasing  polarization  ($P = \frac{D^{\uparrow}_{E_F}-D^{\downarrow}_{E_F}}{D^{\uparrow}_{E_F}+D^{\downarrow}_{E_F}}$) as a function of disorder, as shown in Fig.\ref{para_diso}.b. Since there is a small amount of states present in the calculated spin-down band of the $x=0$ case, the simulated system does not  strictly display a half-metallic behavior ($P=1$). However, the polarization is high ($P=0.8$) and the variation of $P$ with $x$ shows that the characteristic half-metallic properties decrease as the disorder increases.

\section{Concluding remarks on Co$_2$FeSi atomic disorder}

From a single phase sample of Co$_2$FeSi , we explored experimentally the influence of annealing on atomic disorder in the sample, and showed that disorder suppresses the half-metallic properties of this Heusler alloy. DFT calculations provided theoretical information about the electronic states at the Fermi level in this compound.  The polarization $P$ is smaller than the on obtained for a totally ordered compound. This work provides further knowledge on new strategies to help optimization developments of Heusler alloys for spintronics devices.

\chapter[Effects of Ga substitution on Fe$_2$MnSi]{Effects of Ga substitution on the structural, magnetic and magnetocaloric properties of half metallic Fe$_2$MnSi Heusler compound.}\label{capFMS}

Fe$_2$MnSi is a Heusler compound that crystallizes in Cu$_2$MnAl-type cubic structure and $T_C$ = 224 K, and it does not follow the Slater-Pauling rule (explained in \ref{capitulo6}). On the other hand, Fe$_2$MnGa crystallizes in  Pt$_2$CuIn tetragonal-type, or Cu$_3$Au type, and $T_C$= 800 K. Despite the fact that these two compounds crystallize in different structures, we found single phase samples for the substitution of Si by Ga in Fe$_2$MnSi$_{1-x}$Ga$_x$ only for the Si-rich side until $x=0.5$. Here, we shall explore  the substitution of Si by Ga in Fe$_2$MnSi$_{1-x}$Ga$_x$ compounds to see how it affects their structural and magnetic properties. This substitution reflects the change of the valence electron number, and therefore change the half-metallic properties and  the magnetic entropy of the samples.

\section{Characterization and crystal structure}\label{crystalFMS}

Fe$_2$MnSi Heusler compound order in the full Heusler $L2_1$ Cu$_2$MnAl-type structure, while Fe$_2$MnGa can be found in two distinct structures: Pt$_3$Au tetragonal-type or the Cu$_3$Au type $L2_1$ phase, presenting half-metallic features only in the $L2_1$ phase. In Fe$_2$MnSi$_{1-x}$Ga$_x$, the Fe atoms occupy the $8c$ Wyckoff position at (0.25, 0.25 0.25); Mn atoms are located in the $4a$ site (0, 0, 0) and the Ga and Si atoms are assumed to occupy randomly the $4b$ site (0.5, 0.5, 0.5). The energy dispersive X-ray spectroscopy (EDS) was used to determine the sample compositions. We performed measurements at several points on the polished surface of each sample. The average values we have found are in very good agreement with the nominal compositions.

\begin{figure}[t!]
\center
\includegraphics[width=10cm]{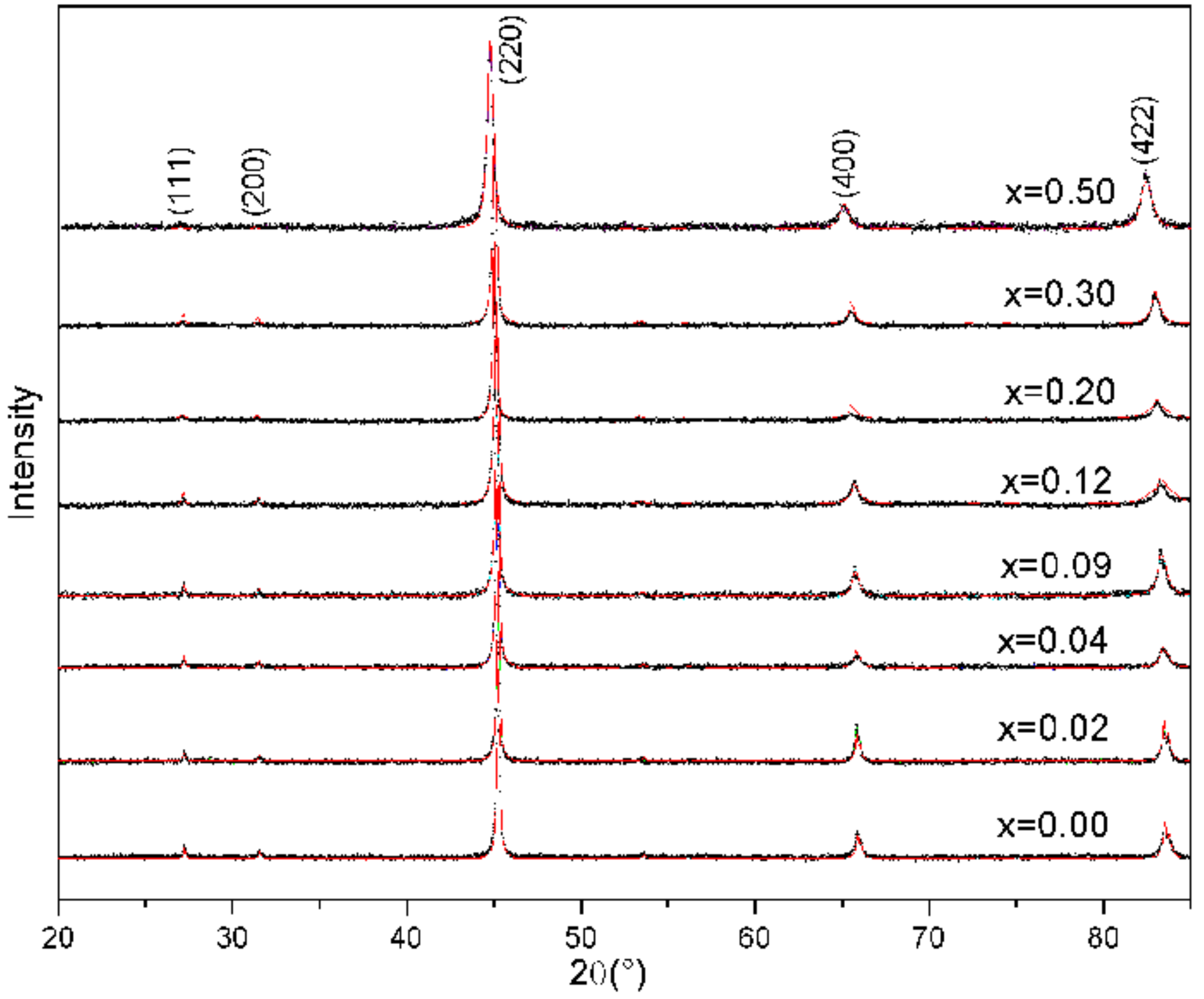}
\includegraphics[width=10cm]{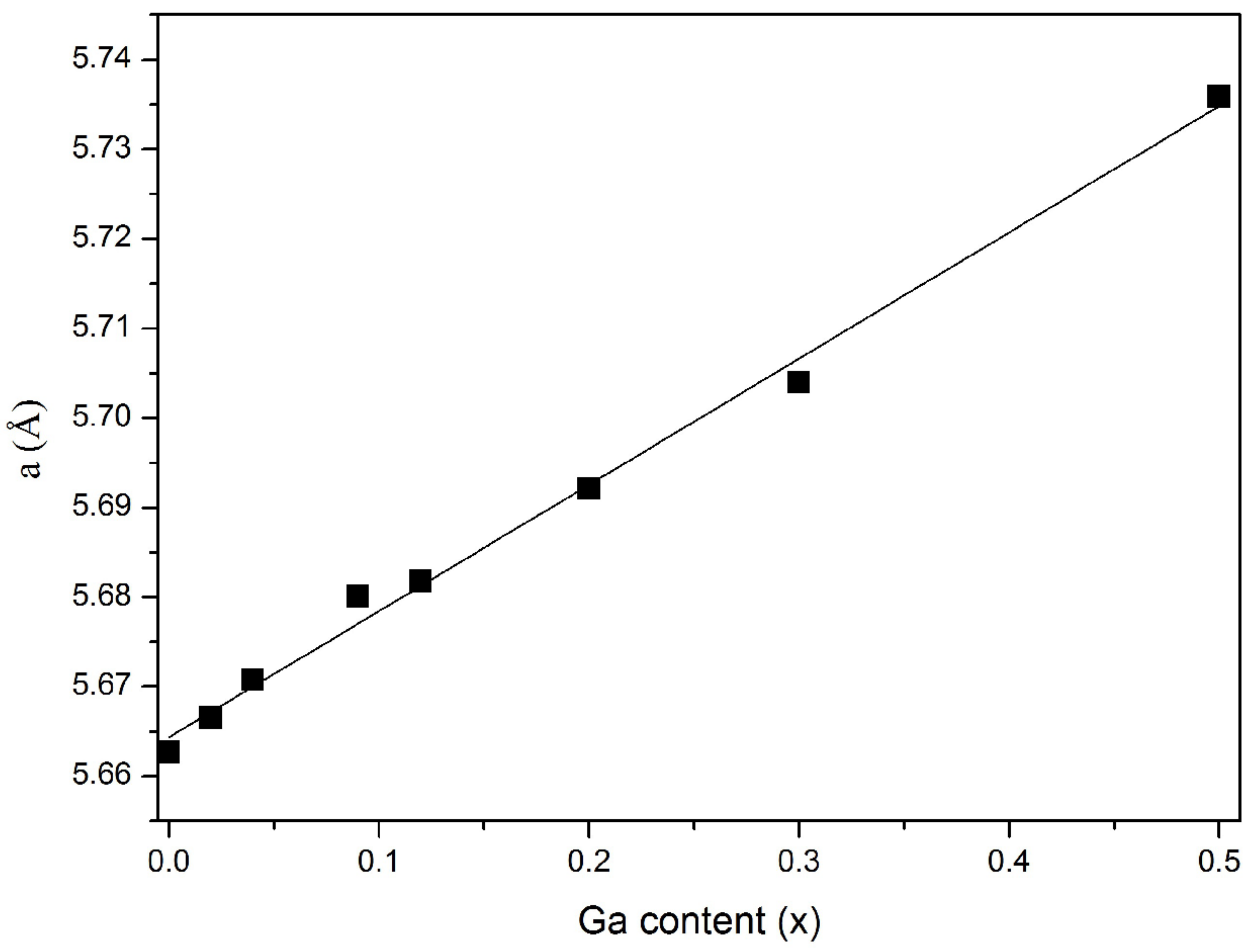}
\caption[Powder X-ray diffraction patterns and lattice parameter $a$ variation vs. Ga content]{(Top) Powder X-ray diffraction patterns of the Fe$_{2}$MnSi$_{1-x}$Ga$_{x}$ and the corresponding refinements. The black lines are the observed patterns and the red lines are the calculated ones by Rietveld refinement. (Bottom) Lattice parameter $a$ variation vs. Ga content. The straight line is the linear fit of the data.}
\label{difratogramsGa}
\end{figure}

Figure \ref{difratogramsGa}-top exhibits the X-ray diffratograms of Fe$_2$MnSi$_{1-x}$Ga$_x$ measured at room temperature. The characteristic reflections in the diffractograms obtained of the samples are in accordance with the $L2_1$ phase. Evidences of secondary phases were not detected. The powder X-ray diffraction data were refined by Rietveld method with the PowderCell software, using the Fe$_2$MnSi structural data (ICSD code number 186061) as basis. The results show that the cell parameter $a$ tends to increase with the substitution of Si by Ga, from $a$=5.6627 \AA{} ($x$ = 0) to 5.7359 \AA{} ($x$ = 0.5). The reason for the increase of the cubic cell parameter is directly related with the atomic radii of Ga and Si, since the atomic radius of Ga is larger than Si.

This tendency can be seen in Figure \ref{difratogramsGa}-bottom, where the lattice parameter $a$ obtained from the Rietveld refinement is shown as a function of Ga content. By increasing the Ga content we observed that the reflections are shifted to lower angles.  The same behavior was observed earlier in Fe$_2$MnSi, where Si was replaced by Ge \cite{zhang2003crystallographic}. From the figure, we can see that the cell parameter $a$ increases linearly with the Ga content within the concentration range considered.  We note that there is no evidence of changes in the crystal structure for Si substitution up to $x$ = 0.50. The value of the cell parameter $a$ for $x$ = 1.0 (Fe$_2$MnGa) estimated by linear extrapolation using the concentration dependence is approximatelly equal to 5.8053 \AA{}. This result is very close to the value of 5.808 \AA{} reported by Kudryavtsev and co-workers \cite{kudryavtsev2012electronic}.

\section{Magnetic properties}

\begin{figure}[t!]
\center
\includegraphics[width=10cm]{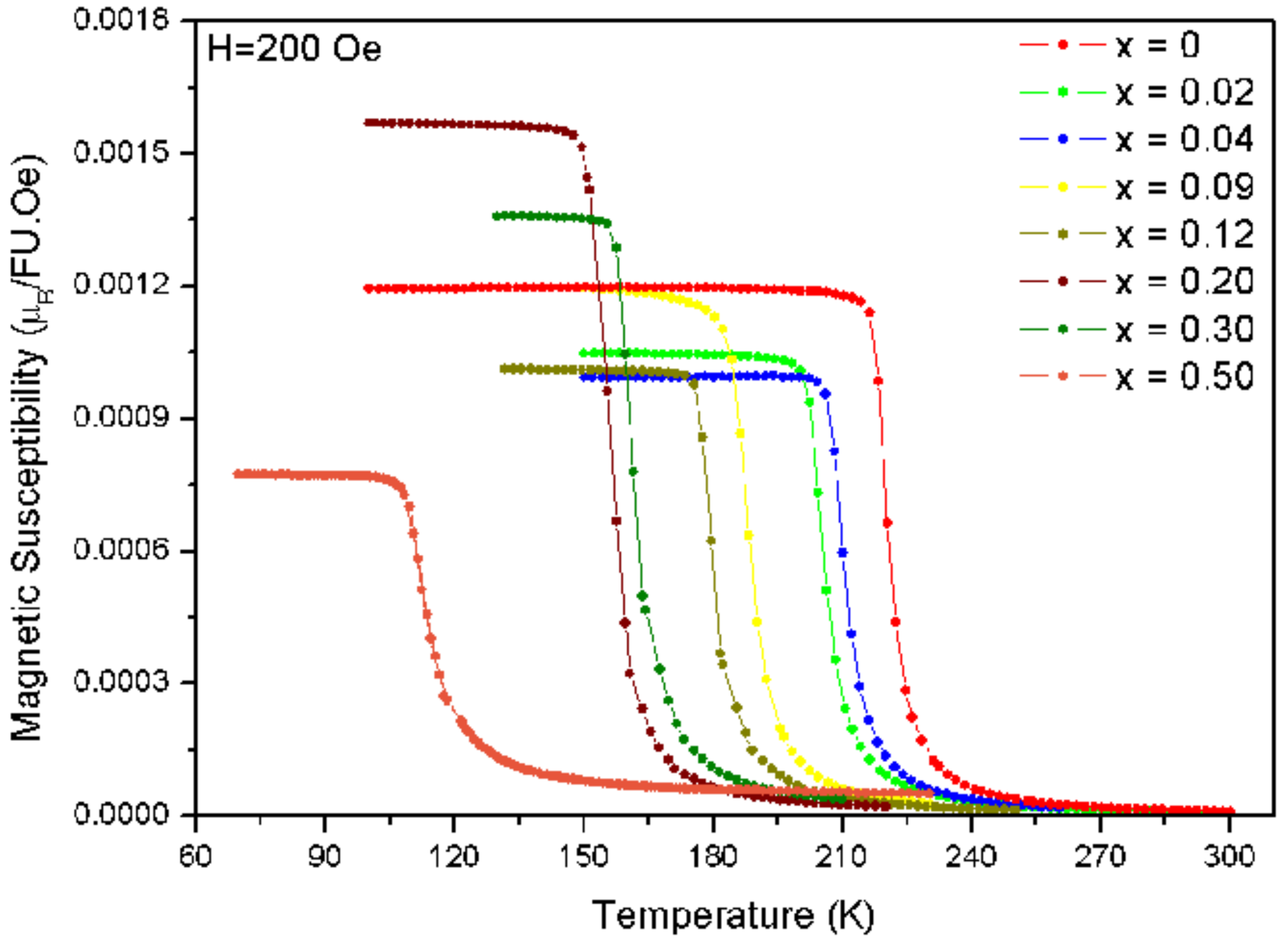}
\includegraphics[width=10cm]{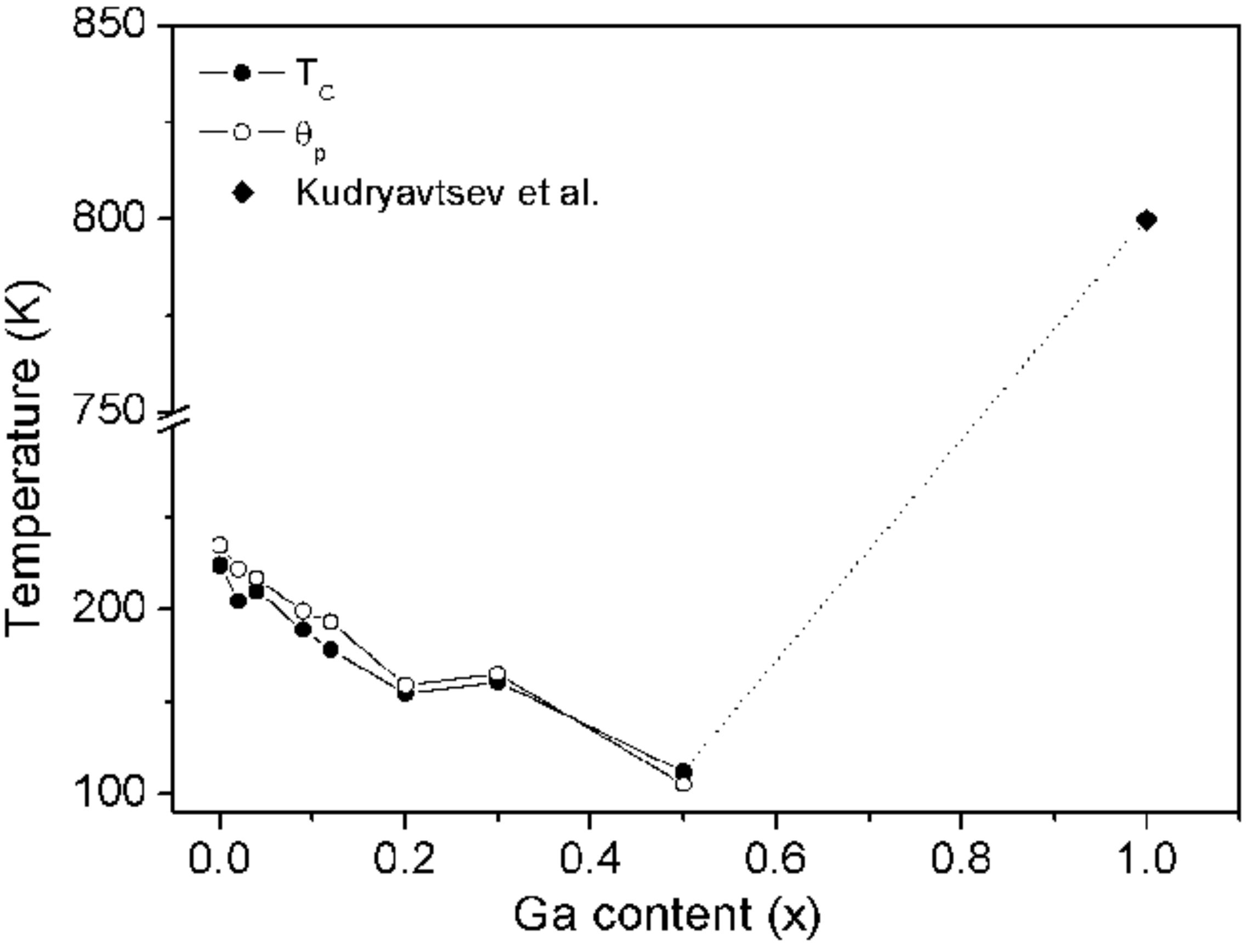}
\caption[Magnetic susceptibility curves]{(Top) Magnetic susceptibility curves for the Fe$_2$MnSi$_{1-x}$Ga$_x$ (x = 0, 0.02, 0.04, 0.09, 0.12, 0.20, 0.30 and 0.50) Heusler compounds measured at 200 Oe. (Bottom) The Curie temperature ($T_C$) and paramagnetic Curie temperature ($\theta_p$) of the Fe$_2$MnSi$_{1-x}$Ga$_x$ Heusler compounds.}
\label{mag}
\end{figure}

Fig. \ref{mag}-top shows the temperature dependence of the magnetic susceptibility ($\chi=M/H$) for the Fe$_2$MnSi$_{1-x}$Ga$_x$ compounds for an applied magnetic field of 200 Oe. These compounds exhibit a magnetic transition from a paramagnetic state at high temperature to a ferromagnetic ordered state, and the transition temperature dependent on the Ga content. The transition does not present detectable thermal hysteresis, and has a second order character, despite being sharp for low magnetic field.

From our data, we have determined how $T_C$ varies with the Ga concentration for the Fe$_2$MnSi$_{1-x}$Ga$_x$ compounds, and the results are shown in Fig. \ref{mag}-bottom. The Curie temperatures were obtained from the first derivative of the magnetization at 200 Oe. At  $x$ = 0, $T_C$ is 224 K. It is clear from Fig. \ref{mag}-top and bottom that the ferro-paramagnetic transition temperature initially decreases (almost linearly) with the increase of Ga content, down to a minimum of 112 K for x = 0.50.  The paramagnetic Curie temperatures ($\theta_p$) were calculated for all samples from the fitting of the linear segment of the inverse susceptibility $1/\chi$ data as a function of temperature. The obtained values are shown in Fig. \ref{mag}-bottom.

\begin{figure}[t!]
\center
\includegraphics[width=12cm]{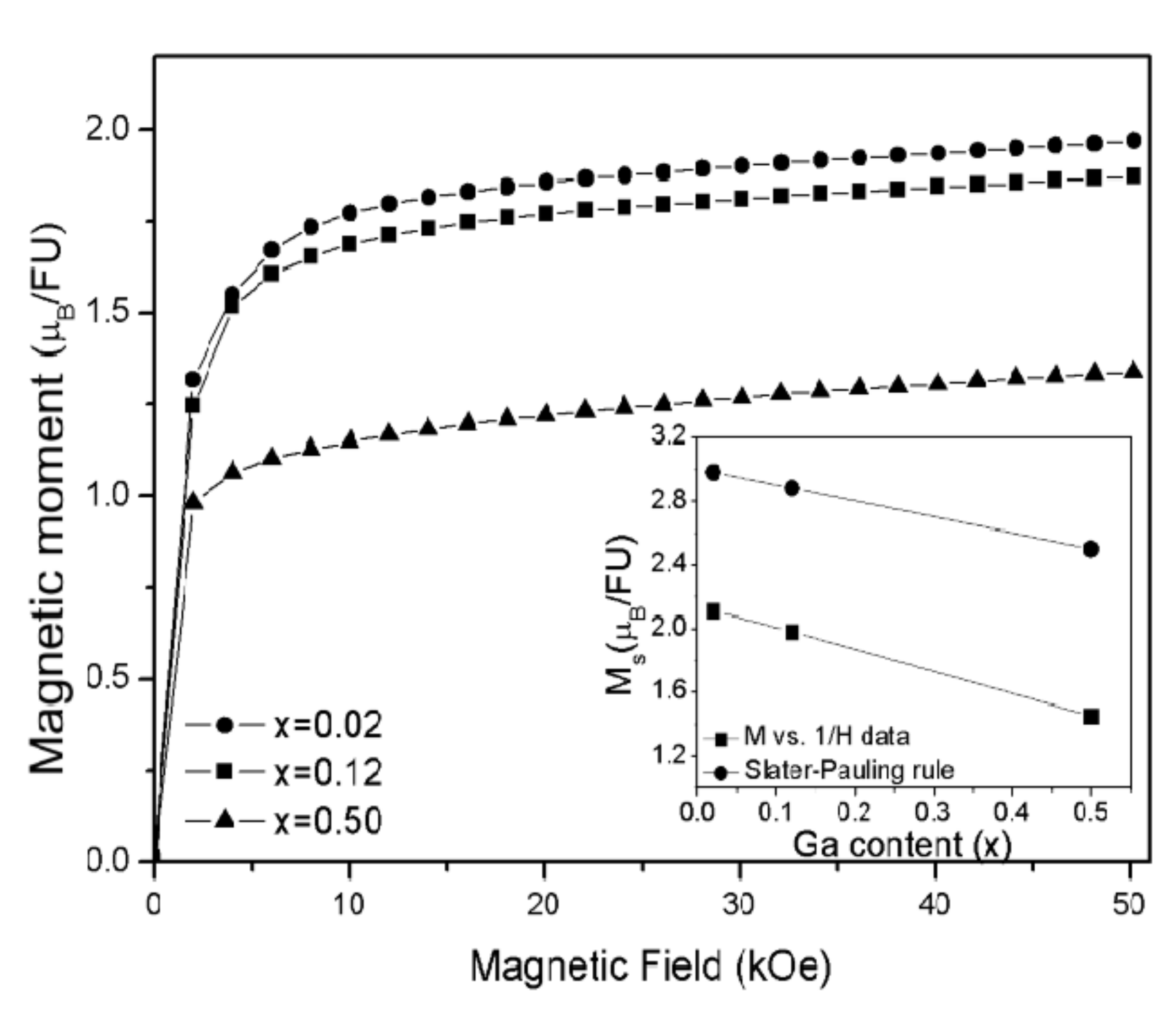}
\caption[Magnetic field dependence on the magnetic moment of Fe$_2$MnSi$_{1-x}$Ga$_x$]{Magnetic field dependence on the magnetic moment of Fe$_2$MnSi$_{1-x}$Ga$_x$ for x = 0.02, 0.12 and 0.50 samples at T = 4 K. Inset: Saturation M$_s$ as a function of the Ga content. The squares are the values estimated from M vs. 1/H experimental data; the circles are the values estimated from the Slater-Pauling rule.}
\label{mvsh4k}
\end{figure}

Fig. \ref{mvsh4k} shows the magnetic field dependence of the magnetization up to 50 kOe for $x$ = 0.02, 0.12 and 0.50 samples at 4 K. There is a clear tendency  of magnetization reduction with increasing the Ga content, but the saturation is not reached  for the field values up to 50 kOe. However, the magnetic saturation can be estimated from M vs. 1/H curves (not shown) and are 2.11, 1.98 and 1.45 $\mu_B/$F.U. for $x$ = 0.02, 0.12 and 0.50 samples, respectively (see inset of Fig. \ref{mvsh4k}).

\section{Half-metallic properties}

Fe$_2$MnSi has a total of $(2 \times 8) + 7 + 4 = 27$ valence electrons in the unit cell and, accordingly, Fe$_2$MnGa has 26 (Ga contributes with 3 valence electrons); for this reason the magnetic moment is expected to vary linearly from 3 $\mu_B$ to 2 $\mu_B$ by increasing Ga content in the Fe$_2$MnSi$_{1-x}$Ga$_x$. As mentioned earlier, the magnetic moments estimated from M \textit{vs.} 1/H curves at 4 K and 50 kOe for x = 0.02, 0.12 and 0.50 samples are 2.11, 1.98 and 1.45 $\mu_B$ per unit formula, respectively. These values are far from the expected ones according to the Slater Pauling rule (see section \ref{SPHA}), of 2.98, 2.88 and 2.50 $\mu_B$/F.U., respectively, but the decrease of the magnetic moment with the Ga concentration is clearly visible, as can be seen in the inset of Fig. \ref{mvsh4k}. The difference between the estimated values and the experimental ones is quite large. Such discrepancies may be attributed to a partial atomic disorder in the structure, since the formed structures do not constitute a superlattice, as confirmed by the low intensity of the (111) and (200) X-ray diffraction peaks. The same behaviour was observed by Nakatani and co-authors  \cite{nakatani2007structure}, who studied the magnetic and structural properties of the Co$_2$FeAl$_x$Si$_{1-x}$ Heusler alloy. In their work, the authors report spin polarization and saturation magnetization dependences on the Al content. Nevertheless, the half-metallicity of the compound is preserved even for a partially disordered state, although the saturation magnetization values do not follow the Slater-Pauling rule. In this way, atomic disorder effects may indicate the Slater-Pauling rule is not followed by our samples, but our system still remains half-metallic, since both parent compounds Fe$_2$MnSi and Fe$_2$MnGa exhibit half-metal behavior.

\begin{figure}[t!]
\center
\includegraphics[width=12cm]{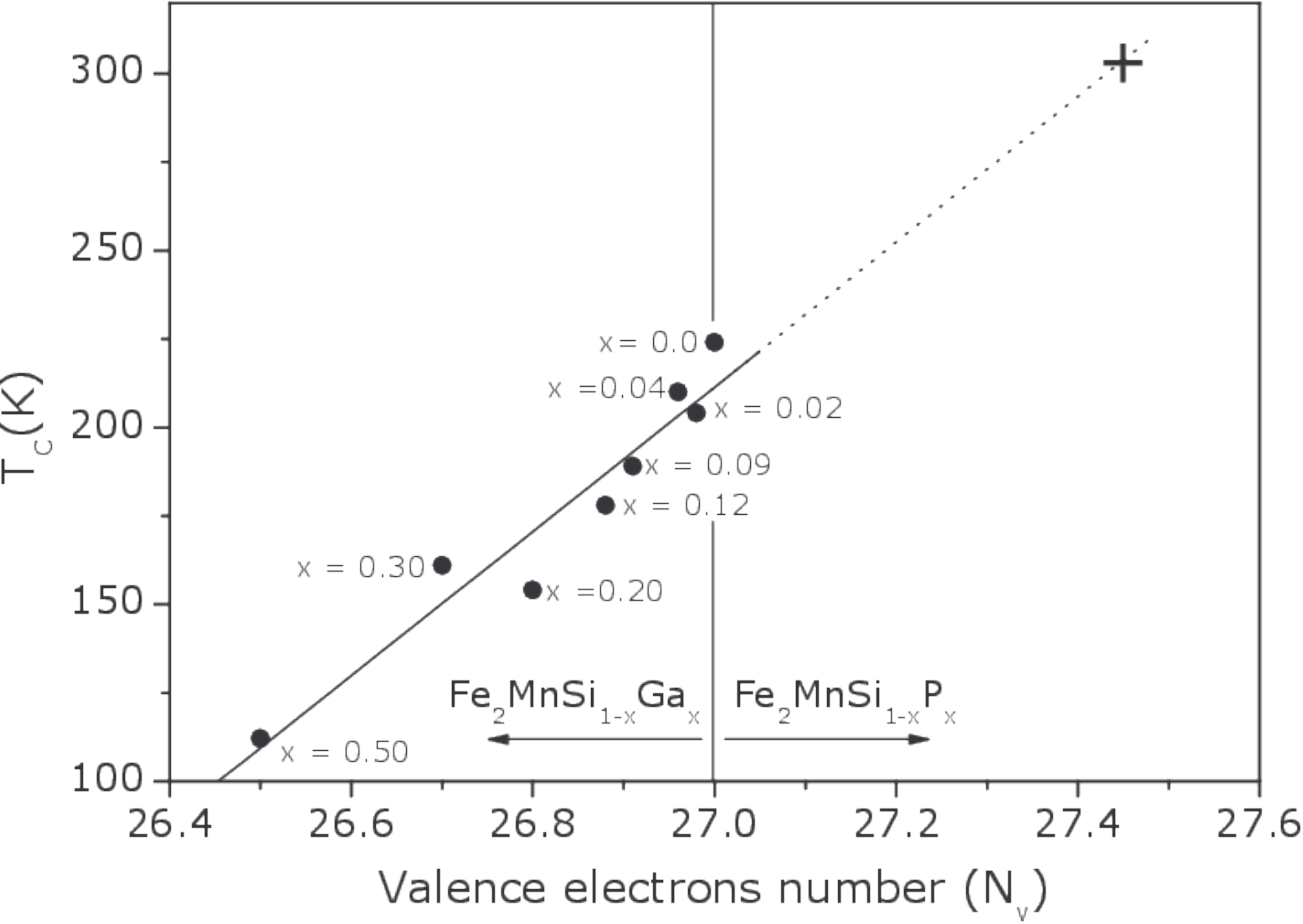}
\caption[Linear behavior of $T_c$ as a function of $N_V$]{Linear behavior of the Curie temperature ($T_c$) as a function of the valence electrons number $N_V$. The  `+' signal marks the necessary $N_V$ value to reach $T_c$ at room temperature, found to be $N_V = 27.44$.}
\label{tc}
\end{figure}

According to a work conducted by Graf and co-authors \cite{graf2011heusler}, there is a linear dependence of the Curie temperature $T_C$ with the magnetic moment on half metallic Heusler alloys. In the same work, the authors also showed that $T_C$ increases for half-metallic compounds with the valence electron number. The decrease of $T_C$ with $x$ is probably related to the reduction of the magnetic moment caused by changes in the number of valence electrons in the system \cite{umetsu2014magnetic}. This tendency was observed in our Fe$_2$MnSi$_{1-x}$Ga$_x$ series, as can be seen in Fig. \ref{tc}, where the magnetic transition temperature $T_C$ presents an almost linear dependence as a function of the number of valence electrons. Such behaviour reinforces the half-metallic behaviour of the Fe$_2$MnSi$_{1-x}$Ga$_x$ series, despite the fact that it does no follow the Slater-Pauling rule due to a minor structural disorder in the system.

Despite this latter fact, it is possible to observe the linear dependence of saturation magnetization and $T_C$ with the $N_V$. Other studies connecting $T_c$ and $N_V$ confirm the linear growth tendency of these quantities for half-metallic Heusler alloys  \cite{graf2011simple,dannenberg2010competing}. Thus, we provide a multifunctional Heusler alloy with enhanced magnetocaloric properties ruled by $N_V$ and half-metallicity.  The figure \ref{tc} the extrapolation the Curie temperature to 300 K and verify that $N_V = 27.44$ would bring the ferromagnetic transition up to room temperature (a desired feature expected to optimize magnetocaloric materials).   

Therefore, one should increase the valence electrons number $N_V$ to further optimize the magnetocaloric properties of half-metal Heusler alloys. To achieve this goal, either Si or Ga may be replaced by other element (or elements) that can contribute with more electrons; i.e., those elements belonging to, group 15 of the periodic table, such as P or As, for instance. These elements contribute with 5 electrons, and can increase the overall valence electrons number of the system. On the other hand, a substitution by elements of group 14, such as Ge and Sn, does not increase the overall valence electrons number for this series, because they have only 4 valence electrons (the same valence electron number of Si). In fact, Zhang \cite{zhang2003crystallographic} found values for $T_C$ varying between 243 and 260 K only by replacing Si by Ge in parental Fe$_2$MnSi compound.

We expect that Fe$_2$MnSi$_{0.56}$P$_{0.44}$ may be an interesting system. Since Kervan and Kervan \cite{kervan2012half} conducted an \textit{ab initio} calculations for Fe$_2$MnP and confirmed the half-metallic features of this system.

\section{Magnetocaloric effect and the valence electrons number}

\begin{figure*}[t!]
\center
\includegraphics[width=16cm]{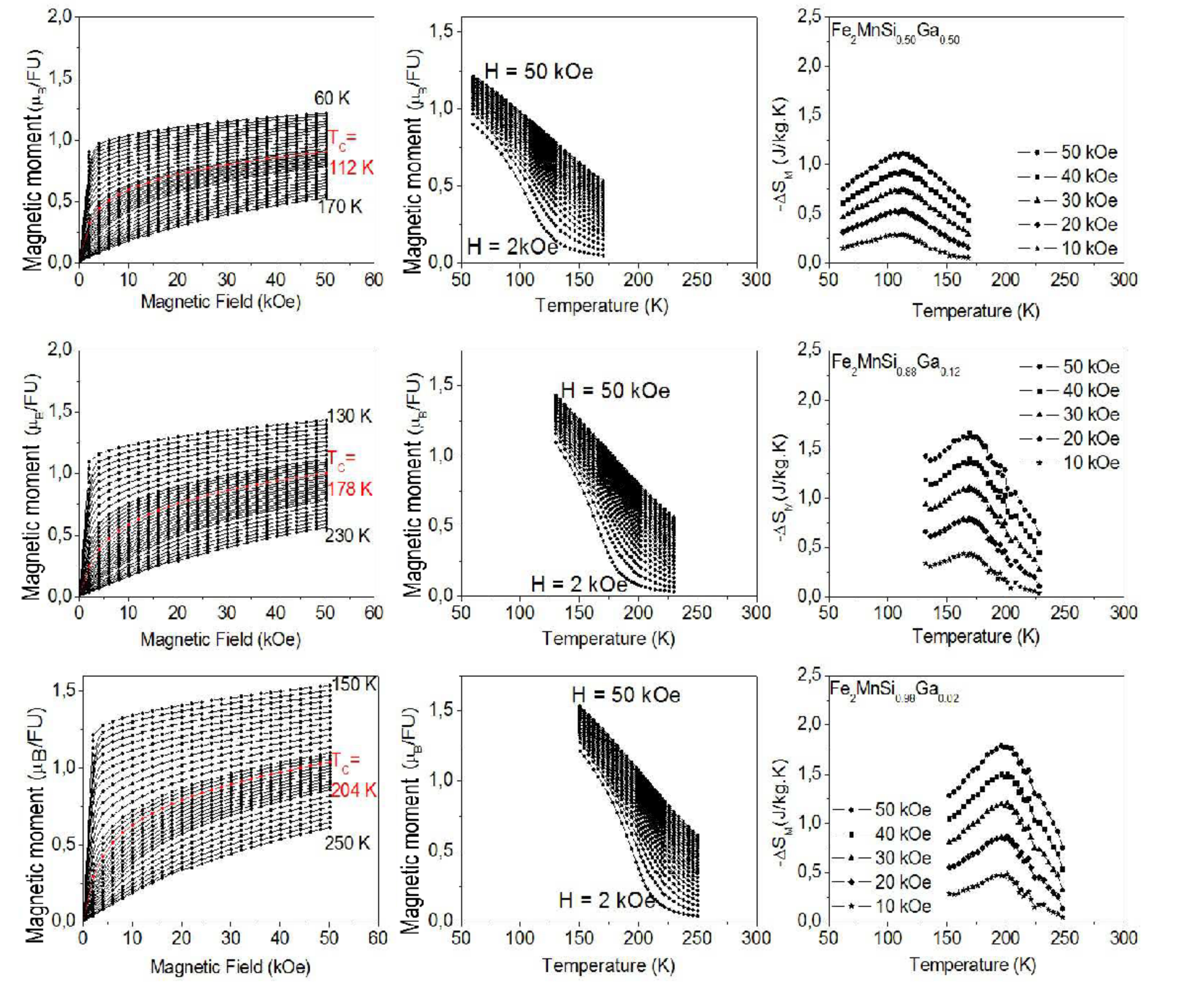}
\caption[Magnetic entropy change as a function of temperature]{Magnetic moment er formula unitary as a function of (left) magnetic field and (center) temperature, as well as the magnetic entropy change (right) as a function of temperature. These results are for the Si-rich side of the Fe$_2$MnSi$_{1-x}$Ga$_x$ series ($x$ = 0.50, 0.12, 0.02).}
\label{mce}
\end{figure*}

We performed magnetization measurements as a function of magnetic field for several temperatures around $T_c$ (see Fig. \ref{mce}-left). We have chosen a few compositions ($x=0.50$, 0.12 and 0.02), among those depicted in Fig. \ref{tc}. Our results are shown in Fig. \ref{mce}-left, thermo-magnetic curves are presented in Fig. \ref{mce}-center. The magnetic entropy change $\Delta S(T,\Delta H)$ calculated by:

\begin{equation}
 \Delta S(T,\Delta H)=\int^{H}_{0}\left(\frac{\partial M(T,H)}{\partial T}\right)_{H}dH,
\label{equ1}
\end{equation}

\noindent and the results are shown in Fig. \ref{mce}-right for $\Delta H$ = 10, 20, 30, 40 and 50 kOe. An interesting result is found by increasing $N_V$ (replacing Ga by Si), showing that the maximum magnetic entropy change also increases. As mentioned above, the Curie temperature increases when $N_V$ rises (see Fig. \ref{tc}), and therefore, a shift towards higher temperatures on the magnetic entropy change peak is expected as observed. 

\begin{figure}[t!]
\center
\includegraphics[width=12cm]{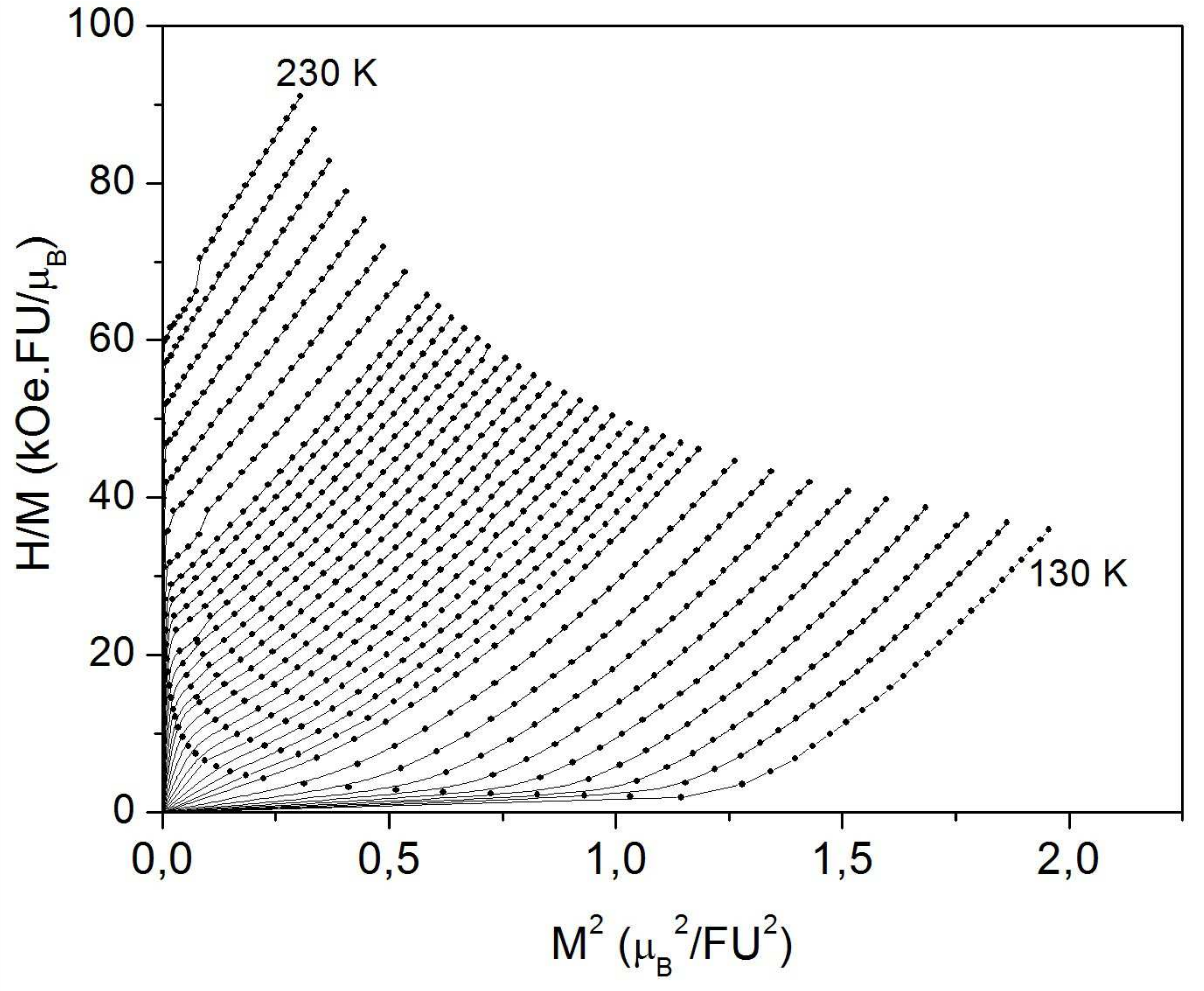}
\caption[Arrot plots for the $x=0.02$ sample]{Arrott plots for the $x=0.02$ sample. These curves were obtained from the corresponding figure \ref{mce}-left and ratifies the second order character of the magnetic transition.}
\label{arrot}
\end{figure}

At this point it is important to explore the character of the magnetic transition; and, to this purpose, we have generated Arrott plots for the sample with $x=0.02$, as shown in Fig. \ref{arrot}. The curves present a positive slope ($B$ parameter of the Landau expansion \cite{marioIntro}), for low values of magnetization, which indicates a second order magnetic transition according to Banerjee's criterion \cite{banerjee}. Similar results were obtained for the other samples.

\begin{figure}[t!]
\center
\includegraphics[width=12cm]{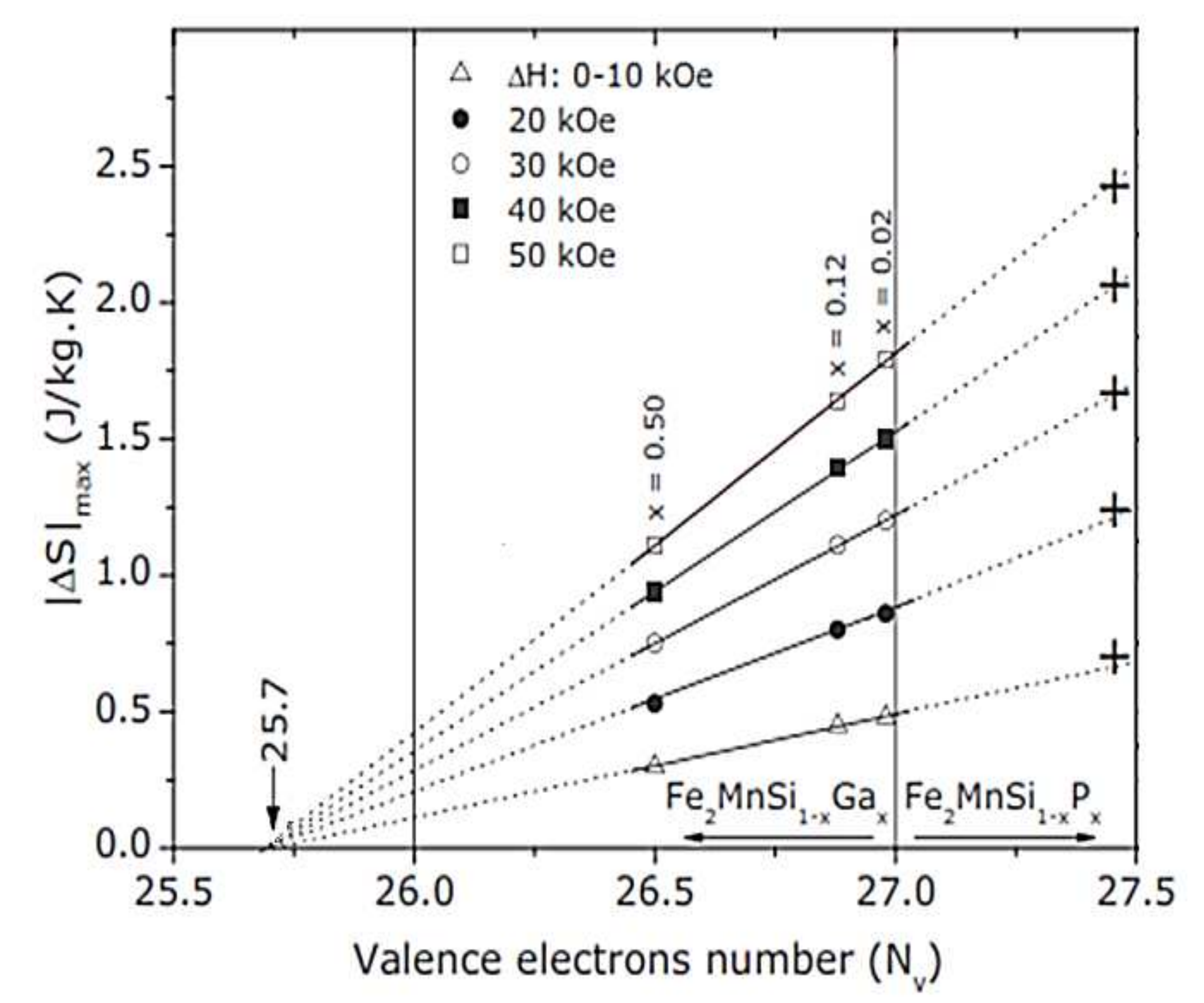}
\caption[Maximum magnetic entropy change $|\Delta S|_{max}$ for several values]{(left axis) Maximum magnetic entropy change $|\Delta S|_{max}$ for several values of $\Delta H$ and as a function of $N_V$ for Fe2MnSi$_{1-x}$Ga$_x$ series. The `+' signals mark the possible $|\Delta S|_{max}$ values of the proposed compound with $T_c$ close to room temperature.}
\label{ds}
\end{figure}

The relative cooling power RCP\footnote{Defined as the maximum magnetic entropy change times the full width at half maximum} (at 50 kOe), ranges from 47 J/kg (for $N_V=26.50$), up to 76 J/kg (for $N_V=26.98$); but we will focus our attention on the maximum magnetic entropy change $|\Delta S|_{max}$, depicted on Fig. \ref{ds}. It shows a linear behavior as a function of $N_V$. For $N_V=27.44$, it is expected to achieve 1.2 J/kg.K@20 kOe, that would indeed be comparable to standard metallic Gd (4 J/kg.K@20 kOe). Thus, the Heusler alloy with $N_V=27.44$ would optimize the Curie temperature, shifting $|\Delta S|_{max}$ towards room temperature, and, in addition, will enhance the magnetocaloric properties (see figure \ref{ds}).

\section{Concluding remarks on Fe$_2$MnSi$_{1-x}$Ga$_x$}

The substitutional series of the Heusler compound Fe$_2$MnSi$_{1-x}$Ga$_{x}$ has been synthesized and investigated experimentally. The phase obtained crystallizes in the cubic $L2_1$ structure and the lattice parameter $a$ increases linearly with the increase Ga content. The Curie temperature changes significantly with Ga content in the 0 $\leq$ $x$ $\leq$ 0.5 range, but the Ga doping interval chosen was not enough to bring $T_C $ to room temperature. The compounds do not follow the Slater-Pauling rule, probably due to a minor degree of disorder in the system, but the almost linear dependence of the magnetic transition temperature with the valence electron number reinforce the half-metallic behavior of the compound. Samples with x $>$ 0.5 were prepared, but x-ray diffraction analysis revealed the appearance of additional peaks, indicating the existence of secondary phases or structural changes in these samples. 

Furthermore, we explored the Si-rich side of Fe$_2$MnSi$_{1-x}$Ga$_{x}$ Heusler alloys, and concluded that the valence electron number $N_V$ plays an important rule on their Curie temperature and magnetic entropy change. Increasing $N_V$ (equivalent to increase the Si content in Fe$_2$MnSi$_{0.5}$Ga$_{0.5}$ compound), leads to a linear increase of both quantities. Our conclusion is that $N_V=27.44$ would bring the Curie temperature of the compound to room temperature, as well as promote an increase in the maximum magnetic entropy change. We also propose to substitute Ga by a group 15 element, like P; and we expect that Fe$_2$MnSi$_{0.56}$P$_{0.44}$ would have a Curie temperatures close to 300 K, with an enhanced magnetocaloric effect. These ideas can indeed lead to new multifunctional materials, opening doors for further researches on this topic.



\part{Perovskite oxides}\label{parte3}

	\chapter{General consideration on perovskite oxide structure}\label{capitulo9}

Perovskite-type oxides belong to a big family that includes $A_{n+1}B_nO$-type perovskites, and they form a type of structure that is very versatile for developments in several technological areas. Perovskites are functional materials that exhibit a range of stoichiometries and crystal structures, and they continue to be actively studied because of their special magnetic and electronic properties. This chapter, we shall focus on the $ABO_3$ perovskite structure type, and explore some of its magnetic properties.

\section{Introduction to the perovskite structure}

Perovskite  $ABX_3$ ($ABO_3$ for us) is the name of a structural family, which the ion $A$ has a larger size than ion $B$, and most of the metallic ions in the periodic table can be used to create $ABX_3$ perovskites \cite{galasso1990perovskites}. Although the majority of the perovskite compounds are oxides or fluorides (for instance K$B$F$_3$ with $B$= Mn,  Fe and Zn \cite{knox1961perovskite}), other forms like heavier halides, sulfides, hydrides, cyanides, oxyfluorides and oxynitrides are also reported \cite{muller1974major}. Perovskites refer to the cubic crystal structure $Pm3m$, which consist of a three-dimensional arrangement of corner-sharing $BO_6$ octahedrons; however, $ABO_3$ perovskites  can crystallize also in other structures.

\begin{figure}[h!]
\begin{center}
\includegraphics[width=13cm]{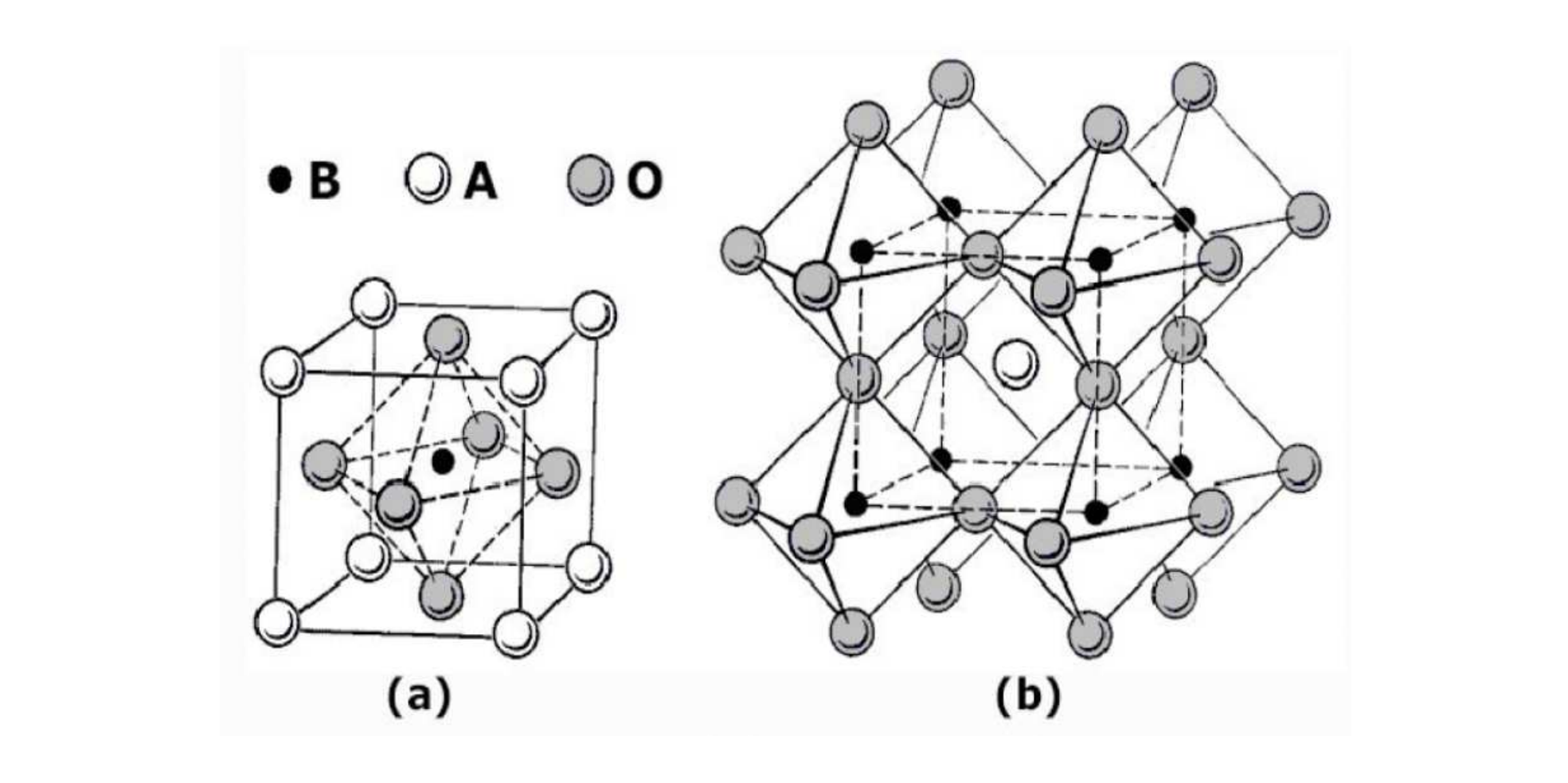}
\caption[Perovskite general structure]{Perovskite general structure. (a) Cubic structure $Pm3m$. b) Octahedral $BO_6$ structure on the corner of the perovskite forming by $3d$ transition metal in cubic configuration.}
\label{estructura_Pero}
\end{center}
\end{figure}

An $A$-site cation fills 12 coordinate cavities formed by the $BO_3$ network and is surrounded by 12 equidistant anions \cite{muller1974major}, as seen in Fig.\ref{estructura_Pero}. According to Lufaso et al. \cite{lufaso2001prediction}, a cubic perovskite can transform into other crystal structures through the distortion of the octahedral $BO_6$, and  the resulting structures are closely related to the cubic perovskite.  In general, the stability of perovskites is often studied in terms of the {\it tolerance factor} ($t$), which was introduced by Goldschmidt in the 1920s \cite{bhalla2000perovskite},

\begin{center}
\begin{equation}
t = \frac{r_A+r_O}{\sqrt{2} (r_B+r_O)}.
\end{equation}
\end{center}

\noindent Here, $r_A$ and  $r_B$ are the ionic radii of cation $A$ and  anion $B$, respectively, while $r_O$ is the ionic oxygen radius. When $t$ is approximately 1, we have a cubic structure. However, if $r_A$ and consequently $t$ decrease, the crystal structure changes to rhombohedral (0.96 < $t$ < 1) or orthorhombic ($t$ < 0.96) structures \cite{tokura1999colossal, jiang2006prediction, li2004formability}. In fact, variations in the $ t$ values define the limits of the crystal structure.

\section{Distortion effects on the perovskite oxide structure}\label{distortion9}

The observed magnetic and electric properties of perovskite oxides  are directly related to their structure , and constituent elements. As mentioned above, site $B$ can be occupied by a  $3d$ transition metal (for example $M$ = Mn and Co ), and site $A$ can be occupied by a trivalent rare earth ($R$) such as  La$^{3+}$, Nd$^{3+}$, and Pr$^{3+}$ or by a divalent alkaline earth ($T$) like Sr$^{2+}$, Ca$^{2+}$ and Ba$^{2+}$. The valence of the $3d$ transition  metal depends directly on what material occupies the site $A$ ($R$ or $T$). When it is occupied by a trivalent element ($R$), the $3d$ transition metal becomes trivalent ($R^{3+}M^{3+}O_3$). However, if $ A $ is occupied by a divalent element ($T$), then $3d$ transition metal becomes tetravalent ($T^{2+}M^{4+}O_3$). For compounds with  partial substitution, $A=(1-x)R + x T$, the $M$ valence is a mixture of $3+$ and $4+$, characterized by: 

\begin{equation}
R_{1-x}^{3+} T_x^{2+} M_{1-x}^{3+} M_x^{4+}O_3.		
\end{equation}

\begin{figure}[h!]
\begin{center}
\includegraphics[width=15cm]{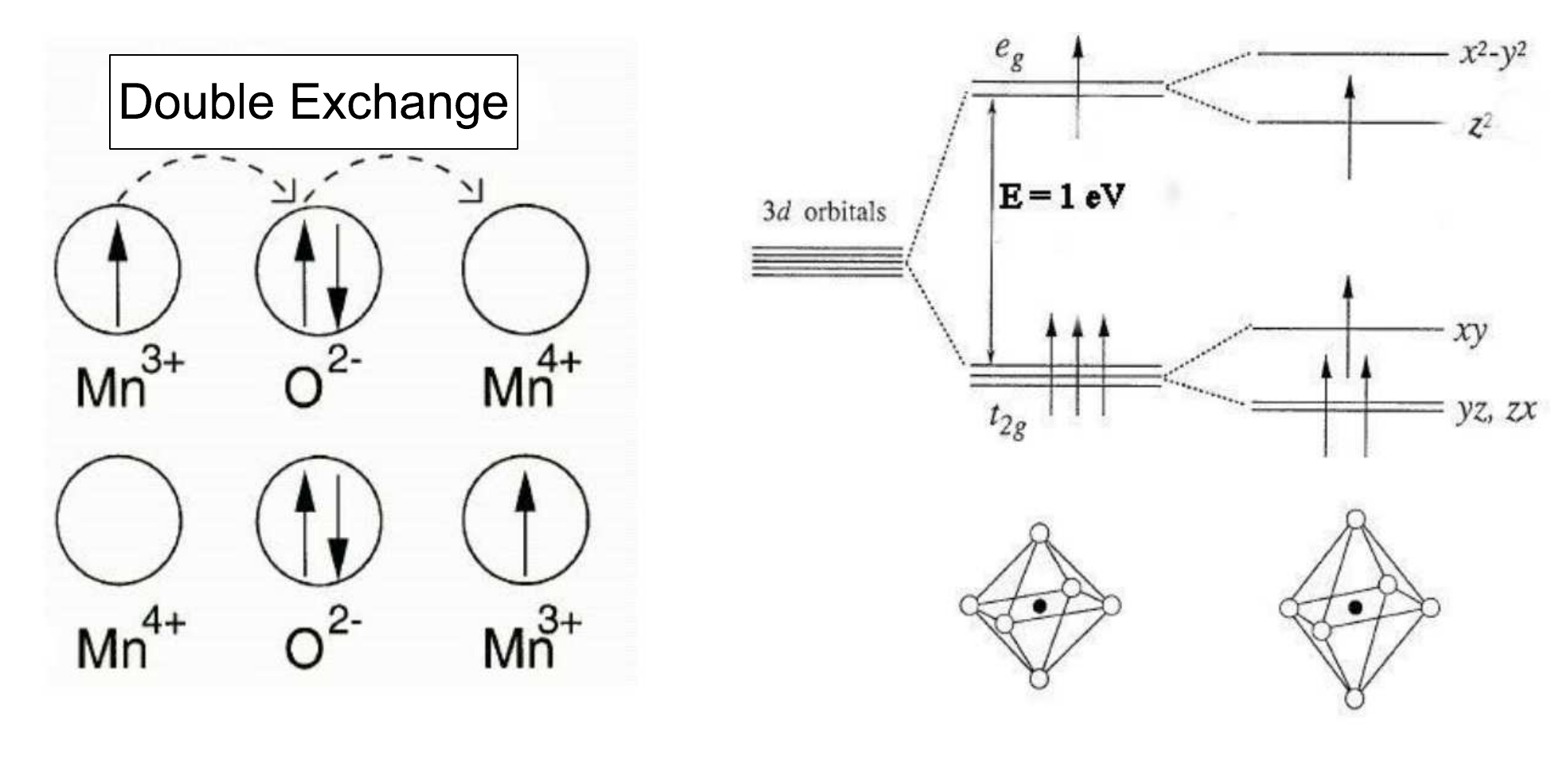}
\caption[Scheme of the double exchange mechanism and Energy diagram of the orbitals]{(Left) Scheme of the double exchange mechanism proposed by C. Zener \cite{zener1951interaction}. (Right) Energy diagram of the orbitals of a free transition metal in an octahedron crystal field formed by O atoms, and in an octahedron with distortion (Jahn-Teller effect).}
\label{manga1}
\end{center}
\end{figure}

This mixture of valence leads to the so-called double exchange interaction, as proposed by Zener using the manganites (where $M$ = Mn) \cite{zener1951interaction}. This mechanism of exchange has its origin in the itinerant character of the electron that occupies the $e_g$ orbital of Mn$^{3+}$, and the hybridization between the  Mn $d$-orbital and the oxygen $p$-orbital. The left panel of Fig. \ref{manga1} illustrated this process. When the initial state has the \{Mn$^{3+}$-O-Mn$^{4+}$\} configuration, one electron passes from the Mn$^{3+}$ to the oxygen, and another (previously in O$^{2-}$) simultaneously migrates to Mn$^{4+}$, leading to the final state \{Mn$^{4+}$-O-Mn$^{3+}$\}. The double exchange occurs when there is a ferromagnetic coupling between Mn ions. The ferromagnetic arrangement observed in these oxides is closely related to the electron movement.

Another important distortion of these perovskites is due to the Jahn-Teller effect \cite{jahn1937stability}, which consists in a distortion caused by the presence of Mn$^{3+}$ ions (manganites case) inside the oxygen octahedrons.  It happens because of the interaction between the $3d$ transition metal orbitals ($e_g$ and $t_{2g}$) and the oxygens at vertices of the octahedral. Electrostatic repulsion induces a distortion of the octahedron in order to decrease the total energy of the system. Therefore, depending on the degree of such lattice distortions, cubic (ideal), orthorhombic and even rhombohedral symmetries may occur        
\cite{bhalla2000perovskite}. This rearrangement of octahedron atoms breaks the degeneracy of the $e_g$ and $t_{2g}$ orbitals, so that the $t_{2g}$ orbitals split into three levels of lower energy, $d_{xy}$, $d_{yz}$, and $d_{zx}$, the orbitals into two levels ($d_{z^2}$ and $d_{x^2-y^2}$) with higher energy \cite{dagotto2001colossal}, as illustrated in the right panel of Fig.\ref{manga1}. 
The splitting occur due to interaction between the transition metal and oxygen orbitals \cite{blundell2001magnetism}. 

A Mn$^{4+}$ ion has three electrons at the $3d$ level so Hund's strong coupling favors population of the orbital $e_g$, and as consequence gives a spin value of 3/2. Since, a Mn$^{3+}$ ion has four electrons at the $3d$ level, an additional electron can occupy either a place in orbital $t_{2g}$ (with spin antiparallel to the others), or in orbital $e_g$ with spin parallel to the others. The latter is energetically more favorable because of the reduction in the Coulomb interaction, leading to a total spin value equal to `2'. The important consequence of the spin separation in different energy levels in the $3d$ orbitals complex is the strong coupling between the $e_g$ conduction electron and those in the  $t_{2g}$ orbitals. In the case of cobaltites ($M$ = Co) this does not always occur, because the energies of the Hund coupling and the crystalline field energy are comparable. This gives to cobaltites an additional degree of freedom, which depends on the spin state of the material; this property will be further explored in Chapter \ref{cap11Coba}.

\section{Perovskites in technology}

Owing to simple preparation, low cost of production, and several interesting physical properties, perovskite oxides are one of the most versatile family of compounds for technological developments. Some of the technologies that are currently being developed that are based on oxides with perovskite structure are the following:

\begin{itemize}

\item {\it Solar cells}: based on organic-inorganic perovskite structured semiconductors that exhibit high charge-carrier mobilities, together with high charge-carrier lifetimes. The light-generated electrons and holes can move relativity long distances, being extracted as current instead of losing their energy as heat within the cell \cite{hodes2013perovskite,niu2015review}.

\item {\it Magnetic hyperthermia}: the magnetic properties of perovskites nanoparticles immersed in a fluid (ferrofluid)  and subjected to external ac external magnetic field, make them useful for local heat treatments. It is possible to heat specific zones affected by cancer and kill the undesired cell. Combined with traditional treatment, magnetic hyperthermia is an alternative tool to help the cure of certain types of cancer \cite{bubnovskaya2012nanohyperthermia}.

\item {\it Spintronics}: several properties of perovskite compounds, such as giant magneto-resistance, colossal magneto-resistance, and their dependence on the electronic charge and spin degrees of freedom, are exploited for efficient use of spintronic devices, and may be efficiently used in advanced technology \cite{ali2013robust,duan2015effects}.

\item {\it Magnetic refrigeration}: as we explained in Chapter \ref{fundaMCE}, the change in the magnetic entropy is one of the qualities that quantify the magnetocaloric effect, and how good a material is for magnetic refrigeration. The magnetic entropy change of perovskite-type manganese oxides is larger than that of Gd (which is the most studied material for this type of application), and therefore  perovskite-type oxides appear to be more suitable candidates for magnetic refrigeration at a high temperature, especially near room temperature \cite{wei2013review,phan2007review}.

\end{itemize}

In the two following chapters, we will explore the magnetic properties of these compounds and focus on half-doped cobaltite (Nd$_{0.5}$Sr$_{0.5}$MnO$_3$) and manganites (La$_{0.6}$Sr$_{0.4}$MnO$_3$).


	\chapter{Spin state and magnetic ordering of half-doped Nd$_{0.5}$Sr$_{0.5}$CoO$_3$ cobaltite}\label{cap11Coba}

Cobaltites show intriguing magnetic and transport properties. When compared with manganites, for instance, they exhibit an additional degree of freedom: the spin state of the Co ions. For $\text{Nd}_{0.5}\text{Sr}_{0.5}\text{CoO}_3$ this spin state  configuration is not well-established, as well as the magnetic ordering below the Curie temperature. In this Chapter, we report magnetization measurements that we have performed in order to understand various aspects of the half-doped $\text{Nd}_{0.5}\text{Sr}_{0.5}\text{CoO}_3$ cobaltite. Our results show that the Co and Nd magnetic sub-lattices couple antiferromagnetically below the Curie temperature $T_c$=215 K, down to very low temperatures. They clarify the presence of the plateau observed at 80 K in the M(T) curve, which in the literature is erroneously attributed, to the onset of an antiferromagnetic ordering. Our magnetization data also clearly shows that Co$^{3+}$ and Co$^{4+}$ are in an intermediate spin state. In addition, we have investigated the magnetic entropy changes in this system. Our experimental results and theoretical analysis allowed a consistent description of this magnetic behavior. Finally, a magnetic phase diagram for $\text{Nd}_{0.5}\text{Sr}_{0.5}\text{CoO}_3$ was built, bored on our results.

Room temperature synchrotron powder diffraction measurements were carried out at beamline 17-BM at the Advanced Photon Source using a 100 $\mu$m monochromated x-ray beam at a wavelength of $\lambda$ = 0.727750 \AA. Magnetization measurements were carried out as a function of temperature and magnetic field using a commercial Superconducting Quantum Interference Device (SQUID) from Quantum Design$^{\circledR}$.

\section{Brief survey on cobaltite compounds}\label{intro}

Physical properties of cobaltites, as well as manganites, depend on the interplay between charge, spin, orbital and lattices degrees of freedom. However, cobaltites have an additional degree of freedom as Co$^{3+}$ and Co$^{4+}$ can occur in three different spin configurations \cite{PhysRevB.63.224416}: low-spin LS ($t^6_{2g}e_g^0$ for Co$^{3+}$ and $t^5_{2g}e_g^0$ for Co$^{4+}$), intermediate-spin IS ($t^5_{2g}e_g^1$ for Co$^{3+}$ and $t^4_{2g}e_g^1$ for Co$^{4+}$) and high-spin HS ($t^4_{2g}e_g^2$ for Co$^{3+}$ and $t^3_{2g}e_g^2$ for Co$^{4+}$). This unusual flexibility arises from the comparable energies between the Hund's coupling and the crystal-field splitting \cite{PhysRevB.63.224416}.

Those intriguing magnetic and transport properties are mainly ruled by the doping level $x$ in  $R_{1-x}T_x\text{CoO}_3$, which induces lattice distortions and changes the valence of the Co ions. Compounds with $x = 0$ contain only Co$^{3+}$ ($3d^6$), while those with $x=1$ have only Co$^{4+}$ ($3d^5$); and, obviously, intermediate values of $x$ lead to mixed valence compounds. Rao et al. \cite{Rao1977353} investigated $R_{1-x}T_x\text{CoO}_3$ (with $R$ = Eu, Nd, Pr, Sm and Gd and $T$ = Sr, Ca and Ba) and show that, in general, lower values of $x$ lead to a very high electrical resistivity with no magnetic ordering; and above a critical value of $x$, a ferromagnetic ordering occurs. The Curie temperature, for this heavy doped region of the phase diagram, decreases with decreasing the size of the rare-earth ion. $\text{La}_{1-x}\text{Sr}_x\text{CoO}_3$ \cite{Hjalmarsson20081422,PhysRevB.54.16044,PhysRevB.57.R3217} and $\text{Pr}_{1-x}\text{Sr}_x\text{CoO}_3$ \cite{PhysRevB.68.024427,jsoldstate1995,Deac20141} series are the most studied cobaltites, and the latter are more interesting due to the non-zero magnetic moment of Pr. Some compounds of $\text{Pr}_{1-x}\text{Sr}_x\text{MnO}_3$ series, for example, exhibit a broad peak in the susceptibility data around 90 K, whose origin  remains  an open question \cite{Deac20141}. Some authors claiming that it is due to a state spin transition from the intermediate to low-spin state \cite{JETPLetters2006}, while others argue that it is caused by a change of the ferromagnetic state associated with orbital ordering \cite{PhysRevB.68.024427}.

$\text{Nd}_{1-x}\text{Sr}_x\text{CoO}_3$ compounds have also been intensely investigated. Fondado and co-workers \cite{Fondado2001444}, for example, studied the compounds in the composition range $0 \leq  x \leq 0.4$. The authors found that, for $x > $0.2, the system is ferromagnetic, but for $x = $0 ($\text{NdCoO}_3$), the material is paramagnetic due to the zero paramagnetic effective magnetic moment of Co ions, which indicates that they are in low spin configuration. However, for $T > 250$ K, a deviation from linearity appears in $\chi^{-1}(T)$ indicating an onset of spin transition in the Co ions. This behavior is in fact observed for samples with $x \leq$ 0.1. Transport measurements \cite{Fondado2001444} show that as $x$ increases from 0.1 to 0.4, the activation energy also increases, which would induce different spin-states in the Co ions for the $\text{Nd}_{1-x}\text{Sr}_x\text{CoO}_3$ compounds. However, the authors were not able to define if the spin state transition occurs due to thermal energy or crystal distortion as $x$ changes. For $x = 0.33$, Paraskevopoulos et al. \cite{PhysRevB.63.224416} showed that for temperatures above $T_C$ ($\sim$ 173 K) the Co$^{3+}$ and Co$^{4+}$ are in the IS and LS configurations, respectively, which confirms the change of the spin-state of the Co ions as $x$ increases. For samples of half-doped $\text{Nd}_{1-x}\text{Sr}_x\text{CoO}_3$, the spin-state is not yet well defined and, the broad maximum around 80 K is also an open question in the literature.

Here we investigated the magnetic investigation of half-doped $\text{Nd}_{0.5}\text{Sr}_{0.5}\text{CoO}_3$ cobaltite. Magnetization measurements were made in order to determine the magnetic arrangement of Co and Nd sub-lattices, as well as the spin-state of the Co ions. In addition, magnetic entropy change was also explored, and those results are also consistent with our other findings. We close this contribution with a complete magnetic phase diagram for half-doped cobaltite.

\section{Magnetostructural relationship}\label{mss}

\begin{figure}
\center
\includegraphics[width=8cm]{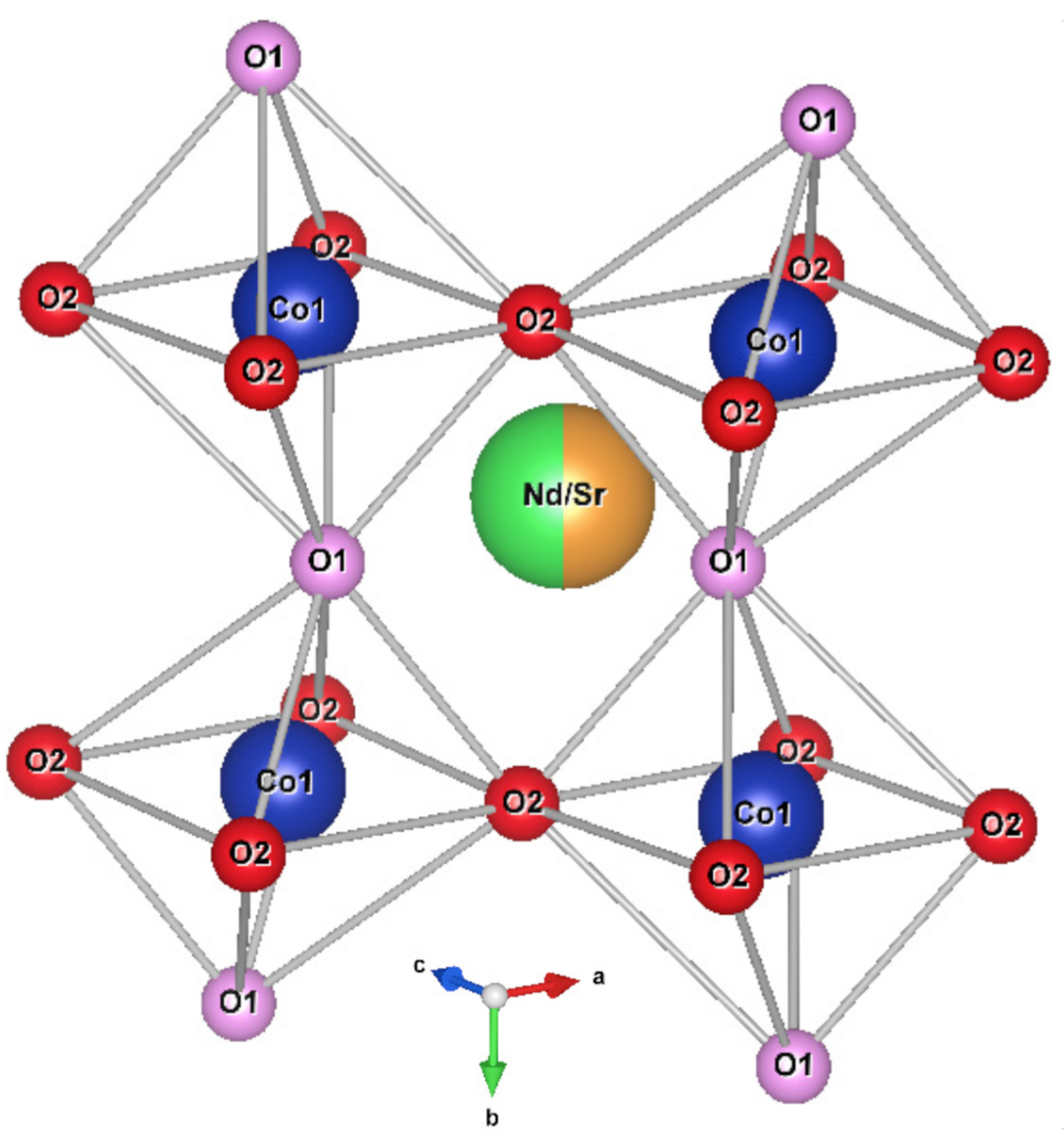}
\caption[Partial crystal structure of Nd$_{0.5}$Sr$_{0.5}$CoO$_3$ cobaltite]{Partial crystal structure of Nd$_{0.5}$Sr$_{0.5}$CoO$_3$ cobaltite, where the Co octahedral environment is highlighted.}
\label{estructure}
\end{figure}

Crystal structure and magnetism are strongly related. In the present case, the Co ions are able to assume low, intermediate and high spin configurations, and identification of this relationship is of great importance. This section discusses the crystal structure of $\text{Nd}_{0.5}\text{Sr}_{0.5}\text{CoO}_3$ cobaltite (see Fig. \ref{estructure}).

\subsection{Crystal structure}

\begin{figure}
\center
\includegraphics[width=10cm]{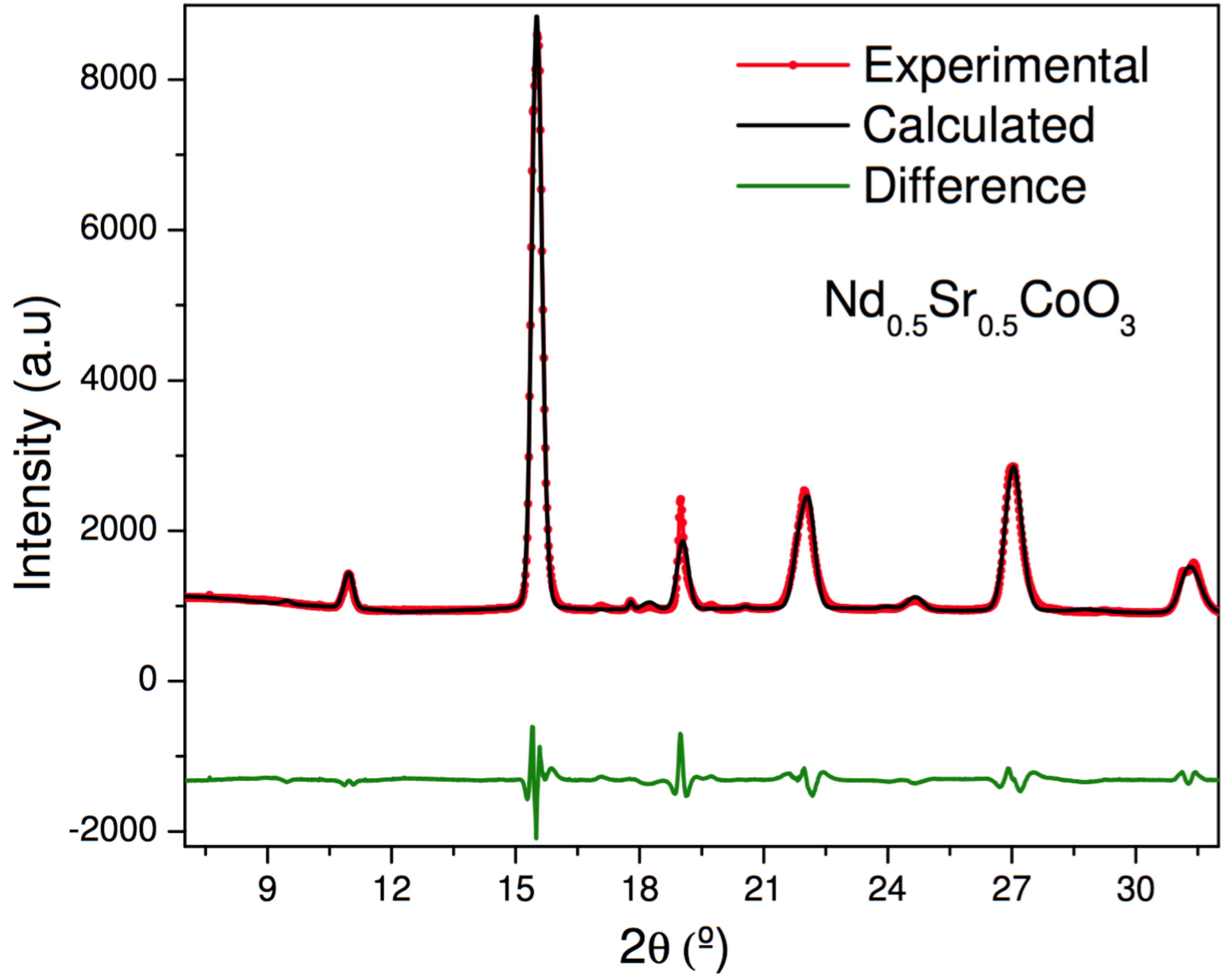}
\caption[X-ray diffraction pattern for Nd$_{0.5}$Sr$_{0.5}$CoO$_3$ cobaltite]{X-ray diffraction pattern for Nd$_{0.5}$Sr$_{0.5}$CoO$_3$ cobaltite using synchrotron radiation with a wavelength of $\lambda$ = 0.727750 \AA. Fitting was performed with the Rietveld method.}
\label{drx}
\end{figure}

Rietveld analysis of  the Nd$_{0.5}$Sr$_{0.5}$CoO$_3$ diffractogram shows a single phase, as can be seen in Fig. \ref{drx}. The system crystallizes in an orthorhombic structure with space group Imma \cite{0953-8984-24-23-236005}. In this structure, the rare-earth metal, Sr and one of the oxygens ions (O1 - apical) occupy the Wyckoff position $4e$ (0,1/4,x), Co occupies the $4b$ site (0,0,1/2) and the other oxygen (O2 - equatorial) occupies the 8g (3/4,y,1/4) positions.  The lattice parameters of our sample are presented in Table \ref{crystaldata10}, and are in very good agreement with other result reported in the literature  \cite{0953-8984-24-23-236005}. The table also shows the interatomic distances, and it is possible to see that O1 - Co distance (apical distance of the octahedron) is larger than the O2-Co basal distance, which indicates that, at least at room temperature, the octahedron is mostly elongated. In other words, the equatorial oxygens (O2) are symmetrically arranged around the cobalt ion. at a distance of 1.898 (\AA{}), while the Co-O1 distance (apical oxygen) is of 1.986 (\AA{}), almost 4.5 \% of difference. These values of the Co-O distances are larger than those reported previously \cite{PhysRevB.64.224404} for the $\text{Nd}_{0.67}\text{Sr}_{0.33}\text{CoO}_3$ compound, where the Co ions are assumed to be in a low-spin configuration. Considering that the ionic radius of high spin Co can be 10 \% larger than the one for the low spin Co \cite{PhysRevB.64.224404}, this indicate that the Co spin state is either on intermediate or high spin configuration. As we shall see, this is an important point in the discussion of the magnetic properties of this system.

\begin{table}
\caption{Refined crystallographic data and reliability factors
for Nd$_{0.5}$Sr$_{0.5}$CoO$_3$.\label{crystaldata10}}
    \center
    \begin{tabular}{|c|c|c|c|}
 \multicolumn{4}{c}{Lattice parameters}                                     \\\hline\hline
$a$ (\AA{})   &$b$ (\AA{})    & $c$ (\AA{})          & $V$(\AA{}$^3$)       \\\hline
5.353(2)      & 7.681(1)      & 5.385(2)             & 221.4(1)             \\\hline\hline
\multicolumn{4}{c}{Interatomic distances}                                      \\\hline\hline
O1-Co(\AA{})  &O2-Co(\AA{})   & Nd-Co(\AA{})         &O1-O1 / O2-O2 (\AA{})  \\\hline
  1.986(5)    & 1.898(1)      & 3.294(7)             & 3.972(6) / 3.797(2)   \\\hline\hline
\multicolumn{4}{c}{Interatomic distances and angles}                          \\\hline\hline
Nd-Nd (\AA{}) &Co-Co(\AA{})   & Co-O1-Co ($^\circ$)  & Co-O2-Co ($^\circ$)  \\\hline
  3.785(8)    & 3.796(8)      & 150.5(5)             & 177(1)               \\\hline\hline
   \multicolumn{4}{c}{Reliability factors}                                  \\\hline\hline
R$_{wp}$ (\%) &R$_{p}$ (\%)   & R$_{exp}$ (\%)       & $\chi^{2}$           \\\hline
  10          & 4             & 2                    & 2.8                  \\\hline\hline
            \end{tabular}
\end{table}

\section{Magnetic phase diagram}

As discussed in section \ref{intro}, the Nd and Co magnetic arrangement, as well as  the Co spin state of the Co ion in cobaltites remains as open questions. In order to draw a reliable phase diagram and be able to specify the Co spin state, we have performed magnetization measurements in this system. To support our findings, we have also investigated (both experimentally and teoretically) the magnetocaloric aspect in this compound. Our results allowed us to achieve our goals of determining the Co spin state and build up a magnetic phase diagram for $\text{Nd}_{0.5}\text{Sr}_{0.5}\text{CoO}_3$.

\subsection{Cobalt spin state}

\begin{figure}
\begin{center}
\includegraphics[width=12cm]{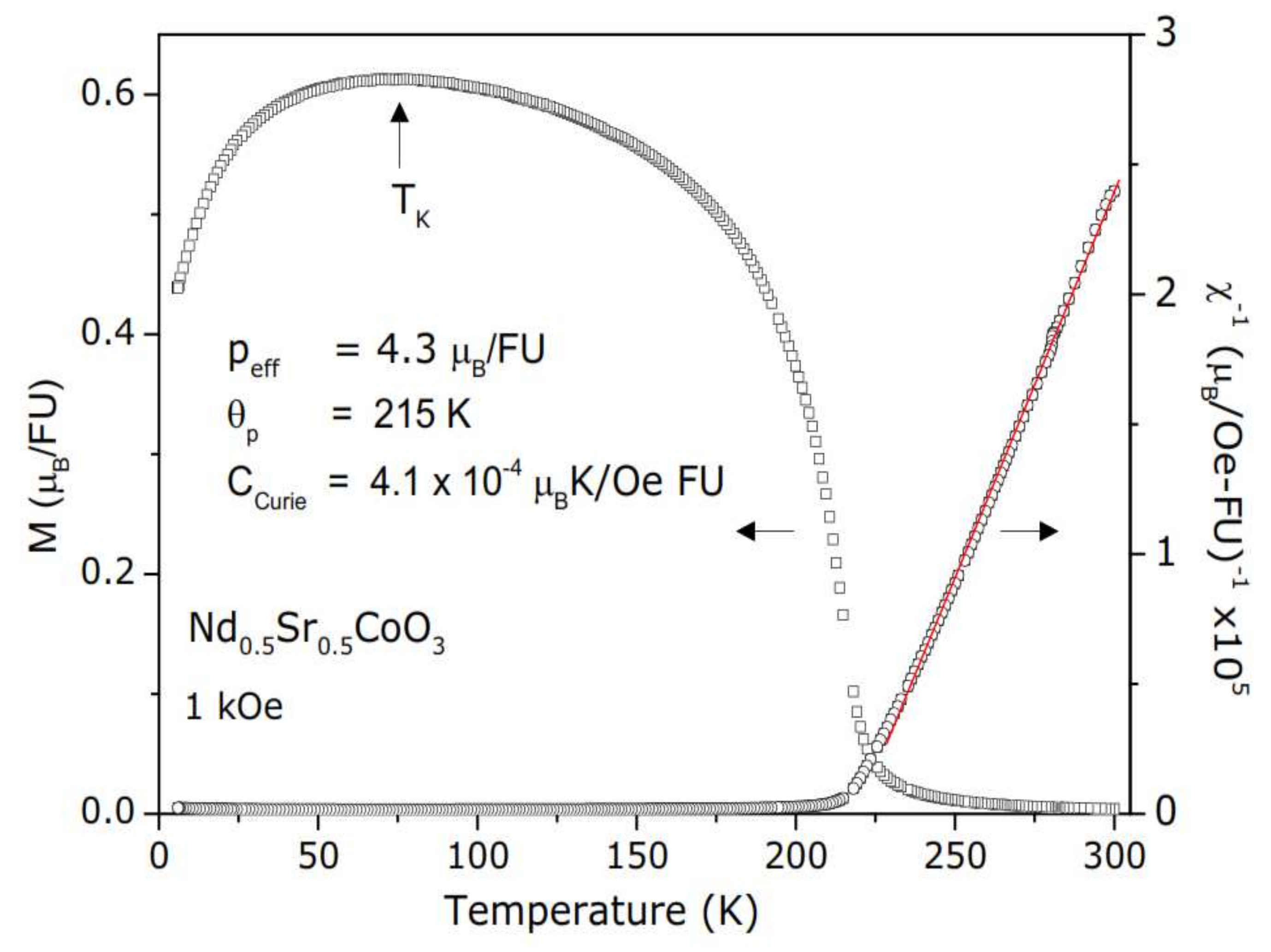}
\end{center}
\caption[Experimental magnetization and reciprocal susceptibility]{Experimental magnetic moment per formula unitary (left axis) and reciprocal susceptibility (right axis) as a function of temperature for $\text{Nd}_{0.5}\text{Sr}_{0.5}\text{CoO}_3$ cobaltite. Solid line is the fitting of equation \ref{CW} to the experimental data. The output parameters of this fitting are displayed on the figure.}\label{mt}
\end{figure}

Magnetization as a function of temperature was obtained under applied magnetic field of 1000 Oe, as shown on Fig. \ref{mt} (left axis). In that curve it is possible to see a clear transition at $T_c=215$ K and a broad maximum peaking at $T_k=80$ K, consistent with previous reports \cite{apl/101/4/10.1063/1.4738889,PhysRevB.63.224416}.

The reciprocal magnetic susceptibility was also be obtained, and is presented on Fig. \ref{mt} (right axis). The linear behavior of this quantity above $T_c$ is a clear indication that the system reached the paramagnetic phase and therefore satisfies the Curie-Weiss law. However, a slight deviation from the genuine linear behavior was detected and a better fit is obtain with the theoretical reciprocal molar magnetic susceptibility given by
\begin{equation}\label{CW}
\chi(T)^{-1}=\left[\chi_0+\frac{C}{T-\theta_P}\right]^{-1},
\end{equation}
where $C=\frac{1}{2}(C_{\text{Nd}}+C_{\text{Co}^{3+}}+C_{\text{Co}^{4+}})$ is the overall molar Curie constant, $\theta_p$ is the paramagnetic Curie constant \cite{marioIntro} and $\chi_0$ was added to take into account some diamagnetic contribution, perhaps from the sample holder. The fitting of the experimental data with Eq. \ref{CW} returns $\chi_0= 6.2\times 10^{-7}$ emu/Oe FU, $\theta_p$= 215 K and $p_{eff}= 4.3 \mu_B$/FU. The latter was calculated considering the contribution of all magnetic atoms (Nd and Co).

\begin{figure}
\begin{center}
\includegraphics[width=10cm]{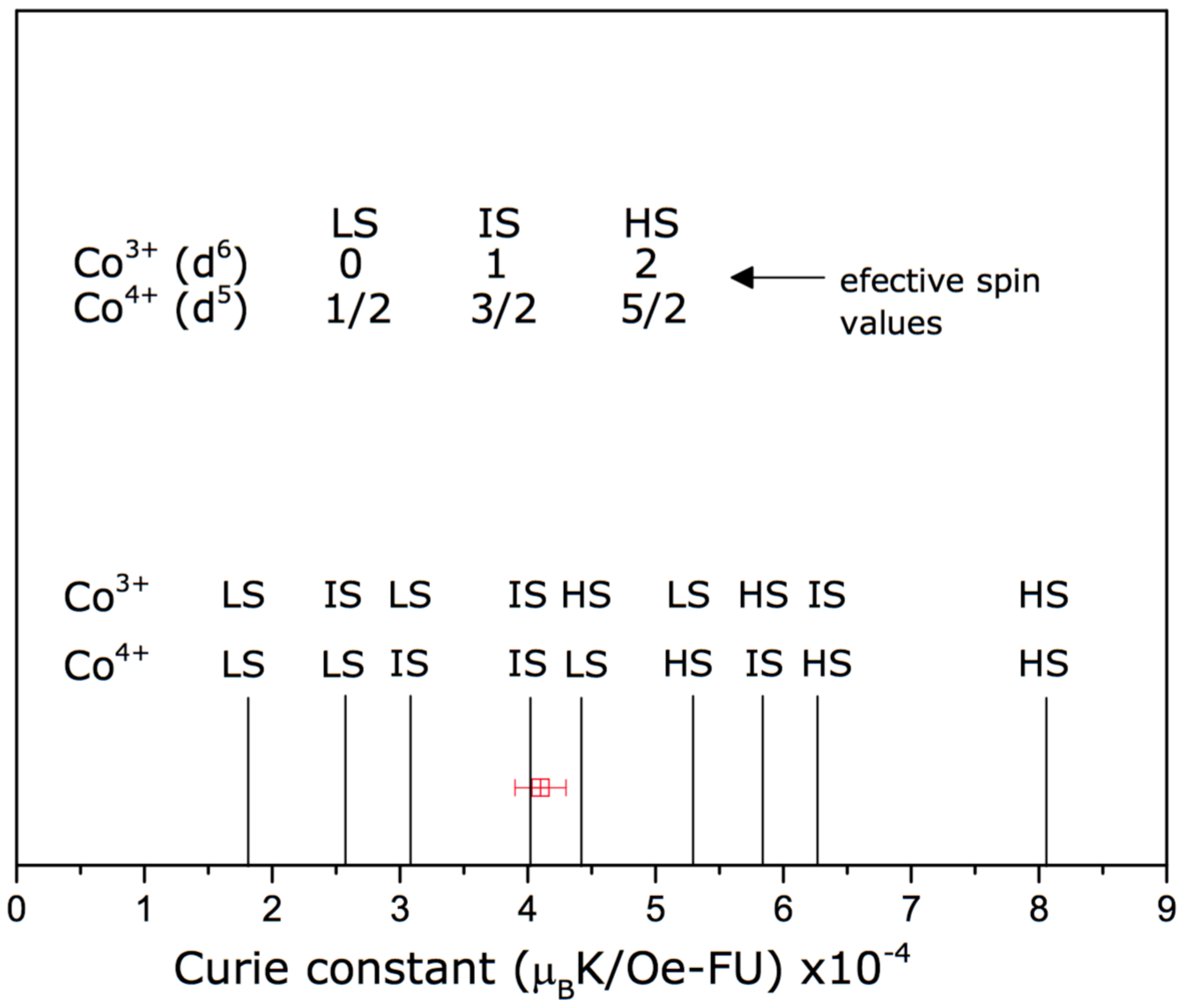}
\end{center}
\caption[Experimental and theoretical Curie constants considering the possible spin states]{Experimental (open square) and theoretical (solid lines) Curie constants considering the possible spin states. This result clearly indicates that both, Co$^{3+}$ and Co$^{4+}$ are on intermediate state.}\label{para}
\end{figure}

The experimental molar Curie constant is depicted in Fig. \ref{para}, together with the theoretical values for several Co spin configurations. The theoretical Curie constants cwere determined the Land\'e factor for both Co sub-lattices $g_\text{Co}=2$ and, for the Nd sub-lattice, $g_\text{Nd}=8/11$ \cite{marioIntro}, while $J_\text{Nd}=9/2$ and $J_\text{Co}=S_\text{Co}$, assuming that there is no orbital contribution. The experimental Curie constant matches the theoretical values obtained for both Co ions (Co$^{3+}$ and Co$^{4+}$) are in an intermediate state; and, have with $g=2$. These results are consistent with our precious hypothesis that the distortions occur in such way that the orbital moment is quenched, and the Co$^{3+}$ octahedral environment becomes elongated, and Co$^{4+}$ octahedral environment is compressed. However, no such distinction is apparent in the diffraction data directly, most likely because it probes long range order and there are no reports in the literature of charge (or orbital) ordering in this family of materials.

\begin{figure}
\begin{center}
\includegraphics[width=10cm]{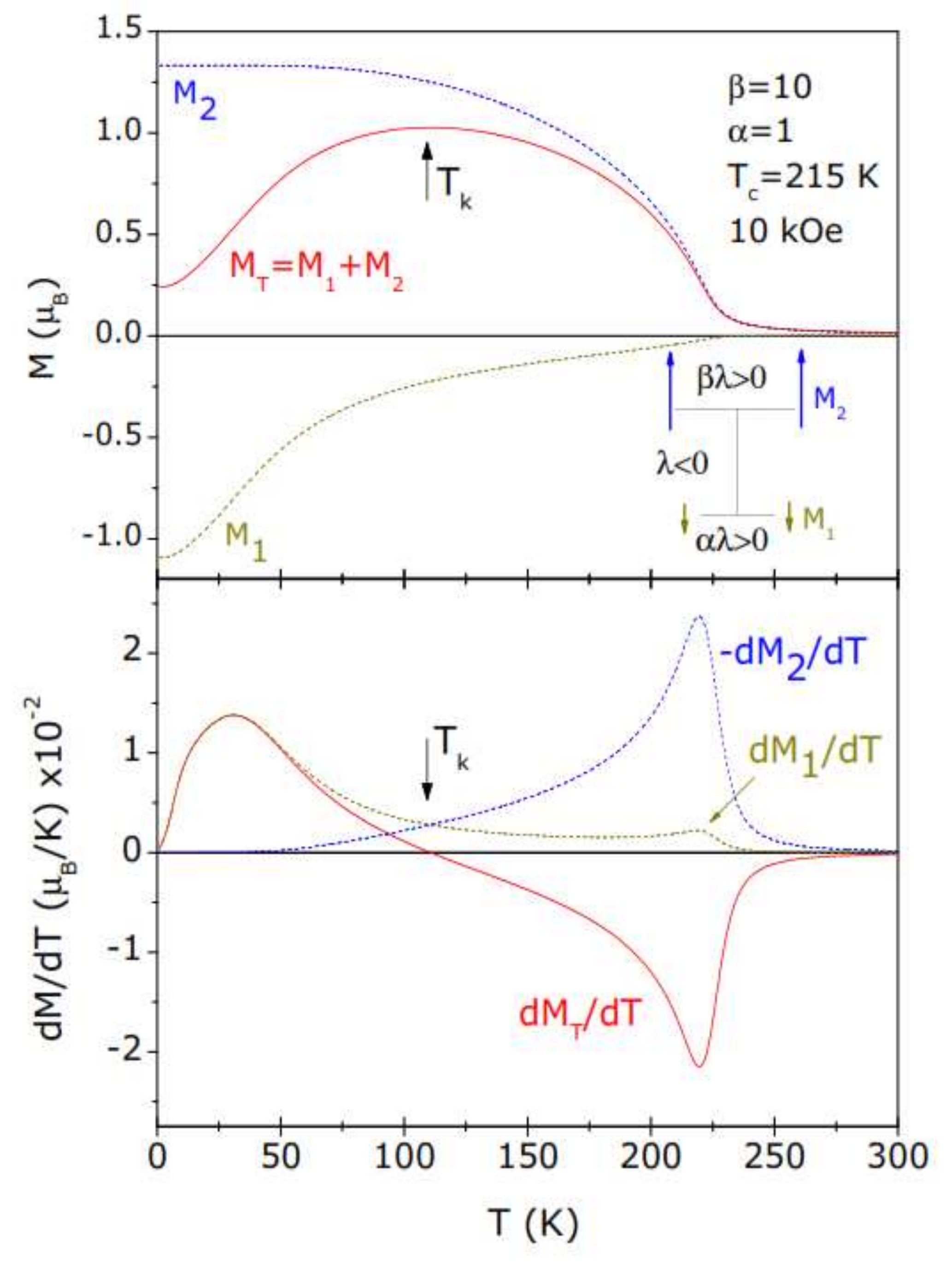}
\end{center}
\caption[Theoretical mean field magnetization curve]{Top: theoretical mean field magnetic moment curve obtained from a simple model of two sub-lattices with ferromagnetic intra-lattice arrangement and ferrimagnetic inter-lattice ordering. Here, $\beta$ ($\alpha$) is refer to interaction intra-molecular between the Nd (Co) ions, while $\lambda$ is refer to the interaction between the two sub-lattice in the system. The aim of this evaluation is to elucidate the Nd and Co ordering by a simple comparison of this result with those experimental data on figure \ref{mt}. Bottom: temperature derivative of each magnetization in order to achieve the physical meaning of $T_k$, the temperature in which the total magnetization peaks. In few words, $T_k$ means the temperature below which the negative magnetization becomes to rule the total magnetization.}\label{teomag}
\end{figure}

To understand the magnetic behavior of this system, Reis et al. \cite{reis2017spin} considered a ferrimagnetic arrangement, between two sub-lattices and a mean field approach to calculate the magnetization, shown in the top panels of Fig. \ref{teomag}. Sub-lattice  
`1' is associated with the Nd sites and sub-lattice `2' with Co sites. $M_1$ and  $M_2$ refer to corresponding sub-lattices magnetization. In the Fig. \ref{teomag}, the calculated values of $M_1$, $M_2$ and $M_T=M_1+M_2$ are plotted as function of the temperature. The `2' sub-lattice rules the critical temperature, as can be seen on figure \ref{teomag}-top. In the high temperature limit both $M_1$ and $M_2$ goes to zero, and do goes $M_T$. On the other hand, as the temperature decreases, the absolute value of $M_1$, due to the low intra-lattice exchange interaction, and as a result, the value of $M_T$ decreases, leading to value of peak at    $T_k$.

$T_k$ is defined as temperature at which $M_T (T)$ is maximum when $dM_T/dT=0$ or, equivalently, when $dM_1/dT=-dM_2/dT$, as the bottom panel of Fig. \ref{teomag}, clearly illustrated. We note that the temperature dependence of the total magnetization is dominated by $M_1$, at low temperatures and by $M_2$ at high temperatures close to and above $T_C$.

At this point we can compare the theoretical (figure \ref{teomag}-top) and experimental (figure \ref{mt}) magnetic moment curve. Indeed, Nd sub-lattice has a ferromagnetic intra-lattice arrangement, as well as the Co sub-lattice. On the other hand,  Nd-Co sub-lattices aligns in an antiparallel to each other; leading to an overall ferrimagnetic arrangement. It is important to note here that previous reports in the literature \cite{PhysRevB.64.224404} ascribe $T_k$ as the onset of antiferromagnetism, i.e., the Nd-Co ordering is antiparallel only below $T_k$. However, from the above theoretical results, it is clear that the magneticzation of the  Nd-Co sub-lattices are always aligned antiparallel for $T$<$T_c$.

\subsection{The magnetocaloric effect}

\begin{figure}
\begin{center}
\includegraphics[width=12cm]{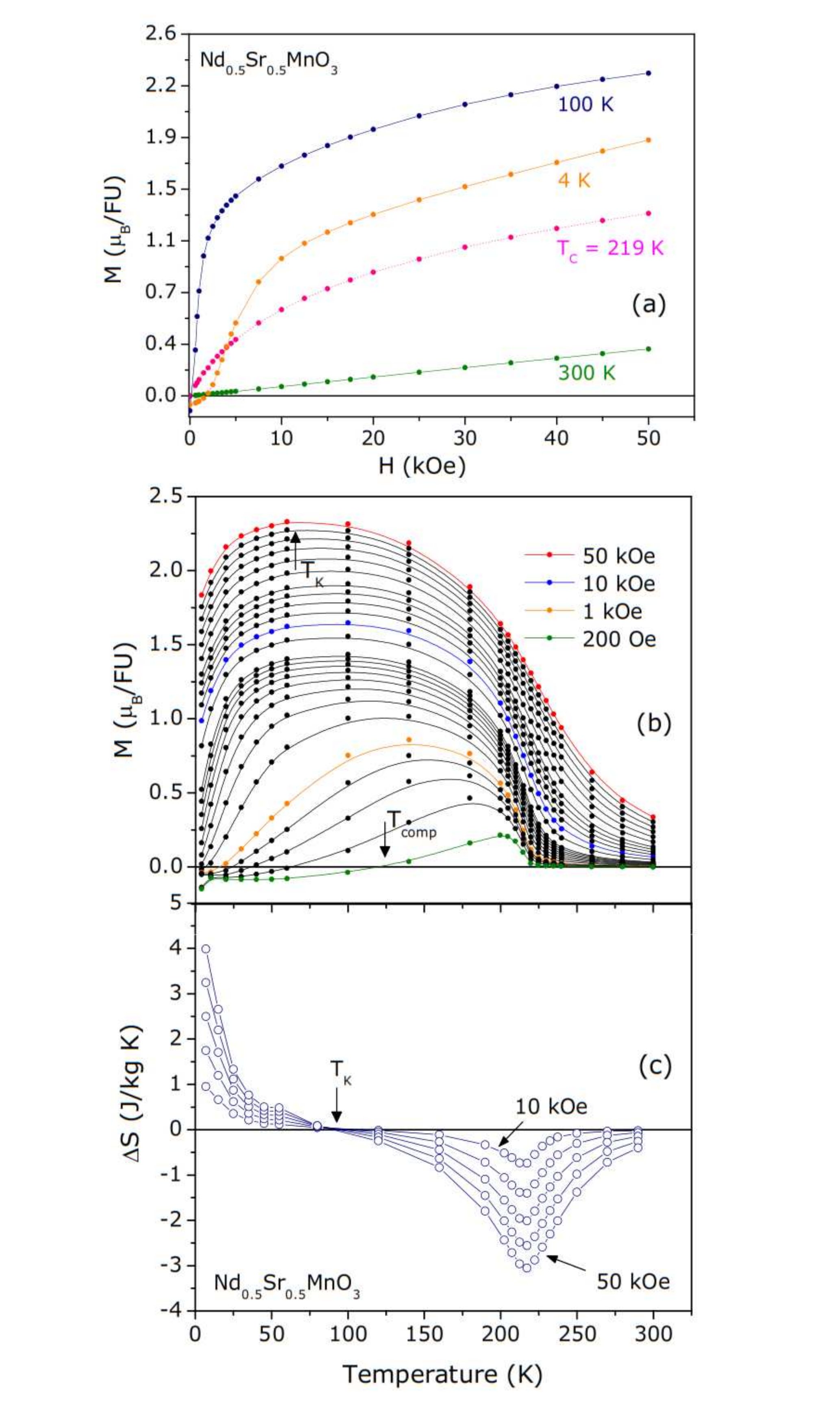}
\end{center}
\caption[Experimental magnetic moment per formula unitary as a function of magnetic field and temperature]{Experimental magnetic moment per formula unitary as a function of (a) magnetic field and (b) temperature; (c) magnetic entropy change as a function of temperature, for several values of applied magnetic field change. $T_k$ and $T_{comp}$ represent, respectively, the temperature in which the magnetization peaks and the one in which Nd and Co sub-lattices are equal in magnitude.}\label{DS}
\end{figure}

We have discussed general aspect of the magnetocaloric effect (MCE) in the Chapter \ref{fundaMCE}. Here, we  measured the  magnetization of Nd$_{0.5}$Sr$_{0.5}$CoO$_3$ as a function of an applied magnetic field, for several temperatures. The result are depicted in Fig. \ref{DS}(a). From the full set of data that we have collected, only a few selected temperatures are shown in this figure. It is noteworthy that the 4 K isothermal curve exhibits  a negative value of magnetization at low values of magnetic field,  characteristic behavior of a ferrimagnetic material \cite{marioIntro}.  In Fig. \ref{DS}(b), we show the magnetization as a function of temperature, for several applied magnetic fields. This data presents another important signature of a ferrimagnetic arrangement: a compensation temperature $T_{comp}$, in which $|M_1|=|M_2|$. From our data, it is possible to extract the magnetic entropy change as a function of temperature, for several values of the applied magnetic field change, and the result are depicted in Fig. \ref{DS}(c). They display some interesting aspects: (i) a peak at $T_c$, similarly to what is observed in ferromagnetic materials, and (ii) a zero crossing at $T_k$. The literature \cite{PhysRevB.64.224404} claims that the system is ferromagnetic above $T_k$ and antiferromagnetic below this temperature, due to the signal (either positive or negative) of the magnetic entropy change\cite{apl/101/4/10.1063/1.4738889}. However, in Ref. \cite{reis2017spin} is our theoretical approach, shows that the system is an ferrimagnetic material and is compatible with the experimental data.

\begin{figure}
\begin{center}
\includegraphics[width=10cm]{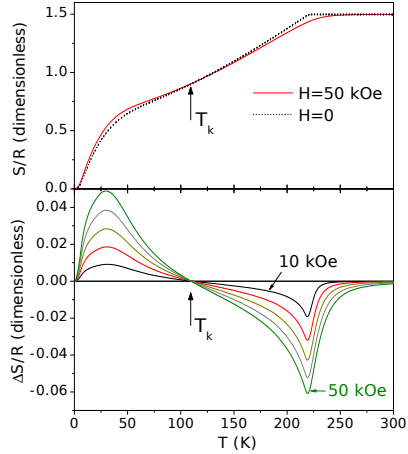}
\end{center}
\caption[Theoretical zero field and field magnetic entropy as a function of temperature]{Top: theoretical zero field and field (50 kOe) magnetic entropy as a function of temperature. Bottom: magnetic entropy change due to a magnetic field change. Comparison of this bottom theoretical figure with the experimental one, on figure \ref{DS}(c), confirms our claim that the system is genuinely ferrimagnetic; and $T_k$ is the crossing temperature of the zero and non-zero field magnetic entropies.}\label{teods}
\end{figure}

Following the stpdf described in Ref. \cite{reis2017spin}, it is possible to obtain the zero field magnetic entropy and, at H=50 kOe; shown i the top panel of the Fig. \ref{teods}. Note that there is a crossing of the curves for H=0 and 50 kOe exactly at $T_k$, emphasizing that this temperature is indeed of extreme importance to the whole magnetic phenomena that occurs in this sample. The difference of these curves, i.e., the magnetic entropy change, way also be evaluated for several values of applied magnetic field and the result are depicted in the bottom panel of the Fig. \ref{teods}. This result qualitatively agree with the ones obtained experimentally and shown in Fig. \ref{DS}(c). It ratifies our claim that this material is indeed ferrimagnetic for the whole temperature range and $T_k$ is the temperature at which the magnetic entropy curves cross, and not the one which the Nd-Co becomes antiparallel, as claimed in the literature \cite{apl/101/4/10.1063/1.4738889}.

\subsection{Phase diagram and concluding remarks}

\begin{figure}
\begin{center}
\includegraphics[width=12cm]{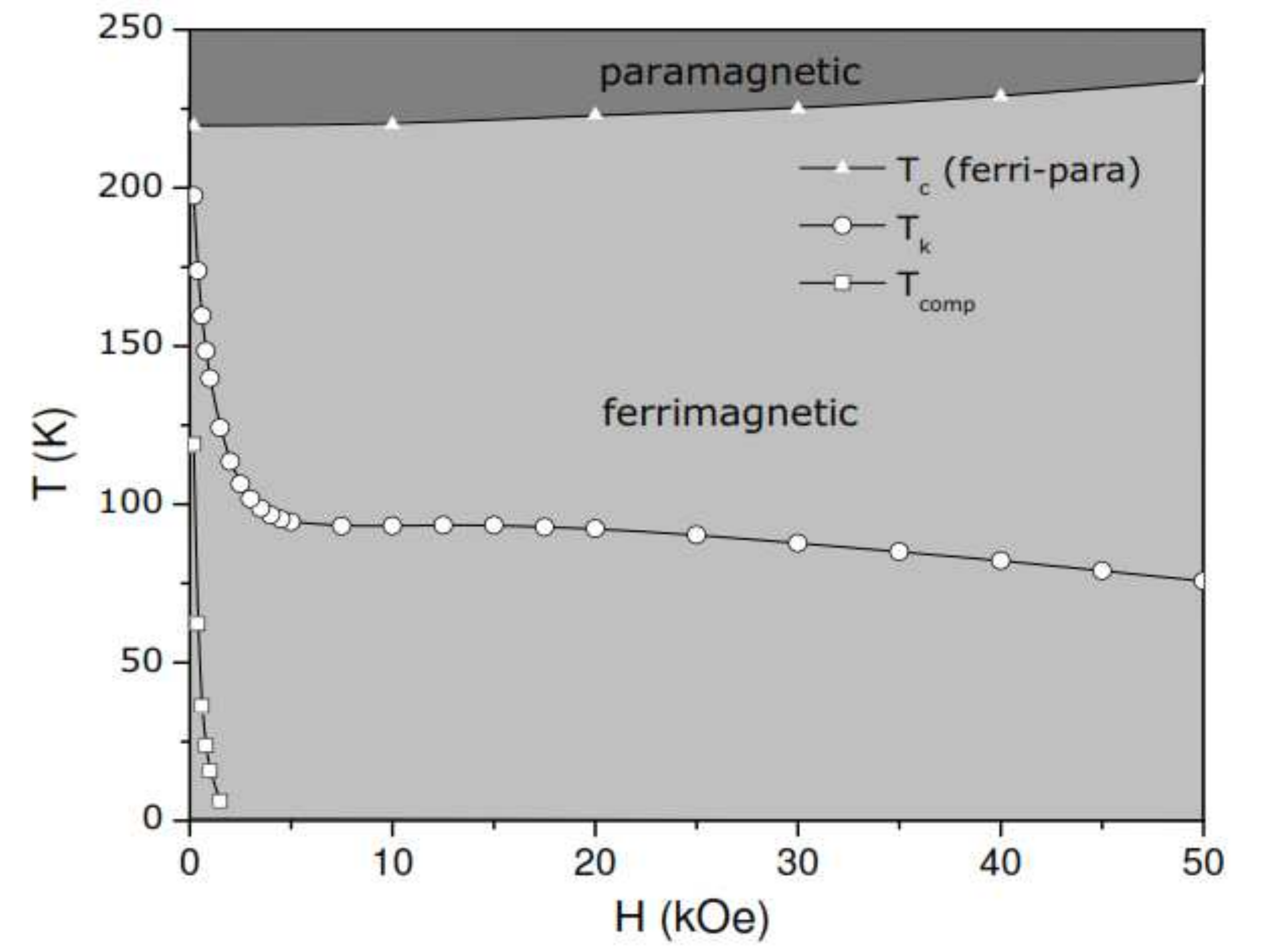}
\end{center}
\caption{Experimental magnetic phase diagram for $\text{Nd}_{0.5}\text{Sr}_{0.5}\text{CoO}_3$.}\label{pd}
\end{figure}

Based on the previous results and analysis, it is possible to construct a phase diagram of $\text{Nd}_{0.5}\text{Sr}_{0.5}\text{CoO}_3$, as depicted on Fig. \ref{pd}. This material has a paramagnetic phase above 215 K, below which the system changes to a ferrimagnetic ordering, with the Nd sub-lattice magnetization in opposition to ones associated with both Co sub-lattices (Co$^{3+}$ and Co$^{4+}$). Due to the difference between the intra-exchange interaction (Co-Co interaction is much stronger than Nd-Nd interaction), the Co sub-lattice dominated the magnetization behavior close to the Curie temperature. At low temperatures, however the Nd sub-lattice controls the total magnetization behavior. $T_k$ in phase diagram is defined as the temperature at which the total magnetization becomes maximum. Finally, at even lower temperatures and low field, due to the ferrimagnetic arrangement, it is convenient to specify a compensation temperature $T_{comp}$, which is also shown in the phase diagram.

X-ray diffraction using synchrotron radiation with Rietveld refinement shows that the $\text{Nd}_{0.5}\text{Sr}_{0.5}\text{CoO}_3$ sample has a single phase, and O1-Co (apical distance) is, on average, larger than O2-Co (equatorial distance). The Curie constant indicates that the system in on intermediate spin state for both, Co$^{3+}$ and Co$^{4+}$, without orbital contribution and, consequently, a Land\'e factor $g=2$, in accordance with the theoretical model of Ref. \cite{reis2017spin}. Magnetization data shows a critical temperature $T_c=215$ K and a broad maximum at $T_k=80$ K. While previous published reports \cite{apl/101/4/10.1063/1.4738889} associate these features to a ferromagnetic transition and the onset below which the Nd and Co sub-lattices become antiparallel. We show that those temperatures are, respectively, the ferrimagnetic transition and the onset below which Nd sub-lattice (negative) magnetization rules the total magnetization. In addition, $T_k$ is the temperature in which (i) the temperature derivative of Nd and Co magnetizations become equal to each other and (ii) the field and zero-field magnetic entropy cross each other. The magnetic entropy change was also experimentally calculated and these results are consistent with  ferrimagnetic system. Finally, from all these results, the magnetic phase diagram for $\text{Nd}_{0.5}\text{Sr}_{0.5}\text{CoO}_3$ cobaltite could be drawn.


	\chapter{Magnetic and structural investigations on nanostructured La$_{0.6}$Sr$_{0.4}$MnO$_3$}\label{manga10}

This Chapter presents the structural and magnetic properties of La$_{0.6}$Sr$_{0.4}$MnO$_3$ nanoparticles with sizes distribution ranging from 21 to 106 nm. They have been prepared using the sol-gel method (see Chapter \ref{capitulo4}). The reduction of the nanoparticles size tends to broaden the paramagnetic to ferromagnetic transition, as well as to promote magnetic hysteresis and a remarkable change on the magnetic saturation. The XRD measurements were carried out at {\it Laborat\'orio de Raios X da UNiversidade Federal Fluminense},  the transmission electron microscopy (TEM) technique was carried out at {\it Inmetro}, and the magnetic measurements were carried out using a commercial superconducting quantum interference device (SQUID) at {\it Unicamp}.

\section{Considerations on manganites nanoparticles}

The physics of nano-materials has been an exciting research field since the 90s.  The magnetic properties of these systems have attracted special attention to fundamental and technological point view \cite{zhang1998magnetization}. They find applications on medicine  \cite{kaman2009silica, chono2008efficient, hu2008facile}, catalysis processes \cite{liu2008highly} and magnetic refrigeration \cite{mathew2011tuning}, for instance. When the size of the magnetic nanoparticles decreases to a few nanometers, these materials exhibit different magnetic behaviors, such as surface spin-glass, superparamagnetism, large coercivities, low-field saturation magnetization, and low Curie temperature, as compared to their bulk counterparts \cite{zhu2001surface, dey2008enhanced, katiyar2001magnetic}. The magnetic behavior of the particle's surface usually differs from that of its to the core, because of the distinct atomic coordination, compositional gradients, concentration and nature of the defects present in both regions \cite{westman2008surface, kodama1999magnetic}. The surface gives rise to a large magnetocrystalline anisotropy due to its low symmetry and may induces different magnetic arrangements due to weakening of exchange interaction \cite{batlle2002finite}.

The magnetic properties of manganite nanoparticles synthesized by sol-gel method present novel features compared with the bulk materials, prepared by conventional solid-state reaction method \cite{mathew2011tuning}. When the particle size is reduced to few nanometers, are observed effects as the broadening of the paramagnetic to ferromagnetic transition, decreasing of the saturation magnetization value, increase the magnetic hysteresis \cite{xi2012surface} and appearance of superparamagnetic (SPM) behavior to very low particle size (< 17 nm) \cite{rostamnejadi2012magnetic}. Generally, the magnetization of ferromagnetic nanoparticles is lower that bulk samples, this is explained usually by the consideration of a not magnetic layer (in the shell) around to magnetic nucleus (core) \cite{xi2012surface}.

In the following sections, the structural and magnetic properties of La$_{0.6}$Sr$_{0.4}$MnO$_3$ nanoparticles produced by sol-gel method will be discuss. X-ray diffraction and transmission electron microscope measurements provide the average diameter of nanoparticles ranges from 21 nm to 106 nm. Magnetic measurements showed a dependence of the nanoparticle size magnetic saturation and transition.

\section{Crystal structure and morphology}

\begin{figure}[h]
\center
\includegraphics[width=14cm]{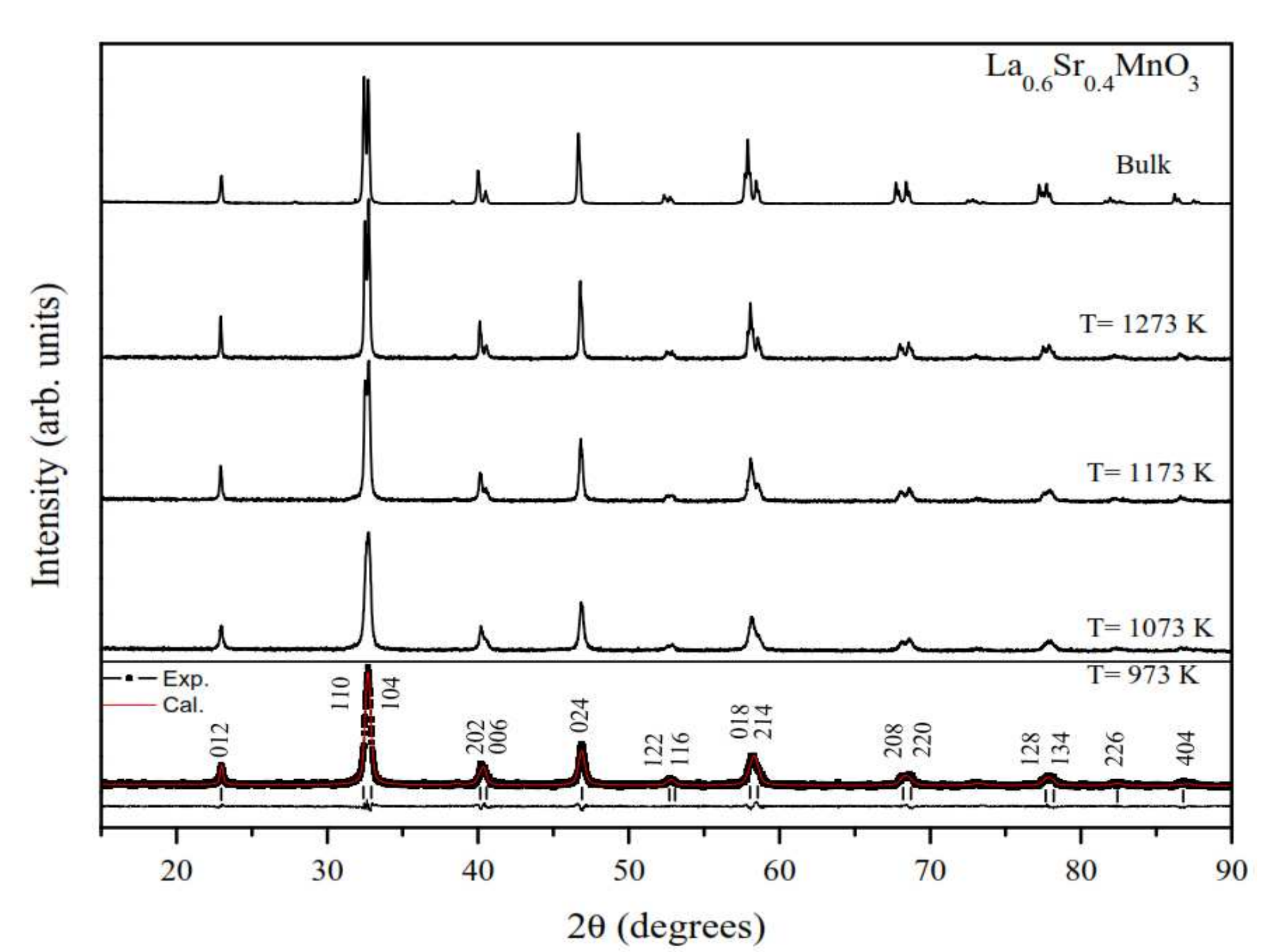}
\caption[Powder diffractograms for the La$_{0.6}$Sr$_{0.4}$MnO$_3$ nanoparticles]{Powder diffractograms for the La$_{0.6}$Sr$_{0.4}$MnO$_3$ nanoparticles, calcinated at several temperatures. (Bottom) Rietveld refinement for sample calcined at 700 $^{o}$C.}\label{difratograms}
\end{figure}

X-ray diffraction data  at room temperature and pressure of the samples submitted to calcination temperatures of 973 K, 1073 K, 1173 K and 1273 K compared with bulk sample is shown in Fig. \ref{difratograms}. Rietveld method using the FullProf software \cite{fullprof} was applied in all La$_{0.6}$Sr$_{0.4}$MnO$_3$. We found that crystallized in perovskite structure space group R$\bar{3}$c (group 167) in a rhombohedral crystal system, as reported by Shen \cite{shen2001internal}. One example of fit is shown in the bottom of Fig.\ref{difratograms}, for sample calcined at 973 K. The crystallographic parameters and reliability factors obtained from refinement can be seen in Table \ref{crystaldata} (only presented for nanoparticles).  The convergence factors $R_p$, $R_{wp}$ and $R_F$ obtained from Rietveld analysis point to the good quality of the refinement.

From XRD data, it was possible to estimate the average particle size $D$ using the Scherrer equation, which we found the average diameter of nanoparticles using the size broadening of the peaks from the relation\cite{dinnebier2008powder},

\begin{equation}
D = k \lambda / \beta_{L}\cos \theta
\label{scherrer}
\end{equation}

\begin{figure}[h]
\center
\includegraphics[width=12cm]{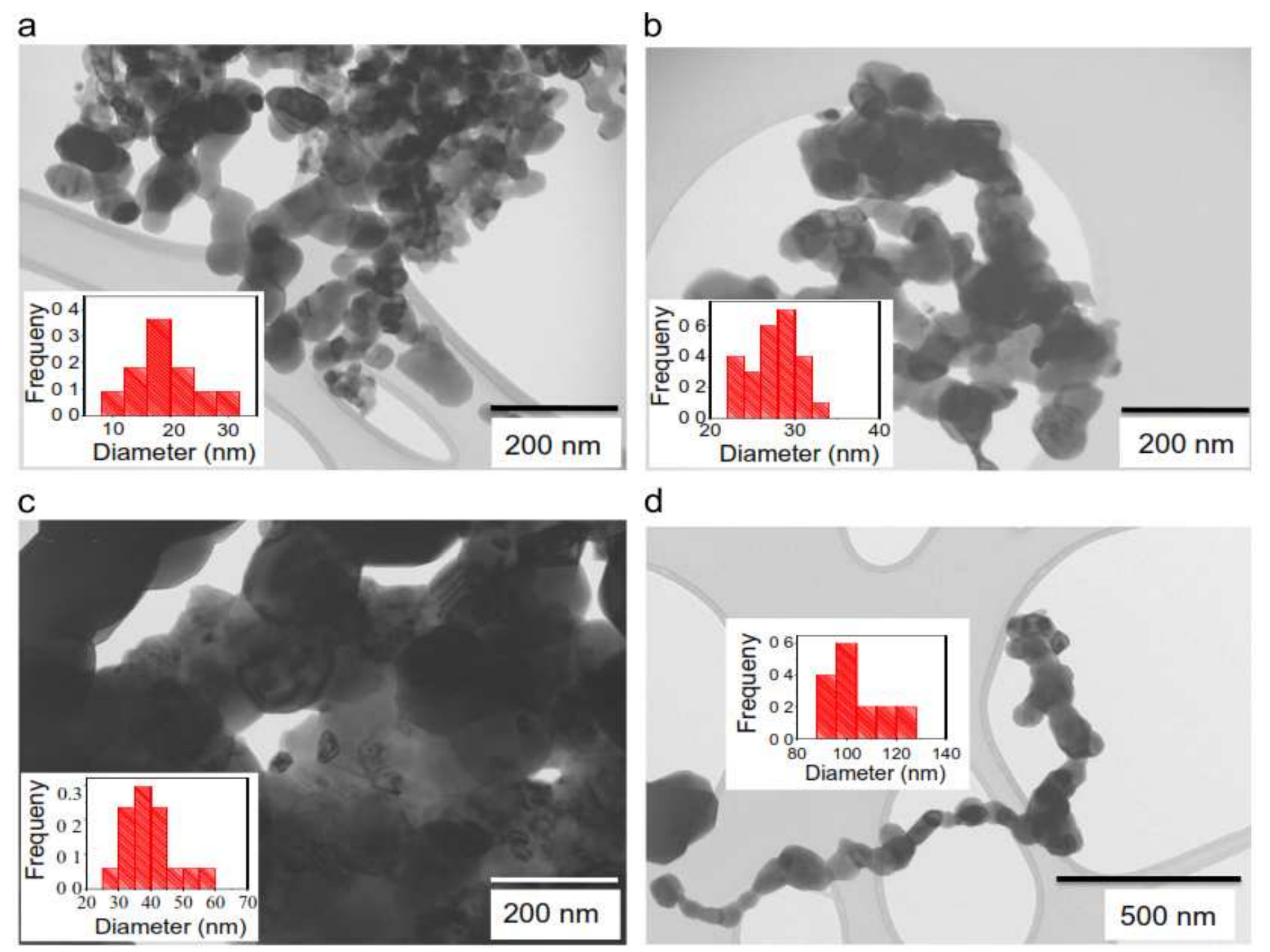}
\caption[TEM micrograph of La$_{0.6}$Sr$_{0.4}$MnO$_3$ samples]{TEM micrograph of La$_{0.6}$Sr$_{0.4}$MnO$_3$ samples annealed at temperatures of (a) 1273 K, (b) 1173 K, (c) 1073 K and (d) 973 K. TEM was carried out using a FEI-LaB$_6$-TECNAI microscope, held at Inmetro, operated at 120 kV, and equipped with a SIS MEGAVIEW III CCD camera. \label{microscopia}}
\end{figure}

\begin{table}
\caption[Refined crystallographic data]{Refined crystallographic data and reliability factors for the La$_{0.6}$Sr$_{0.4}$MnO$_3$ nanoparticles. The nanoparticles size $D$ are also presented.\label{crystaldata}}
    \center
    \begin{tabular}{l|c|c|c|c}
     \multicolumn{2}{l}{Parameter} & \multicolumn{2}{l}{Calcination Temperature}\\\hline\hline
                        & 973 K    & 1073 K     & 1173 K     & 1273 K   \\\hline\hline
    $a$ (\AA{})         & 5.5005        & 5.4999          & 5.5028          & 5.5148  \\
    $c$ (\AA{})         & 13.3575       & 13.3367         & 13.3441         & 13.3586 \\
    O (x)               & 0.5493        & 0.5529          & 0.5415          & 0.5615  \\
    Volume (\AA{}$^{3}$)& 350.00        & 349.37          & 349.94          & 351.85  \\
    R$_{p}$ (\%)        & 10.2          & 11.0            & 11.4            & 9.3     \\
    R$_{wp}$ (\%)       & 13.2          & 14.5            & 14.6            & 12.1    \\
    R$_{F}$ (\%)        & 11.0          & 12.0            & 12.1            & 9.5     \\
    $\chi^{2}$          & 2.7           & 2.8             & 3.0             & 2.5    \\\hline\hline
    $D (nm)$            & 21 $\pm$ 2  & 27  $\pm$ 3     & 41  $\pm$ 4     & 106$\pm$ 10  \\\hline\hline
            \end{tabular}
\end{table}

\noindent where $k$ is a dimensionless constant (close to one), $\lambda$ is the X-ray wavelength, $\theta$ is the diffraction angle for the most intense peaks and $\beta_{L}$ = $\sqrt{U \tan^{2} \theta + V \tan \theta + W}$  is related to the full width at half maximum (FWHM) of the peaks. The broadening of the peaks indicates the formation of a nanocrystalline compound \cite{dey2006effect,lu2008magnetocaloric}. When the calcination temperature arises, the diffraction peaks become sharper, resulting in larger particle sizes \cite{dey2006effect}. The particle sizes estimated from powder X-ray data are shown in table \ref{crystaldata}. In order to confirm the average grain size, we used the TEM images showed in Fig.\ref{microscopia}. It is possible to verify that the average nanograin size decreases from approximately 100 nm down to 20 nm for calcination temperatures from 1273 down to 973 K. These results show a very good agreement between the observed particle size from TEM images and the one from X-ray diffraction analysis. The TEM images also show that the particles (nanograins) exhibit a spherical morphology and that are slightly connected to each other. 

\section{Magnetic properties}

\begin{figure}
\begin{center}
\includegraphics[width=15cm]{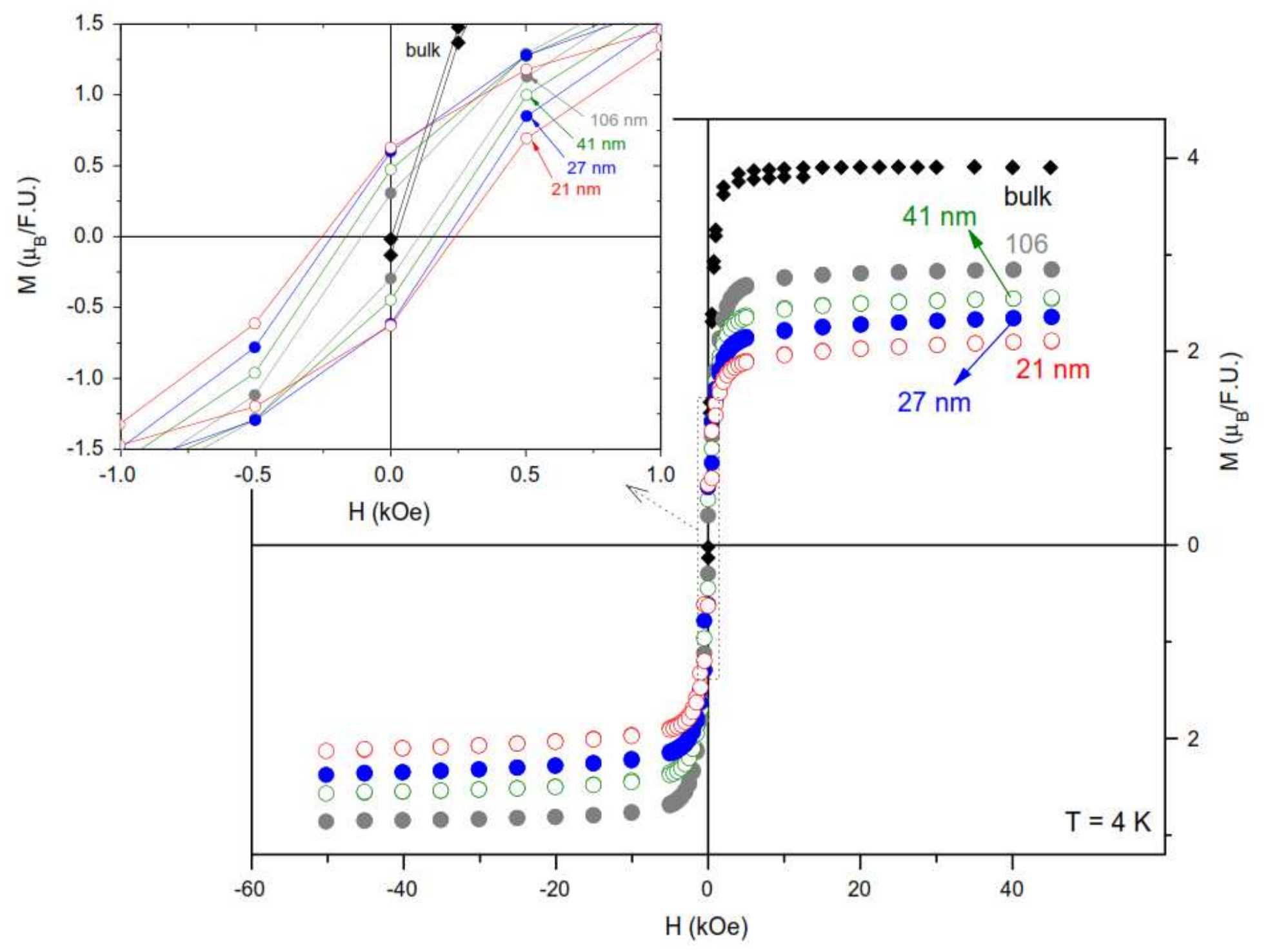}
\end{center}
\caption[Magnetic moment per formula unitary as a function of external magnetic field]{Magnetic moment per formula unitary  as a function of external magnetic field. Note that the magnetic saturation strongly depends on the nanoparticle size. Inset: magnification at the origin, for low values of magnetic field. The bulk sample has no hysteresis, while, on the contrary, hysteresis is observed for all nanostructured manganites.\label{MHpdf}}
\end{figure}

Magnetic moment per formula unitary as a function of applied magnetic field, at 4 K, for all studied samples are presented in Fig. \ref{MHpdf}. We found that the magnetic moment of saturation is strongly depends on the nanoparticle size. The bulk sample has no hysteresis, while, the nanoparticle samples presents increased of remanence with the decrease of the nanoparticle size (see inset of Fig.\ref{MHpdf}). This is a common result \cite{xi2012surface} and it is due to the large surface anisotropy. On the other hand, we can estimate the Land\'{e} factor considering the saturation as $M_{S}=g(0.6\times2+0.4\times\frac{3}{2}) \mu_B$. The bulk sample has a magnetic saturation close to 3.9 $\mu_B$, and the Land\'{e} factor as $g=2.17$, which is a reasonable result for manganites when the quench of the angular moments is not perfect and a really minor orbital contribution takes place \cite{marioIntro}. For manganite nanoparticles, the magnetic saturation ranges from 2.9 $\mu_B$ with $g=1.6$(106 nm), down to 2.1 $\mu_B$ with $g=1.16$ (21 nm), which are non-acceptable Land\'{e} factor for manganites without orbital contribution. Thus, other mechanisms are reducing the magnetic saturation of those nanoparticles by almost half. See Fig. \ref{MHpdf} for further details.

\begin{figure}
\center
\includegraphics[width=12cm]{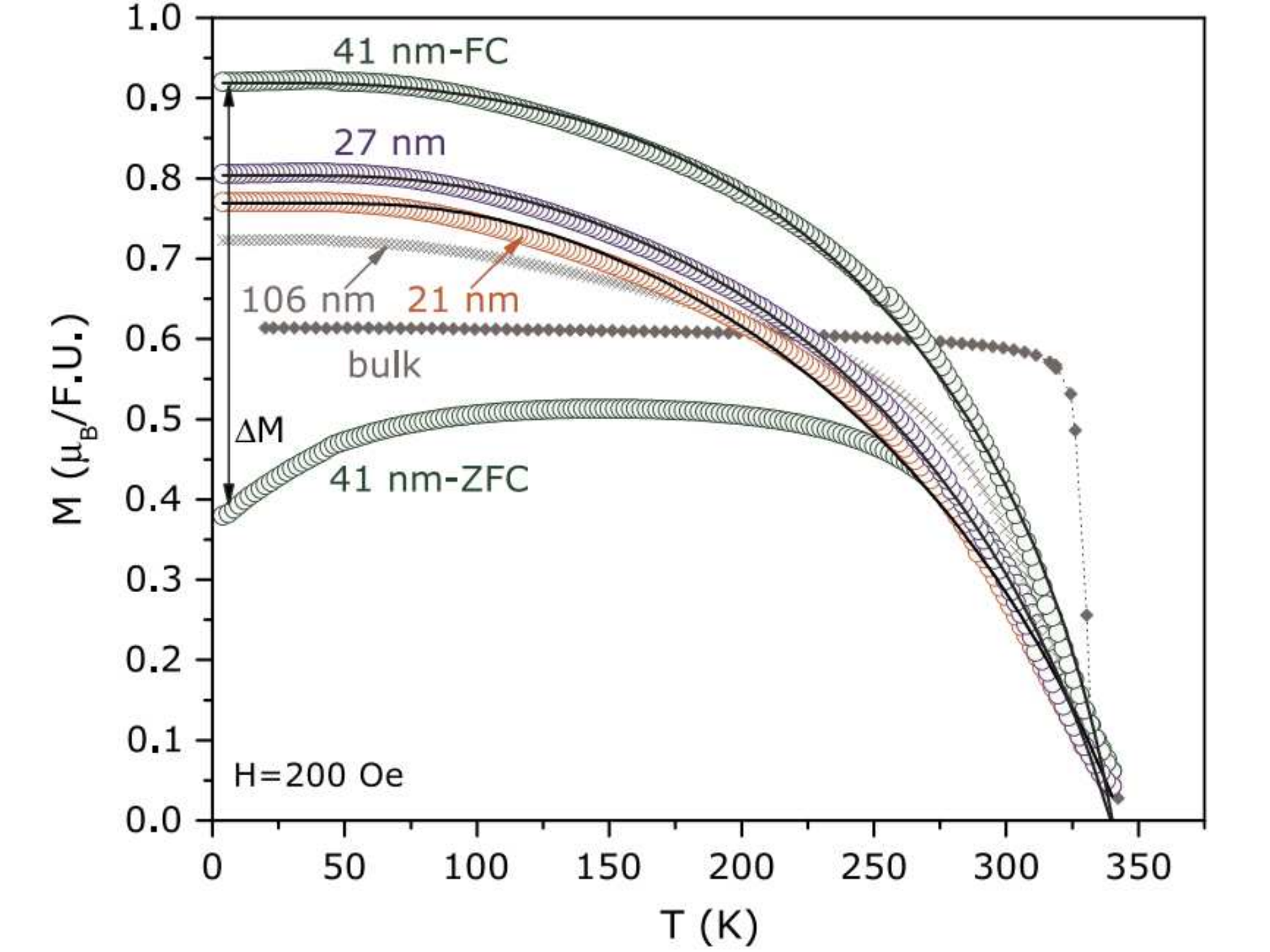}
\caption[Magnetic moment as a function of nanoparticle size]{Magnetic moment as a function of nanoparticle size. Note the strong dependence of the magnetic behavior with nanoparticle size. Circles are experimental data and solid line are fittings of the model proposed on Ref. \cite{andrade2014magnetic}. Dotted line on the bulk sample is only a guide to the eyes. For the sake of clearness, only one ZFC curve is presented, but the irreversibility $\Delta M$ at 4 K is presented in Fig. \ref{MTpdf-nova} for all samples.  \label{MTpdf}}
\end{figure}

\begin{figure}
\center
\includegraphics[width=12cm]{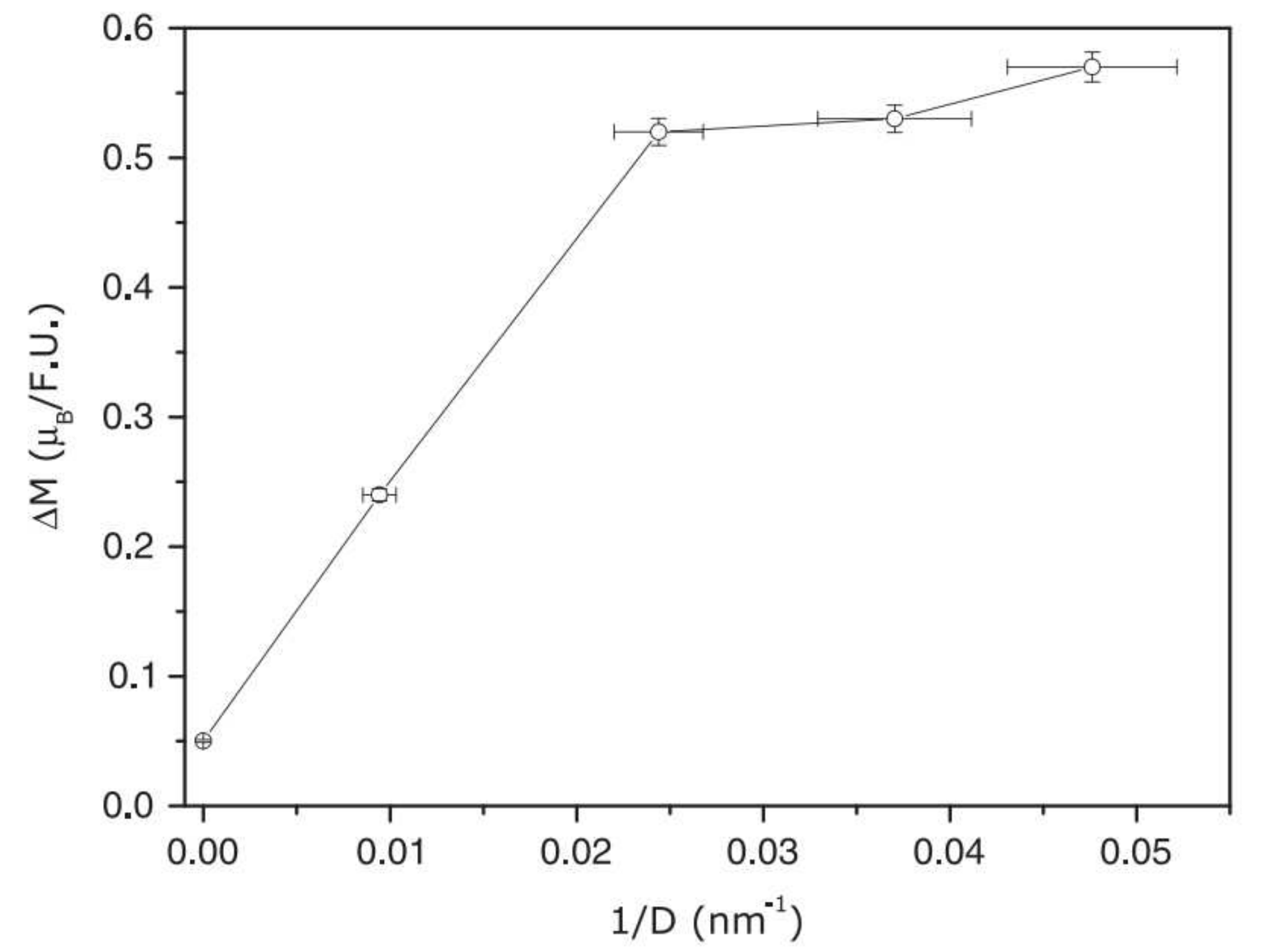}
\caption[Irreversibility magnetization as a function of inverse nanoparticle diameter]{Irreversibility between ZFC and FC magnetization at $4 K$ as a function of inverse nanoparticle diameter $D$. \label{MTpdf-nova}}
\end{figure}

ZFC and FC magnetization measurements the all samples are shown in Fig. \ref{MTpdf}. It is possible observe the influenced of the surface on the magnetization curves, where a large irreversibility $\Delta M$ at $4 K$ between the ZFC and FC magnetization can be seen. Fig. \ref{MTpdf-nova} shows $\Delta M$  decreases as the nanoparticles size increases (in order to include the bulk value we considered $1/D$, instead of $D$). The bulk sample shows no irreversibility in the $M(T)$ curves, as well as hysteresis in the $M(H)$ isotherms.

To further understand the magnetic behavior of those manganite nanoparticles, in the Ref. \cite{andrade2014magnetic} is considered a simple model, based on a ferromagnetic core (since the bulk sample is ferromagnetic), and a ferrimagnetic shell (the only way to justify those values of magnetic saturation for these manganites). Is based in that, the core has two sub-lattices (Mn$^{3+}$ with $s=2$ and 3/5 of molar fraction; and  Mn$^{4+}$ with $s=3/2$ and 2/5 of molar fraction) aligned in a parallel fashion; and the shell has the same kind of two sub-lattices, however, aligned in an antiparallel way.  This model considers single-domain particles (and thus works for the smaller nanoparticles); and increasing the nanoparticle size, magnetic domains appear and the model loses validity. Indeed, $M(T)$ of 106 nm nanoparticle does not follow the tendency of the other nanoparticles and starts to approach to the bulk behavior, that, on its turn, has magnetic domains. Some works \cite{xi2012surface} also propose that the single/multi-domain crossover lies close to 70 nm.

\section{Concluding remarks on La$_{0.6}$Sr$_{0.4}$MnO$_3$}

From standard sol-gel techniques we produced La$_{0.6}$Sr$_{0.4}$MnO$_3$ nanoparticles by controlling the annealing temperature. From the X-ray diffraction and Scherrer equation we obtained the nanoparticle size that ranges from 21 nm up to 106 nm; and these values were confirmed from transmission electron microscopy. Magnetization measurements show that those nanoparticles have a remarkable different behavior compared with bulk sample.

\chapter{General conclusions}\label{final}

We synthesized several intermetallic alloys by arc melting furnace in order to explore the structural and magnetic properties of YNi$_{4-x}$Co$_x$B alloys and X$_2$YZ Heusler compounds, obtaining the following conclusions:

\begin{itemize}
\item[$\bullet$] Intermetallic alloys with B

By XRD, we found that the YNi$_{4-x}$Co$_x$B samples crystallize in CeCo$_4$B type structure (space group P6/mmm), and the lattice parameters of these compounds change almost linearly as a function of Co content. These structures contain two Wyckoff sites ($2c$ and $6i$) for $3d$ metal transition in anisotropy competitions by Co substitution. Thus, we developed a preferential and statistical occupation models for Co ions into the $3d$ sub-lattice of the YNi$_{4-x}$Co$_x$B samples. The experimental data of magnetization as a function of magnetic field and temperature indicate that our model of preferential occupation fits the physical mechanism that rules the $3d$ magnetic anisotropies.

\item[$\bullet$] Heusler alloys

- For Co$_2$FeSi, we performed AXRD measurements and M\"ossbauer spectroscopy in order to found the atomic disorder in this sample. The annealing time in the sample preparation induces different atomic disorder degrees in this compound. Finally, DFT calculations demonstrated that the  atomic disorder affects the half-metallic properties of Co$_2$FeSi.

- Fe$_2$MnSi$_{1-x}$Ga$_{x}$ crystallize in the cubic $L2_1$ structure and the lattice parameter $a$ increases linearly with increasing Ga content. Through magnetization as a function of magnetic field and temperature, we found that  the valence electron number ($N_V$) plays an important role on the magnetic parameters of these compounds, such as Curie temperature, saturation magnetization and magnetic entropy change. Increasing $N_V$ (equivalent to increase the Si content from Fe$_2$MnSi$_{0.5}$Ga$_{0.5}$ compound), leads to a linear increase of those magnetic quantities. In addition, we also propose to substitute Ga by a group 15 element, like P; and we expect that Fe$_2$MnSi$_{0.56}$P$_{0.44}$ would have its Curie temperature close to 300 K, with an enhanced magnetocaloric effect.

\end{itemize}

Another research line which we developed were the studies on the perovskite oxides, and here are the concluding remarks:

\begin{itemize}

\item[$\circ$] $\text{Nd}_{0.5}\text{Sr}_{0.5}\text{CoO}_3$ cobaltite

We synthesized a single phase sample of $\text{Nd}_{0.5}\text{Sr}_{0.5}\text{CoO}_3$ by sol-gel method with nitrate reagents, and it crystallized in an orthorhombic structure with space group Imma, critical temperature $T_C=215$ K and a broad maximum of magnetization at $T_k=80$ K. Through of the magnetization as a function of magnetic field and temperature, we found that these temperatures are the ferrimagnetic transition ($T_C$) and the temperature ($T_k$) below which Nd sub-lattice (negative) magnetization rules the total magnetization. Also, the Curie constant indicates that the system is in an intermediate spin state for both Co$^{3+}$ and Co$^{4+}$. 

\item[$\circ$] La$_{0.6}$Sr$_{0.4}$MnO$_3$ manganites

From standard sol-gel method with oxides reagents, we produced La$_{0.6}$Sr$_{0.4}$MnO$_3$ nanoparticles  with sizes that range from 21 nm up to 106 nm,  due to different annealing temperatures.  These sizes were determinated by  XRD measurement and  transmission electron microscopy. Magnetization measurements show that those nanoparticles have a remarkably different behavior compared with bulk sample, due to increase of the ratio surface/volume. 

\end{itemize}

In general, we explored several magnetic phenomena, through synthesis, and study of  structural and magnetic properties of intermetallic alloys and perovskite oxides. From the intermetallic alloys, the phenomena studied include the competition of anisotropies in materials with $3d$ elements, the magnetocaloric effect and half-metallicity. From the perovskite oxides, we were able to study the spin state in the cobaltite  and finally observe the effect of nanostructured materials of manganite compounds.

\appendix 

\chapter{List of publications}

\section{ Associated with the thesis}

\begin{itemize}

\item[1] {\bf R. J. Caraballo Vivas}, D. L. Rocco, T. Costa-Soares, L. Caldeira, A. A. Coelho, M. S. Reis. {\it Competing anisotropies on $3d$ sub-lattice of YNi$_{4-x}$Co$_x$B compounds}. Journal of Applied Physics, v. 116, p. 063907, 2014.

\item[2] {\bf R. J. Caraballo Vivas}, N. R. Checca, J. C. G. Tedesco, N. M. Fortunato, J. N. Goncalves, R. D. Candela, A. A. Coelho, A. Magnus. G. Carvalho, M. S. Reis. {\it Experimental and theoretical evidences that atomic disorder suppress half-metallic porperties of Heusler alloys}. In redaction (2017).

\item[3] S. S. Pedro, {\bf R. J. Caraballo Vivas}, V. M. Andrade, C. Cruz, L. S. Paix\~ao, C. Contreras, T. Costa-Soares, L. Caldeira, A. A. Coelho, A. M. G. Carvalho, D. L. Rocco, M. S. Reis. {\it Effects of Ga substitution on the structural and magnetic properties of half metallic Fe2MnSi Heusler compound}. Journal of Applied Physics, v. 117, p. 013902, 2015.

\item[4] {\bf R.J. Caraballo Vivas}, S. S. Pedro, C. Cruz, J. C. G. Tedesco, A. A. Coelho, A. M. G. Carvalho, D. L. Rocco, M. S. Reis. {\it Experimental evidences of enhanced magnetocaloric properties at room temperature and half-metallicity on Fe$_2$MnSi-based Heusler alloys}. Materials Chemistry and Physics, v. 174, p. 23-27, 2016.

\item[5] M.S. Reis, D.L. Rocco, {\bf R.J. Caraballo Vivas}, B. Pimentel, N.R. Checca, R. Torr\~ao, L. Paix\~ao, A.M. dos Santos. {\it Spin state and magnetic ordering of half-doped Nd$_{0.5}$Sr$_{0.5}$CoO$_3$ cobaltite} Journal of Magnetism and Magnetic Materials , v. 422, p. 197-203, 2017.

\item[6] V. M. Andrade, {\bf R. J. Caraballo-Vivas}, T. Costa-Soares, S. S. Pedro, D. L. Rocco, M. S. Reis, A. P. C. Campos, A. A. Coelho. {\it Magnetic and structural investigations on La$_{0.6}$Sr$_{0.4}$MnO$_3$ nanostructured manganite: Evidence of a ferrimagnetic shell}. Journal of Solid State Chemistry, v. 219, p. 87-92, 2014.

\end{itemize}

\section{Not associated with the thesis}

\begin{itemize}

\item[7] R. Ribeiro-Palau, {\bf R. Caraballo}, P. Rogl, E. Bauer, I. Bonalde. {\it Strong-coupling BCS superconductivity in noncentrosymmetric BaPtSi$_3$: a low-temperature study}. Journal of Physics. Condensed Matter, v. 26, p. 235701, 2014.

\item[8] V. M. Andrade, {\bf R.J. Caraballo Vivas}, S. S. Pedro, J. C. G. Tedesco, A. L Rossi, A. A. Coelho, D. L. Rocco, M. S. Reis. {\it Magnetic and magnetocaloric properties of La$_{0.6}$Ca$_{0.4}$MnO$_3$ tunable by particle size and dimensionality}. Acta Materialia, v. 102, p. 49-55, 2016.

\item[9] V. M. Andrade, S. S. Pedro, {\bf R.J. Caraballo Vivas}, D. L. Rocco, M. S. Reis, A. P. C. Campos, A. A. Coelho, M. Escote, A. Zenatti, A. L Rossi. {\it Magnetocaloric functional properties of Sm$_{0.6}$Sr$_{0.4}$MnO$_3$ manganite due to advanced nanostructured morphology}. Materials Chemistry and Physics, v. 172, p. 20-25, 2016.

\item[10] J. C. G. Tedesco, S. S. Pedro, {\bf R.J. Caraballo Vivas}, C. Cruz, V. M. Andrade, A.M. dos Santos, A. M. G. Carvalho, M. Costa, P. Venezuela, D. L. Rocco, M. S. Reis. Journal of Physics. Condensed Matter, v. 28, p. 476002, 2016.

\end{itemize}

{\scriptsize
\bibliographystyle{unsrt}

\bibliography{cond-matter}}

\end{document}